\documentclass[pdflatex,aos,preprint]{imsart}
\RequirePackage[OT1]{fontenc}
\RequirePackage{amsthm,amsmath}
\RequirePackage{natbib}
\RequirePackage[colorlinks,citecolor=blue,urlcolor=blue]{hyperref}
\usepackage{amsfonts}
\usepackage{times}
\usepackage{graphicx}
\usepackage{mathrsfs}
\usepackage[scaled=0.90]{helvet}
\usepackage{mathtools}
\usepackage{enumitem}

\arxiv{arXiv:0000.0000}

\startlocaldefs
\graphicspath{{./figs/}}
\newtheoremstyle{mytheorem}{}{}{\itshape}{}{\scshape}{}{2mm}{}
\newtheoremstyle{mytheorem2}{}{}{}{}{\scshape}{}{2mm}{}
\theoremstyle{mytheorem}
\newtheorem{theorem}{Theorem}
\newtheorem{lemma}{Lemma}
\newtheorem{definition}{Definition}
\theoremstyle{mytheorem2}
\newtheorem{remark}{Remark}

\newcommand{\iid}{\stackrel{\text{iid}}{\sim}}
\newcommand{\Real}{\mathbb{R}}
\newcommand{\cas}{\buildrel \text{a.s.} \over \longrightarrow}
\newcommand{\casleft}{\buildrel \text{a.s.} \over \longleftarrow}
\newcommand{\cni}{\buildrel n \rightarrow \infty \over \longrightarrow}
\newcommand{\cmi}{\buildrel m \rightarrow \infty \over \longrightarrow}
\newcommand{\sX}{\textsf{X}}
\newcommand{\sP}{\textsf{P}}
\newcommand{\cd}{\buildrel d \over \rightarrow}
\newcommand{\cP}{\buildrel P \over \rightarrow}
\newcommand{\bbeta}{\boldsymbol{\beta}}
\newcommand{\bpsi}{\boldsymbol{\psi}}
\newcommand{\btheta}{\boldsymbol{\theta}}
\newcommand{\bw}{\boldsymbol{w}}
\newcommand{\bz}{\boldsymbol{z}}
\newcommand{\simplex}{\mathbb{S}}
\DeclareMathOperator{\Dir}{Dir}
\DeclareMathOperator{\Mult}{Mult}

\newcommand{\commentt}[1]{}

\newcommand{\itemEqstar}[1]{%
\begingroup%
\setlength{\abovedisplayskip}{0mm}%
\setlength{\belowdisplayskip}{0mm}%
\parbox[c]{\linewidth}{
\begin{equation*}
  #1
\end{equation*}}%
\endgroup}

\newcommand{\itemEq}[2]{%
\begingroup%
\setlength{\abovedisplayskip}{-2.1mm}%
\setlength{\belowdisplayskip}{0mm}%
\parbox[c]{\linewidth}{
\begin{equation} #1 #2 \end{equation}}%
\endgroup}

\DeclareMathOperator{\Var}{Var}
\DeclareMathOperator{\Cov}{Cov}
\DeclareMathOperator{\argmax}{arg\,max}

\newcounter{itemnum}

\numberwithin{equation}{section}
\newcommand{\pushright}[1]{\ifmeasuring@#1\else\omit\hfill$\displaystyle#1$\fi\ignorespaces}
\endlocaldefs

\begin{document}
\begin{frontmatter}
  \title{An MCMC Approach to Empirical Bayes Inference and Bayesian
  Sensitivity Analysis via Empirical Processes}

  \runtitle{MCMC Approach to Empirical Bayes Inference}

  \begin{aug}
    \author{\fnms{Hani} \snm{Doss} \thanksref{t1}
    \ead[label=e1]{doss@stat.ufl.edu}} \and
    \author{\fnms{Yeonhee} \snm{Park}
    \ead[label=e2]{ypark3@mdanderson.org}}

    \thankstext{t1}{Supported by NSF Grant DMS-11-06395 and NIH
    grant P30 AG028740}

    \runauthor{H. Doss and Y. Park}

    \affiliation{University of Florida and MD Anderson Cancer
    Center}

    \address{Department of Statistics \\
    University of Florida \\
    Gainesville, FL 32611 \\ 
    USA \\
    \printead{e1}}
    \address{Department of Biostatistics \\
    MD Anderson Cancer Center \\
    Houston, TX 77030 \\
    USA \\
    \printead{e2}}
  \end{aug}

  \begin{abstract}
    Consider a Bayesian situation in which we observe $Y \sim
    p_{\theta}$, where $\theta \in \Theta$, and we have a family $\{
    \nu_h, \, h \in \mathcal{H} \}$ of potential prior distributions
    on $\Theta$.  Let $g$ be a real-valued function of $\theta$, and
    let $I_g(h)$ be the posterior expectation of $g(\theta)$ when
    the prior is $\nu_h$.  We are interested in two problems: (i)
    selecting a particular value of $h$, and (ii) estimating the
    family of posterior expectations $\{ I_g(h), \, h \in
    \mathcal{H} \}$.  Let $m_y(h)$ be the marginal likelihood of the
    hyperparameter $h$: $m_y(h) = \int p_{\theta}(y) \,
    \nu_h(d\theta)$.  The empirical Bayes estimate of $h$ is, by
    definition, the value of $h$ that maximizes $m_y(h)$.  It turns
    out that it is typically possible to use Markov chain Monte
    Carlo to form point estimates for $m_y(h)$ and $I_g(h)$ for each
    individual $h$ in a continuum, and also confidence intervals for
    $m_y(h)$ and $I_g(h)$ that are valid pointwise.  However, we are
    interested in forming estimates, with confidence statements, of
    the entire families of integrals $\{ m_y(h), \, h \in
    \mathcal{H} \}$ and $\{ I_g(h), \, h \in \mathcal{H} \}$: we
    need estimates of the first family in order to carry out
    empirical Bayes inference, and we need estimates of the second
    family in order to do Bayesian sensitivity analysis.  We
    establish strong consistency and functional central limit
    theorems for estimates of these families by using tools from
    empirical process theory.  We give two applications, one to
    Latent Dirichlet Allocation, which is used in topic modelling,
    and the other is to a model for Bayesian variable selection in
    linear regression.
  \end{abstract}


\end{frontmatter}

\section{Introduction}
\label{sec:intro}
This paper is concerned with two related problems.  In the first,
there is a function $B \colon \mathcal{H} \rightarrow \Real$, where
$\mathcal{H}$ is a subset of some Euclidean space, and we wish to
obtain confidence sets for $\argmax_{h \in \mathcal{H}} B(h)$.  For
each $h$, the expression for $B(h)$ is analytically intractable;
however, we have at our disposal a family of functions $\{ f_h, \, h
\in \mathcal{H} \}$ and a sequence of random variables $\theta_1,
\ldots, \theta_n$ (these are iid or the initial segment of an
ergodic Markov chain) such that the random function $B_n(h)
\coloneqq (1/n) \sum_{i=1}^n f_h(\theta_i)$ satisfies $B_n(h) \cas
B(h)$ for each $h$.  We are interested in how we can use $B_n$ to
form both a point estimate and a confidence set for $\argmax_{h \in
\mathcal{H}} B(h)$.

This problem appears in empirical Bayes analysis and under many
forms in likelihood inference.  In empirical Bayes analysis, the
application that is the focus of this paper, it arises as follows.
Suppose we are in a standard Bayesian situation in which we observe
a data vector $Y$ whose distribution is $P_{\theta}$ (with density
$p_{\theta}$ with respect to some dominating measure) for some
$\theta \in \Theta$.  We have a family of potential prior densities
$\{ \nu_h, \, h \in \mathcal{H} \}$, and because the hyperparameter
$h$ can have a great impact on subsequent inference, we wish to
choose it carefully.  Selection of $h$ is often guided by the
marginal likelihood of the data under the prior $\nu_h$, given by
\begin{equation}
  \label{eq:ml}
  m_y(h) = \int p_{\theta}(y) \nu_h(\theta) \, d\theta, \qquad h \in
  \mathcal{H}.
\end{equation}
By definition, the empirical Bayes choice of $h$ is $\argmax_h
m_y(h)$.  Unfortunately, analytic calculation of $m_y(h)$ is not
feasible except for a few textbook examples, and estimation of
$m_y(h)$ via Monte Carlo is notoriously difficult---for example, the
``harmonic mean estimator'' introduced by \citet{NewtonRaftery:1994}
typically converges at a rate which is much slower than $n^{1/2}$
\citep{WolpertSchmidler:2012}.

It is very interesting to note that if $c$ is a constant, then the
information regarding $h$ given by the two functions $m_y(h)$ and $c
m_y(h)$ is the same: the same value of $h$ maximizes both functions,
and the second derivative matrices of the logarithm of these two
functions are identical.  In particular, the Hessians of the
logarithm of these two functions at the maximum (i.e.\ the observed
Fisher information) are the same and, therefore, the standard point
estimates and confidence regions based on $m_y(h)$ and $c m_y(h)$
are identical.  This is a very useful observation because it turns
out that it is usually easy to estimate the entire family $\{ c
m_y(h), \, h \in \mathcal{H} \}$ for a suitable choice of $c$.
Indeed, for any $h \in \mathcal{H}$, let $\nu_{h,y}$ denote the
posterior corresponding to $\nu_h$, let $h_1$ be fixed but
arbitrary, and suppose that $\theta_1, \ldots, \theta_n$ are either
independent and identically distributed according to the posterior
$\nu_{h_1,y}$, or are the initial segment an ergodic Markov chain
with invariant distribution $\nu_{h_1,y}$.  Let $\ell_y (\theta) =
p_{\theta}(y)$ be the likelihood function.  Note that $m_y(h)$ given
by~\eqref{eq:ml} is the normalizing constant in the statement ``the
posterior is proportional to likelihood times the prior,'' i.e.
\begin{equation}
  \label{eq:post}
  \nu_{h,y}(\theta) = \ell_y (\theta) \nu_h(\theta) / m_y(h).
\end{equation}
We have
\begin{equation}
  \label{eq:main-conv}
  \begin{split}
  \frac{1}{n} \sum_{i=1}^n \frac{\nu_h(\theta_i)} {\nu_{h_1}
  (\theta_i)} \cas & \int \frac{\nu_h (\theta)} {\nu_{h_1} (\theta)}
  \nu_{h_1,y} (\theta) \, d\theta \\  &= \frac{m_y(h)} {m_y(h_1)} \int
  \frac{\nu_{h,y} (\theta)} {\nu_{h_1,y} (\theta)} \, \nu_{h_1,y}
  (\theta) \, d\theta   = \frac{m_y(h)} {m_y(h_1)},
  \end{split}
\end{equation}
in which the first equality follows from~\eqref{eq:post} and
cancellation of the likelihood.  Let $B(h) = m_y(h) / m_y(h_1)$.
Since $m_y(h_1)$ is a fixed constant, as noted above, the two
functions $m_y(h)$ and $B(h)$ give exactly the same information
about $h$.  If we let $f_h = \nu_h / \nu_{h_1}$, then $B_n(h) =
(1/n) \sum_{i=1}^n (\nu_h(\theta_i) / \nu_{h_1}(\theta_i))$---this
quantity is computable, since it involves only the priors and not
the posteriors---so we have precisely the situation discussed in the
first paragraph of this paper.  Other examples of this situation
arising in frequentist inference, and in particular in missing data
models, are given in \citet{SungGeyer:2007} and
\citet{DossTan:2014}.

In Bayesian applications it is rare that Monte Carlo estimates of
posterior quantities can be based on iid samples; in the vast
majority of cases they are based on Markov chain samples, and that
is the case that is the focus of this paper.  We show that under
suitable regularity conditions,
\begin{equation}
  \label{eq:conv-as-argmax}
  \argmax_h B_n(h) \cas \argmax_h B(h)
\end{equation}
and
\begin{equation}
  \label{eq:conv-d-argmax}
  n^{1/2} \bigl( \argmax_h B_n(h) - \argmax_h B(h) \bigr) \cd
  \mathcal{N}(0, \Sigma),
\end{equation}
where $\Sigma$ can be estimated consistently.  Now, in general,
almost sure convergence of $B_n(h)$ to $B(h)$ pointwise is not
enough to imply that $\argmax_h B_n(h)$ converges to $\argmax_h
B(h)$ under any mode of convergence, and in fact it is trivial to
construct a counterexample in which $g_n$ and $g$ are deterministic
functions defined on $[0, 1]$, $g_n(h) \cni g(h)$ for every $h \in
[0, 1]$, but $\argmax_h g_n(h)$ does not converge to $\argmax_h
g(h)$.  To obtain results~\eqref{eq:conv-as-argmax}
and~\eqref{eq:conv-d-argmax} above, some uniformity in the
convergence is needed.  We establish the necessary uniform
convergence and show that~\eqref{eq:conv-as-argmax}
and~\eqref{eq:conv-d-argmax} are true under certain regularity
conditions on the sequence $\theta_1, \theta_2, \ldots$, the
functions $f_h$, and the function $B$.
Result~\eqref{eq:conv-d-argmax} enables us to obtain confidence sets
for $\argmax_h B(h)$.

The second problem we are interested in pertains to the Bayesian
framework discussed earlier and is described as follows.  Suppose
that $g$ is a real-valued function of $\theta$, and consider $I_g
(h) = \int g(\theta) \nu_{h,y}(\theta) \, d\theta$, the posterior
expectation of $g(\theta)$ given $Y = y$, when the prior is $\nu_h$.
Suppose that $h_1 \in \mathcal{H}$ is fixed but arbitrary, and that
$\theta_1, \theta_2, \ldots$ is an ergodic Markov chain with
invariant distribution $\nu_{h_1,y}$.  A very interesting and
well-known fact, which we review in Section~\ref{sec:pe}, is that
for any $h \in \mathcal{H}$, if we define
\begin{equation*}
  w_i^{(h)} = \frac{[\nu_h(\theta_i) / \nu_{h_1}(\theta_i)]}
  {\sum_{l=1}^n [\nu_h(\theta_l) / \nu_{h_1}(\theta_l)]},
\end{equation*}
then
\begin{equation}
  \label{eq:ihat-orig}
  \hat{I}_g (h) = \sum_{i=1}^n g(\theta_i) w_i^{(h)}
\end{equation}
is a consistent estimate of $I_g(h)$.  Clearly $\hat{I}_g (h)$ is a
weighted average of the $g(\theta_i)$'s.  Under additional
regularity conditions on the Markov chain and the function $g$, we
even have a central limit theorem (CLT): $n^{1/2} \bigl( \hat{I}_g
(h) - I_g(h) \bigr) \cd \mathcal{N}(0, \sigma^2(h))$, and we can
consistently estimate the limiting variance.  Thus, with a single
Markov chain run, using knowledge of only the priors and not the
posteriors, we can estimate and form confidence intervals for
$I_g(h)$ for any particular value of $h$.  Now in Bayesian
sensitivity analysis applications, we will be interested in viewing
$I_g(h)$ for many values of $h$.  For example, in prior elicitation
settings, we may wish to find those aspects of the prior that have
the biggest impact on the posterior, so that the focus of the effort
is spent on those important aspects.  We may also want to determine
whether differences in the prior opinions of many experts have a
significant impact on the conclusions.  (For a discussion of
Bayesian sensitivity analysis see \citet{Berger:1994} and
\citet{KadaneWolfson:1998}.)  In these cases we will be interested
in forming confidence bands for $I_g (\cdot)$ that are valid
globally, as opposed to pointwise.

A common feature of the two problems we study in this paper is the
need for uniformity in the convergence: to obtain confidence
intervals for $\argmax_{h \in \mathcal{H}} B(h)$ we need some
uniformity in the convergence of $B_n(\cdot)$ to $B(\cdot)$, and to
obtain confidence bands for $I_g (\cdot)$ we need functional CLT's
for the stochastic process $\hat{I}_g (\cdot)$.  Empirical process
theory is a body of results that can be used to establish uniform
almost sure convergence and functional CLT's in very general
settings.  However, the results hold only under strong regularity
conditions; and these conditions are often hard to check in
practical settings---indeed the results can easily be false if the
conditions are not met.  Empirical process theory is fundamentally
based on an iid assumption, whereas in our setting, the sequence
$\theta_1, \theta_2, \ldots$ is a Markov chain.  In this paper we
show how empirical process methods can be applied to our two
problems when the sequence $\theta_1, \theta_2, \ldots$ is a Markov
chain, and we also show how the needed regularity conditions can be
established.

The rest of the paper is organized as follows.  In
Section~\ref{sec:theoretical-results} we state our theoretical
results, the main ones---those that pertain to the Markov chain
case---being as follows.  Theorem~\ref{thm:gc-mc} asserts uniform
convergence of $B_n$ to $B$ when the sequence $\theta_1, \theta_2,
\ldots$ is a Harris ergodic Markov chain, under certain regularity
conditions on the family $\{ f_h, \, h \in \mathcal{H} \}$ (the
precise details are spelled out in the statement of the theorem),
and we show how these regularity conditions can be checked with
relative ease in standard settings.  We then give a simple result
which says that under a mild regularity assumption on $B$, the
condition $\sup_h |B_n(h) - B(h)| \cas 0$ entails $\argmax_h B_n(h)
\cas \argmax_h B(h)$.  Theorem~\ref{thm:an} establishes that under
certain regularity conditions, we have asymptotic normality of
$n^{1/2} \bigl( \argmax_h B_n(h) - \argmax_h B(h) \bigr)$.
Theorem~\ref{thm:functional-wc-mc} establishes almost sure uniform
convergence of $\hat{I}_g (\cdot)$ to $I_g (\cdot)$, and also
functional weak convergence: the process $\bigl\{ n^{1/2} \bigl(
\hat{I}_g (h) - I_g(h) \bigr), \, h \in \mathcal{H} \bigr \}$
converges weakly to a mean $0$ Gaussian process indexed by $h \in
\mathcal{H}$.  We also show how this result can be used to construct
confidence bands for $I_g(\cdot)$ that are valid globally.  A
by-product is functional weak convergence of $\bigl\{ n^{1/2}(B_n(h)
- B(h)), \, h \in \mathcal{H} \bigr \}$ to a mean $0$ Gaussian
process indexed by $h \in \mathcal{H}$, and construction of
corresponding globally valid confidence bands for $B(\cdot)$.  In
Section~\ref{sec:illustrations} we give two illustrations on
Bayesian models in which serious consideration needs to be given to
the effect of the hyperparameter and its choice.  The first is to
the Latent Dirichlet Allocation topic model, where we show how our
methodology can be used to do sensitivity analysis, and the second
is to a model for Bayesian variable selection in linear regression,
where we show how our methodology can be used to select the
hyperparameter.  In the Appendix we provide the proofs of all the
theorems except for Theorem~\ref{thm:gc-mc}; additionally, we show
how the regularity conditions in Theorem~\ref{thm:gc-iid} and
Theorem~\ref{thm:gc-mc} would typically be checked, and we verify
these conditions in a simple setting.

\section{Convergence of $B_n(\cdot)$ as a Process and Convergence of
the Empirical Argmax}
\label{sec:theoretical-results}
This section consists of three parts.  Section~\ref{sec:unif-conv}
deals with uniform convergence of $B_n$ for the iid case, and
introduces the framework that will enable us to obtain results for
the Markov chain case; this framework will be used in
Section~\ref{sec:unif-conv} and in the rest of the paper.
Section~\ref{sec:argmax} deals with point estimates and confidence
sets for $\argmax_h B(h)$, and Section~\ref{sec:pe} deals with
uniform convergence and functional CLT's for estimates of posterior
expectations.  Throughout, uniformity refers to a class of functions
indexed by $h \in \mathcal{H}$.

\subsection{Uniform Convergence of $B_n(\cdot)$}
\label{sec:unif-conv}
Let $\Theta$ be a measurable subset of $\Real^d$ for some $d \geq
1$, and let $P$ be a probability measure on $(\Theta, \mathcal{B})$,
where $\mathcal{B}$ is the Borel sigma-field on $\Theta$.  We assume
that $\theta_1, \ldots, \theta_n$ are independent and identically
distributed according to $P$, and we let $P_n$ be the empirical
measure that they induce.  We assume that $\mathcal{H}$ is a convex
compact subset of $\Real^k$ for some $k \geq 1$, and that for each
$h \in \mathcal{H}$, $f_h \colon \Theta \rightarrow \Real$ is
measurable.  The strong law of large numbers (SLLN) states that
\begin{equation}
  \label{eq:slln}
  \frac{1}{n} \sum_{i=1}^n f_h(\theta_i) \cas \int f_h \, dP \qquad
  \text{if } \int |f_h| \, dP < \infty.
\end{equation}
Since we will be interested in versions of~\eqref{eq:slln} that are
uniform in $h$, there will exist measurability difficulties, so we
have to be careful in dealing with measurability issues.  Before
proceeding, we review some terminology and standard facts from the
theory of empirical processes.  We will use the following standard
empirical process notation: for a signed measure $\mu$ on $\Theta$
and a $\mu$-integrable function $f \colon \Theta \rightarrow \Real$,
$\mu(f)$ denotes $\int f \, d\mu$.  Let $Q$ be an arbitrary
probability measure on $\Theta$, suppose that $\xi_1, \xi_2, \ldots$
are independent and identically distributed according to $Q$, and
let $Q_n$ be the empirical measure induced by $\xi_1, \ldots,
\xi_n$.  If $\mathcal{V}$ is a class of functions mapping $\Theta$
to $\Real$, and $\mu$ is a signed measure on $\Theta$, we use the
notation ${\| \mu \|}_{\mathcal{V}} = \sup_{v \in \mathcal{V}}
|\mu(v)|$.  We say that $\mathcal{V}$ is \textsl{Glivenko-Cantelli}
if ${\| Q_n - Q \|}_{\mathcal{V}}$ converges to $0$ almost surely;
sometimes we will say $\mathcal{V}$ is
$Q$-\textsl{Glivenko-Cantelli}, to emphasize the dependence on $Q$.
Let $\mathcal{F} = \{ f_h, \, h \in \mathcal{H} \}$.  Our goal is to
establish that $\mathcal{F}$ is $P$-Glivenko-Cantelli, which is
exactly equivalent to the statement that the convergence
in~\eqref{eq:slln} holds uniformly in $h$.

\subsubsection*{The IID Case}
\begin{theorem}[Theorem 6.1 and Lemma 6.1 in \citet{Wellner:2005}]
  \label{thm:gc-iid}
  \ \commentt{I put a space here}Suppose that $\theta_1, \theta_2,
  \ldots$ are independent and identically distributed according to
  $P$.  Suppose that $f_{\cdot} (\cdot) \colon \mathcal{H} \times
  \Theta \rightarrow \Real$ is continuous in $h$ for $P$-almost all
  $\theta$.  If $\sup_h |f_h|$ is measurable and satisfies $\int
  \sup_h |f_h| \, dP < \infty$, then the class $\mathcal{F}$ is
  $P$-Glivenko-Cantelli.
\end{theorem}
\noindent %
Let $B_n(h) = (1/n) \sum_{i=1}^n f_h(\theta_i)$ and $B(h) =
E_P(f_h(\theta))$ (the subscript to the expectation indicates that
$\theta \sim P$).  Then the conclusion of the theorem is the
statement $\sup_{h\in\mathcal{H}} |B_n(h) - B(h)| \cas 0$.

The integrability condition $\int \sup_h |f_h| \, dP < \infty$ seems
strong, and an even stronger integrability condition is imposed in
Theorem~\ref{thm:gc-mc}.  We discuss this issue in
Remark~\ref{rem:int-cond} following the statement of
Theorem~\ref{thm:gc-mc}, where we explain that in fact the two
conditions are fairly easy to check in practice.

The next theorem also establishes that the class $\mathcal{F}$ is
Glivenko-Cantelli.  In the theorem, the integrability condition on
$\sup_h |f_h|$ is replaced by an integrability condition on $\sup_h
\| \nabla_h f_h \|$ (here, $\nabla_h f_h$ is the gradient vector of
$f_h$ with respect to $h$, and $\| \cdot \|$ is Euclidean norm).
The condition on the gradient is sometimes easier to check.  We
include the theorem in part because a component of its proof is a
key element in the proofs of Theorems~\ref{thm:functional-wc-iid}
and~\ref{thm:functional-wc-mc} of this paper.
\begin{theorem}
  \label{thm:gc-PakesPollard}
  Suppose that $\theta_1, \theta_2, \ldots$ are independent and
  identically distributed according to $P$, and that for each $h \in
  \mathcal{H}$, $\int |f_h| \, dP < \infty$.  Assume also that for
  $P$-almost all $\theta \in \Theta$, $\nabla_h f_h$ exists and is
  continuous on $\mathcal{H}$.  If $\sup_h \| \nabla_h f_h \|$ is
  measurable and satisfies $\int \sup_h \| \nabla_h f_h \| \, dP <
  \infty$, then the class $\mathcal{F}$ is $P$-Glivenko-Cantelli.
\end{theorem}

\subsubsection*{The Markov Chain Case}
Suppose now that the sequence $\theta_1, \theta_2, \ldots$ is a
Mar\-kov chain with invariant distribution $P$, and that it is
Harris ergodic (that is, it is irreducible, aperiodic, and Harris
recurrent; see \citet[chapter~17]{MeynTweedie:1993} for
definitions).  Suppose also that $\int |f_h| \, dP < \infty$ for all
$h \in \mathcal{H}$.  The best way to deal with the family of
averages $(1/n) \sum_{i=1}^n f_h(\theta_i), \, h \in \mathcal{H}$,
is through the use of ``regenerative simulation.''  A
\textsl{regeneration} is a random time at which a stochastic process
probabilistically restarts itself; therefore, the ``tours'' made by
the process in between such random times are iid.  For example, if
the stochastic process is a Markov chain on a discrete state space
$\Theta$, and if $\theta_0 \in \Theta$ is any point to which the
chain returns infinitely often with probability one, then the times
of return to $\theta_0$ form a sequence of regenerations.  This iid
structure will enable us to establish uniform convergence of the
family $(1/n) \sum_{i=1}^n f_h(\theta_i), \, h \in \mathcal{H}$.
Before we explain this, we first note that for most of the Markov
chains used in MCMC algorithms, the state space is continuous, and
there is no point to which the chain returns infinitely often with
probability one.  Fortunately, \citet{MyklandTierneyYu:1995}
provided a general technique for identifying a sequence of
regeneration times $1 = \tau_0 < \tau_1 < \tau_2 < \cdots$ that is
based on the construction of a \textsl{minorization condition}.
This construction is reviewed at the end of this subsection, and
gives rise to regeneration times with the property that
\begin{equation}
  \label{eq:EN-finite}
  E(\tau_r - \tau_{r-1}) < \infty.
\end{equation}

Suppose now that there exists a regeneration sequence $1 = \tau_0 <
\tau_1 < \tau_2 < \cdots$ which satisfies~\eqref{eq:EN-finite}.
Such a Markov chain will be called regenerative.  For any $h \in
\mathcal{H}$, consider $(1/n) \sum_{i=1}^n f_h(\theta_i)$.  Let
\begin{equation}
  \label{eq:Sh}
  S_r^{(h)} = \sum_{i=\tau_{r-1}}^{\tau_r-1} f_h(\theta_i), \qquad r
  = 1, 2, \ldots
\end{equation}
be the sum of $f_h$ over the $r^{\text{th}}$ tour.  Also, let $N_r =
\tau_r - \tau_{r-1}, \, r = 1, 2, \ldots$, denote the length of the
$r^{\text{th}}$ tour.  The $N_r$'s do not involve $h$.  Note that
the pairs $\{ (N_r, S_r^{(h)}) \}_{r=1}^{\infty}$ are iid.  If we
run the chain for $R$ regenerations, then the total number of cycles
is given by
\begin{equation*}
  n = \sum_{r=1}^R N_r = \tau_R.
\end{equation*}
Also, $\sum_{i=1}^n f_h(\theta_i) = \sum_{r=1}^R S_r^{(h)}$.  We
have
\begin{equation}
  \label{eq:EY}
  E_P(f_h(\theta)) \casleft \frac{\sum_{i=1}^n f_h(\theta_i)}{n} =
  \frac{ \sum_{r=1}^R S_r^{(h)} } { \sum_{r=1}^R N_r } =
  \frac{\bigl( \sum_{r=1}^R S_r^{(h)} \bigr) / R} {\bigl(
  \sum_{r=1}^R N_r \bigr) / R} \cas \frac{E(S_1^{(h)})} {E(N_1)}.
\end{equation}
In~\eqref{eq:EY}, the convergence statement on the left follows from
Harris ergodicity of the chain.  The convergence statement on the
right follows from two applications of the SLLN:
By~\eqref{eq:EN-finite}, $(1/R) \sum_{r=1}^R N_r \cas E(N_1)$ and
this, together with the convergence statement on the left, entails
convergence of $(1/R) \sum_{r=1}^R S_r^{(h)}$.  The SLLN then
implies that $E(|S_1^{(h)}|) < \infty$ (if $E(|S_1^{(h)}|) = \infty$
then the SLLN implies that $\limsup (1/R) \sum_{r=1}^R S_r^{(h)} =
\infty$ with probability one).  We conclude that $E(S_1^{(h)}) =
E_P(f_h(\theta)) E(N_1)$.  Note that continuity in $h$ of
$S_1^{(h)}$ for almost all sequences $\theta_1, \theta_2, \ldots$
follows from continuity in $h$ of $f_h$ for almost all $\theta \in
\Theta$, since with probability one, $S_1^{(h)}$ is a finite sum.
Suppose in addition that $\sup_h |S_1^{(h)}|$ is measurable and
satisfies $E\bigl( \sup_h |S_1^{(h)}| \bigr) < \infty$.  Then by
Theorem~\ref{thm:gc-iid} we have $\sup_h \big| \bigl( \textstyle
\sum_{r=1}^R S_r^{(h)} \bigr) / R - E(S_1^{(h)}) \big| \cas 0$.
Since $\bigl( \sum_{r=1}^R N_r \bigr) / R \cas E(N_1)$, we obtain
\begin{equation*}
  \sup_h \Bigg| \frac{\bigl( \textstyle \sum_{r=1}^R S_r^{(h)}
  \bigr) / R} {\bigl( \sum_{r=1}^R N_r \bigr) / R} -
  \frac{E(S_1^{(h)})}{E(N_1)} \Bigg| \cas 0,
\end{equation*}
i.e.\
\begin{equation}
  \label{eq:gc-mc}
  \sup_h \Bigg| \frac{\sum_{i=1}^n f_h(\theta_i)}{n} -
  E_P(f_h(\theta)) \Bigg| \cas 0.
\end{equation}
We summarize this in the following theorem.
\begin{theorem}
  \label{thm:gc-mc}
  Suppose that $\theta_1, \theta_2, \ldots$ is a Harris ergodic
  Markov chain with invariant distribution $P$ for which there
  exists a regeneration sequence $1 = \tau_0 < \tau_1 < \tau_2 <
  \cdots$ satisfying $E(\tau_1 - \tau_0) < \infty$.  Suppose also
  that $f_{\cdot} (\cdot) \colon \mathcal{H} \times \Theta
  \rightarrow \Real$ is continuous in $h$ for $P$-almost all
  $\theta$.  For each $h \in \mathcal{H}$, let $S_r^{(h)}, \, r = 1,
  2, \ldots$ be defined by~\eqref{eq:Sh}.  If $\sup_h |S_1^{(h)}|$
  is measurable and satisfies $E\bigl( \sup_h |S_1^{(h)}| \bigr) <
  \infty$, then~\eqref{eq:gc-mc} holds.
\end{theorem}
\begin{remark}
  \label{rem:int-cond}
  We now discuss the integrability condition $E\bigl( \sup_h
  |S_1^{(h)}| \bigr) < \infty$, and our discussion encompasses the
  weaker condition $\int \sup_h |f_h| \, dP < \infty$ assumed in
  Theorem~\ref{thm:gc-iid}.  Suppose that $\int |f_h| \, dP <
  \infty$ for all $h \in \mathcal{H}$.  In the Appendix we show
  that, because $\mathcal{H}$ is assumed to be compact, it is often
  possible to prove that for some $d \geq 1$,
  \begin{equation}
    \label{eq:bound-on-sup}
    \begin{split}
      \text{there exist } & h_1, \ldots, h_d \in \mathcal{H} \text{
      and constants } c_1, \ldots, c_d \text{ such that} \\
      & \sup_h |f_h(\theta)| \leq \sum_{j=1}^d c_j
      |f_{h_j}(\theta)| \qquad \text{ for all } \theta \in \Theta.
    \end{split}
  \end{equation}
  In this case, since $|S_1^{(h)}| \leq \sum_{i=\tau_0}^{\tau_1-1}
  |f_h(\theta_i)|$, we obtain
  \begin{equation*}
    \sup_h |S_1^{(h)}| \leq \sum_{i=\tau_0}^{\tau_1-1} \sup_h
    |f_h(\theta_i)| \leq \sum_{i=\tau_0}^{\tau_1-1} \sum_{j=1}^d
    c_j |f_{h_j}(\theta_i)|.
  \end{equation*}
  Hence,
  \begin{equation*}
    E\Bigl( {\sup_h} |S_1^{(h)}| \Bigr) \leq \sum_{j=1}^d E\Biggl(
    \sum_{i=\tau_0}^{\tau_1-1} c_j |f_{h_j}(\theta_i)| \Biggr) =
    \sum_{j=1}^d c_j E_P(|f_{h_j}(\theta)|) E(N_1),
  \end{equation*}
  which is finite.  Thus, checking that $E\bigl( \sup_h | S_1^{(h)}|
  \bigr) < \infty$ reduces to establishing~\eqref{eq:bound-on-sup}.
  In the Appendix we consider the Bayesian framework discussed in
  Section~\ref{sec:intro}, in which $f_h = \nu_h / \nu_{h_*}$, where
  $\{ \nu_h, \, h \in \mathcal{H} \}$ is a family of priors, and $P
  = \nu_{h_*,y}$, the posterior distribution corresponding to the
  prior $\nu_{h_*}$, where $h_* \in \mathcal{H}$ is fixed.  We show
  that if $\{ \nu_h, \, h \in \mathcal{H} \}$ is an exponential
  family, then condition~\eqref{eq:bound-on-sup} holds.  Therefore,
  the integrability condition $E\bigl( \sup_h |S_1^{(h)}| \bigr) <
  \infty$ is satisfied in a large class of examples.  Moreover, the
  method we use for establishing~\eqref{eq:bound-on-sup} can be
  applied to other examples as well.
\end{remark}

\begin{remark}
  The idea to transform results for the iid case to the Markov chain
  case via regeneration has been around for many decades.
  \citet{Levental:1988} also obtained a Glivenko-Cantelli theorem
  for the Markov chain setting.  In essence, the difference between
  his approach and ours is that his starting point is a
  Glivenko-Cantelli theorem for the iid case which requires a
  condition involving the minimum number of balls of radius
  $\epsilon$ in $L_1(P)$ that are needed to cover $\mathcal{F}$---he
  is using metric entropy.  This condition is very hard to check.
  By contrast, our starting point is a Glivenko-Cantelli theorem for
  the iid case which is based on bracketing entropy---in brief, the
  main regularity condition is implied by the continuity condition
  in Theorem~\ref{thm:gc-mc}.  This continuity condition is trivial
  to verify: the parametric families that we are working with in our
  Bayesian setting satisfy it automatically.
\end{remark}

\paragraph{The Minorization Construction}
We now describe a minorization condition that can sometimes be used
to construct regeneration sequences.  Let $K_{\theta}(A)$ be the
transition function for the Markov chain $\theta_1, \theta_2,
\ldots$.  The construction described in
\citet{MyklandTierneyYu:1995} requires the existence of a function
$s \colon \Theta \rightarrow [0, 1)$, whose expectation with respect
to $P$ is strictly positive, and a probability measure $Q$ on
$(\Theta, \mathcal{B})$, such that $K$ satisfies
\begin{equation}
  \label{eq:min-cond}
  K_{\theta}(A) \ge s(\theta) Q(A) \qquad \text{for all } \theta \in
  \Theta \text{ and } A \in \mathcal{B}.
\end{equation}
This is called a minorization condition and, as we describe below,
it can be used to introduce regenerations into the Markov chain
driven by $K$.  Define the Markov transition function
$G_{\cdot}(\cdot)$ by
\begin{equation*}
  G_{\theta}(A) = \frac{K_{\theta}(A) - s(\theta) Q(A)}{1 -
  s(\theta)}.
\end{equation*}
Note that for fixed $\theta \in \Theta$, $G_{\theta}$ is a
probability measure.  We may therefore write
\begin{equation*}
  K_{\theta} = s(\theta) Q + (1 - s(\theta)) G_{\theta},
\end{equation*}
which gives a representation of $K_{\theta}$ as a mixture of two
probability measures, $Q$ and $G_{\theta}$.  This provides an
alternative method of simulating from $K$.  Suppose that the current
state of the chain is $\theta_n$.  We generate $\delta_n \sim
\text{Bernoulli} (s(\theta_n))$.  If $\delta_n = 1$, we draw
$\theta_{n+1} \sim Q$; otherwise, we draw $\theta_{n+1} \sim
G_{\theta_n}$.  Note that if $\delta_n = 1$, the next state of the
chain is drawn from $Q$, which does not depend on the current state.
Hence the chain ``forgets'' the current state and we have a
regeneration.  To be more specific, suppose we start the Markov
chain with $\theta_1 \sim Q$ and then use the method described above
to simulate the chain.  Each time $\delta_n = 1$, we have
$\theta_{n+1} \sim Q$ and the process stochastically restarts
itself; that is, the process regenerates.
\citet{MyklandTierneyYu:1995} provided a very widely applicable
method, the so-called ``distinguished point technique'', for
constructing a pair $(s, Q)$ that can be used to form a minorization
scheme which satisfies~\eqref{eq:EN-finite}.

For any fixed $h \in \mathcal{H}$, consider now the expression
\begin{equation*}
  \frac{\bigl( \sum_{r=1}^R S_r^{(h)} \bigr) / R} {\bigl(
  \sum_{r=1}^R N_r \bigr) / R}
\end{equation*}
in~\eqref{eq:EY}.  The bivariate CLT gives
\begin{equation}
  \label{eq:bv-clt}
  R^{1/2}
  \begin{pmatrix}
    \bigl( \sum_{r=1}^R S_r^{(h)} \bigr) / R - E_P(f_h(\theta))
    E(N_1) \\[1mm]
    \bigl( \sum_{r=1}^R N_r \bigr) / R - E(N_1)
  \end{pmatrix}
  \cd \mathcal{N}(0, \Sigma_h),
\end{equation}
where $\Sigma_h = \Cov\bigl( (S_1^{(h)}, N_1)^{\top} \bigr)$.  (We
have ignored the moment conditions on $S_1^{(h)}$ and $N_1$ that are
needed, but we will return to these conditions in
Section~\ref{sec:pe}, where we give a rigorous development of a
functional version of the CLT~\eqref{eq:bv-clt}, in which the left
side of~\eqref{eq:bv-clt} is viewed as a process in $h$.)  The delta
method applied to the function $g(x, y) = x/y$ gives the CLT
\begin{equation*}
  R^{1/2} \Biggl( \frac{\bigl( \sum_{r=1}^R S_r^{(h)} \bigr) / R}
  {\bigl( \sum_{r=1}^R N_r \bigr) / R} - E_P(f_h(\theta)) \Biggr)
  \cd \mathcal{N}(0, \sigma_h^2),
\end{equation*}
where $\sigma_h^2 = (\nabla g)^{\top} \, \Sigma_h \, \nabla g$ (and
$\nabla g$ is evaluated at the vector of means
in~\eqref{eq:bv-clt}).  Moreover, $\sigma_h^2$ can be estimated in a
simple manner using a plug-in estimate.  \linebreak \commentt{xxxI
forced a line break here}Whether or not this method gives estimates
of variance that are useful in the practical sense depends on
whether or not the minorization condition we construct yields
regenerations which are sufficiently frequent.  Successful
constructions of minorization conditions have been developed for
widely used chains in many papers (we mention in particular
\citet{MyklandTierneyYu:1995}, \citet{RoyHobert:2007},
\citet{TanHobert:2009}, and \citet{DossEtal:2014}); nevertheless,
successful construction of a minorization condition is the exception
rather than the norm.  In this context, we point out that here
regenerative simulation is notable primarily as a device that
enables us to prove the theoretical results in the present paper and
to arrive at informative expressions for asymptotic variances, but
it may be possible to estimate these variances by other methods;
this point is discussed further in Section~\ref{sec:argmax}.
\begin{remark}
  The main regularity assumption in Theorem~\ref{thm:gc-mc} is the
  condition \linebreak \commentt{xxxI forced a line break
  here}$E\bigl( \sup_h |S_1^{(h)}| \bigr) < \infty$.  Without giving
  the details, we mention that in analogy with
  Theorem~\ref{thm:gc-PakesPollard}, it is possible to give a
  version of Theorem~\ref{thm:gc-mc} in which this condition is
  replaced with the condition $E\bigl( \sup_h \| \nabla_h S_1^{(h)}
  \| \bigr) < \infty$.
\end{remark}

\subsection{A Consistent Estimator and Confidence Sets for
$\argmax_h B(h)$}
\label{sec:argmax}
This section pertains to $\argmax_h B_n(h)$ as an estimator of
$\argmax_h B(h)$.  After establishing that~\eqref{eq:gc-mc} entails
that $\argmax_h B_n(h)$ is consistent, we show that under additional
regularity conditions, (i) $n^{1/2} \bigl( \argmax_h B_n(h) -
\argmax_h B(h) \bigr)$ is asymptotically normal, and (ii) we can
consistently estimate the asymptotic variance.  Results (i) and (ii)
enable us to form asymptotically valid confidence sets for
$\argmax_h B(h)$.
\begin{lemma}
  \label{lem:conv-emp-argmax}
  Suppose that $H$ is a compact subset of Euclidean space, and let
  $f_n, \, n = 1, 2, \ldots$ and $f$ be deterministic real-valued
  functions defined on $H$.  Suppose further that $f$ is continuous
  and has a unique maximizer, and that for each $n$ the maximizer of
  $f_n$ exists and is unique.  If $f_n$ converges to $f$ uniformly
  on $H$, then the maximizer of $f_n$ converges to the maximizer of
  $f$.
\end{lemma}
The proof of Lemma~\ref{lem:conv-emp-argmax} is routine and is given
in the Appendix.  Consider now $B_n(h) = (1/n) \sum_{i=1}^n
f_h(\theta_i)$ and $B(h) = E_P(f_h(\theta))$.  By
Lemma~\ref{lem:conv-emp-argmax}, if $B$ is continuous and its
maximizer is unique, then $\sup_h |B_n(h) - B(h)| \cas 0$ implies
$\argmax_h B_n(h) \cas \argmax_h B(h)$.  Thus, under continuity of
$B$ and uniqueness of its maximizer, any conditions that
imply~\eqref{eq:gc-mc}---in particular the conditions of
Theorems~\ref{thm:gc-iid},~\ref{thm:gc-PakesPollard},
or~\ref{thm:gc-mc}---are also conditions that imply strong
consistency of $\argmax_h B_n(h)$ as an estimator of $\argmax_h
B(h)$.

Before stating the next theorem, we need to set some notation and
assumptions.  We assume that each of $B$ and $B_n, \, n = 1, 2,
\ldots$ has a unique maximizer, and we denote $h_0 = \argmax_h B(h)$
and $h_n = \argmax_h B_n(h)$.  For a function $g \colon \mathcal{H}
\rightarrow \Real$, $\nabla_h g(h)$ denotes the gradient vector and
$\nabla_h^2 g(h)$ denotes the Hessian matrix.  We will assume that
for every $\theta$, $\nabla_h f_h(\theta)$ and $\nabla_h^2
f_h(\theta)$ exist and are continuous for all $h$.  Recall that
$S_r^{(h)}$ is defined by~\eqref{eq:Sh}.  The Markov chain will be
run for $R$ regenerations, and in the asymptotic results below, $R
\rightarrow \infty$.  We will use the notation $\bar{N} = \bigl(
\sum_{r=1}^R N_r \bigr) / R$, $\bar{S}^{(h)} = \bigl( \sum_{r=1}^R
S_r^{(h)} \bigr) / R$, $\nabla_h \bar{S}^{(h)} = \bigl( \sum_{r=1}^R
\nabla_h S_r^{(h)} \bigr) / R$, etc.  For almost any realization
$\theta_1, \theta_2, \ldots$, the random variable $S_r^{(h)}$ is a
finite sum, and therefore $\nabla_h S_r^{(h)} =
\sum_{i=\tau_{r-1}}^{\tau_r-1} \nabla_h f_h(\theta_i)$.  Similarly,
$\nabla_h^2 S_r^{(h)} = \sum_{i=\tau_{r-1}}^{\tau_r-1} \nabla_h^2
f_h(\theta_i)$.  We will assume that the family $\{ f_h, \, h \in
\mathcal{H} \}$ is such that the interchange of the order of
integration and either first or second order differentiation is
permissible, i.e.
\begin{equation}
  \label{eq:interchange-d-i}
  \nabla_h \int f_h \, dP = \int \nabla_h f_h \, dP \qquad
  \text{and} \qquad \nabla_h^2 \int f_h \, dP = \int \nabla_h^2 f_h
  \, dP.
\end{equation}
For $h \in \mathcal{H}$, let
\begin{equation*}
  J(h) = \nabla_h^2 B(h), \qquad J_n(h) = \nabla_h^2 B_n(h),
\end{equation*}
\begin{equation*}
  \tau^2(h) = [E(N_1)]^{-2} E\Bigl( \bigl[ \nabla_h S_1^{(h)} - N_1
  E_P(\nabla_h f_h(\theta)) \bigr] \bigl[ \nabla_h S_1^{(h)} - N_1
  E_P(\nabla_h f_h(\theta)) \bigr]^{\top} \Bigr),
\end{equation*}
and
\begin{equation*}
  \tau_n^2(h) = \frac{1}{R \bar{N}^2} \sum_{r=1}^{R} \bigl( \nabla_h
  S_r^{(h)} - N_r \nabla_h \bar{S}^{(h)} / \bar{N} \bigr) \bigl(
  \nabla_h S_r^{(h)} - N_r \nabla_h \bar{S}^{(h)} / \bar{N}
  \bigr)^{\top}.
\end{equation*}

Suppose that $X_1, X_2, \ldots$ is a Markov chain on the measurable
space $(\sX, \mathcal{B})$ and has $\pi$ as an invariant probability
measure.  Let $K^n(x, A)$ be the $n$-step Markov transition
function.  Recall that the chain is called \textsl{geometrically
ergodic} if there exist a constant $c \in [0, 1)$ and a function $M
\colon \sX \rightarrow [0, \infty)$ such that for $n = 1, 2,
\ldots$,
\begin{equation*}
  \sup_{A \in \mathcal{B}}|K^n(x, A) - \pi(A)| \leq M(x) c^n \qquad
  \text{for all } x \in \sX.
\end{equation*}
If $Q(\theta)$ is a $k \times k$ matrix, then a statement of the
sort $E(|Q(\theta)|) < \infty$ will mean $E(|Q_{i,j}(\theta)|) <
\infty$ for $i, j = 1, \ldots, k$.  We will refer to the following
conditions.
\begin{list}{A\arabic{itemnum} }{
  \usecounter{itemnum}
  \setlength{\topsep}{2mm}
  \setlength{\itemsep}{0mm}
  \setlength{\leftmargin}{7.4mm}
  }
\item The chain $\{ \theta_i \}_{i=0}^{\infty}$ is geometrically
  ergodic.
  \label{ass:A1}
\item For every $h \in \mathcal{H}$, there exists $\epsilon > 0$
  such that $E_P \bigl( \| \nabla_h f_h(\theta) \|^{2+\epsilon}
  \bigr) < \infty$.
  \label{ass:A2}
\item The function $B$ is twice continuously differentiable and the
  $k \times k$ matrix $J(h_0)$ is nonsingular.
  \label{ass:A3}
\item $\sup_h |S_1^{(h)}|$ is measurable and $E\bigl( \sup_h
  |S_1^{(h)}| \bigr) <\infty$.
  \label{ass:A4}
\item $\sup_h |\nabla_h^2 S_1^{(h)}|$ is measurable and $E\bigl(
  \sup_h |\nabla_h^2 S_1^{(h)}| \bigr) < \infty$.
  \label{ass:A5}
\item $\sup_h |\nabla_h f_h|$ is measurable and $E(\sup_h |\nabla_h
  f_h|) <\infty$.
  \label{ass:A6}
\item $\bigl( \sup_h |\nabla_h S_1^{(h)}| \bigr) \bigl( \sup_h
  |\nabla_h S_1^{(h)}| \bigr)^{\top}$ is measurable and has finite
  expectation.
  \label{ass:A7}
\end{list}
\begin{theorem}
  \label{thm:an}
  Suppose that $\theta_1, \theta_2, \ldots$ is a regenerative Markov
  chain with invariant distribution $P$.  Let
  \begin{equation}
    \label{eq:vsquared}
    v^2 = J(h_0)^{-1} \tau^2(h_0) J(h_0)^{-1}.
  \end{equation}
  \begin{enumerate}[itemsep=0mm,topsep=2mm,leftmargin=4.5mm]
  \item Under A\ref{ass:A1}--A\ref{ass:A5}
    \begin{equation}
      \label{eq:an-argmax-R}
      R^{1/2} (h_n - h_0) \cd \mathcal{N}(0, v^2) \qquad \text{as }
      R \rightarrow \infty,
    \end{equation}
    and consequently
    \begin{equation}
      \label{eq:an-argmax-n}
      n^{1/2} (h_n - h_0) \cd \mathcal{N}\bigl( 0, E(N_1) v^2 \bigr)
      \qquad \text{as } R \rightarrow \infty.
    \end{equation}
  \item Under A\ref{ass:A1}--A\ref{ass:A7}, for large $R$ the matrix
    $J_n(h_n)$ is invertible, and the variance estimate
    \begin{equation*}
      v_n^2 = \bigl[ J_n(h_n) \bigr]^{-1} \tau_n^2(h_n) \bigl[
      J_n(h_n) \bigr]^{-1}
    \end{equation*}
    is a strongly consistent estimate of $v^2$.
  \end{enumerate}
\end{theorem}
\begin{remark}
  In the expression for the asymptotic variance given
  by~\eqref{eq:vsquared}, the term $\tau^2(h_0)$ is the variance of
  a certain function of the Markov chain, and the term $J(h_0)^{-1}$
  measures the inverse of the curvature of $B$ at its maximum ($B$
  is a deterministic function and does not involve the Markov
  chain): the flatter the surface $B$ at its maximum, the higher is
  the asymptotic variance.
\end{remark}
\begin{remark}
  \label{rem:int-cond-2}
  The integrability condition in Assumption~A\ref{ass:A4} was
  discussed in Remark~\ref{rem:int-cond}, where we showed that it is
  satisfied whenever there exist $h_1, \ldots, h_d \in \mathcal{H}$
  such that $\sup_h |f_h(\theta)| \leq \sum_{j=1}^d
  |f_{h_j}(\theta)|$ for all $\theta \in \Theta$
  (cf.~\eqref{eq:bound-on-sup}, in which without loss of generality
  we take the constants $c_j$ to be equal to $1$.)  The
  integrability conditions in~A\ref{ass:A5}--A\ref{ass:A7} are
  satisfied under~\eqref{eq:bound-on-supdf}
  and~\eqref{eq:bound-on-supd2f} below, which are very similar
  to~\eqref{eq:bound-on-sup}.  To make our explanation notationally
  less cumbersome and easier to follow, we will assume that
  $\dim(\mathcal{H}) = 1$, so that $\nabla_h S_1^{(h)}$, $\nabla_h
  f_h(\theta)$, $\nabla_h^2 S_1^{(h)}$, and $\nabla_h^2 f_h(\theta)$
  are all scalars.  Assume that there exist $h_1, \ldots, h_d \in
  \mathcal{H}$ and constants $c_1, \ldots, c_d$ such that
  \begin{align}
    \label{eq:bound-on-supdf}
    & \sup_h |\nabla_h f_h(\theta)| \leq \sum_{j=1}^d c_j
    |\nabla_h f_{h_j}(\theta)| \qquad \text{ for all } \theta \in
    \Theta, \\
    \label{eq:bound-on-supd2f}
    & \sup_h |\nabla_h^2 f_h(\theta)| \leq \sum_{j=1}^d c_j
    |\nabla_h^2 f_{h_j}(\theta)| \qquad \text{ for all } \theta \in
    \Theta.
  \end{align}

  The integrability condition in~A\ref{ass:A5}, $E\bigl( \sup_h
  |\nabla_h^2 S_1^{(h)}| \bigr) < \infty$, follows
  from~\eqref{eq:bound-on-supd2f} using an argument identical to the
  one we used to show that the integrability condition
  in~A\ref{ass:A4} follows from~\eqref{eq:bound-on-sup}.
  Clearly,~A\ref{ass:A6} follows immediately
  from~\eqref{eq:bound-on-supdf}.

  We now deal with~A\ref{ass:A7} and consider $\bigl( \sup_h
  |\nabla_h S_1^{(h)}| \bigr)^2 = \sup_h \bigl( \nabla_h S_1^{(h)}
  \bigr)^2$.  Let $F(\theta) = \sum_{j=1}^d c_j |\nabla_h
  f_{h_j}(\theta)|$, and let $\mathcal{T}_1$ denote the set of
  indices that comprise the first tour.  Since $\nabla_h S_1^{(h)} =
  \sum_{i \in \mathcal{T}_1} \nabla_h f_h(\theta_i)$, we have
  \begin{equation*}
    |\nabla_h S_1^{(h)}| \leq \sum_{i \in \mathcal{T}_1} |\nabla_h
    f_h(\theta_i)| \leq \sum_{i \in \mathcal{T}_1} F(\theta_i),
  \end{equation*}
  where the second inequality is from~\eqref{eq:bound-on-supdf}.
  Therefore $\bigl( \nabla_h S_1^{(h)} \bigr)^2 \leq \bigl( \sum_{i
  \in \mathcal{T}_1} F(\theta_i) \bigr)^2$, and hence
  \begin{equation}
    \label{eq:supdelSh}
    \sup_h \bigl( \nabla_h S_1^{(h)} \bigr)^2 \leq \bigl(
    \textstyle{\sum_{i \in \mathcal{T}_1}} F(\theta_i) \bigr)^2.
  \end{equation}
  Now by~A\ref{ass:A2} and the Minkowski inequality, $E_P\bigl(
  F^{2+\epsilon}(\theta) \bigr) < \infty$.  This integrability
  condition, together with geometric ergodicity of the chain
  (cf.~A\ref{ass:A1}), enables us to apply Theorem~$2$ of
  \citet{HobertEtal:2002} to conclude that $E\bigl[ \bigl( \sum_{i
  \in \mathcal{T}_1} F(\theta_i) \bigr)^2 \bigr] < \infty$ which,
  by~\eqref{eq:supdelSh}, implies that $E\bigl[ \sup_h \bigl(
  \nabla_h S_1^{(h)} \bigr)^2 \bigr] < \infty$, which is the
  integrability condition in~A\ref{ass:A7}.
\end{remark}
\begin{remark}
  To see why convergence statement~\eqref{eq:an-argmax-n} is a
  consequence of~\eqref{eq:an-argmax-R}, note that $n = \sum_{r=1}^R
  N_r$, so $n/R = \bigl( \sum_{r=1}^R N_r \bigr) / R \cas E(N_1)$.
  So from~\eqref{eq:an-argmax-R} and Slutsky's theorem, we have
  $(n/R)^{1/2} R^{1/2} (h_n - h_0) \cd \mathcal{N} \bigl( 0, E(N_1)
  v^2 \bigr)$, which is statement~\eqref{eq:an-argmax-n}.
\end{remark}
\begin{remark}
  We now step back and put Theorem~\ref{thm:an} in the context of
  frequentist inference.  We do not require that the number of
  components of our data vector $Y$ goes to infinity, or even that
  the components are iid.  We observe $Y = y$, which induces a
  marginal likelihood surface $m_y(\cdot)$, and Theorem~\ref{thm:an}
  pertains to estimation of this surface and its argmax, with the
  asymptotics referring to the Markov chain length $n$ going to
  infinity.  In this regard, it is natural to ask what are the
  frequentist properties of inference based on this argmax.  A very
  general result, known as the Bernstein-von Mises Theorem, asserts
  that under certain regularity conditions, if $Y_1, Y_2, \ldots$
  are iid with distribution $Q_{\theta_0}$, and if $\hat{\theta}_m$
  is the maximum likelihood estimate of $\theta$ based on $Y_{(m)} =
  (Y_1, \ldots, Y_m)$, then for any $h \in \mathcal{H}$, ${\big\|
  \nu_{h,y_{(m)}} - \phi_{\hat{\theta}_m, i^{-1} (\theta_0) / m}
  \big\|}_{\text{TV}} \cmi 0, \, [Q_{\theta_0}]$-a.s.  Here,
  $\phi_{a,V}$ denotes the normal distribution with mean vector $a$
  and covariance matrix $V$, $i(\theta)$ is the Fisher information
  at $\theta$, and the subscript TV denotes total variation norm.
  In particular, the usual Bayesian $95\%$ credible region coincides
  with the usual $95\%$ confidence region, and therefore has
  asymptotic frequentist coverage probability equal to $.95$.
  Theorem~1 of \citet{PetroneRousseauScricciolo:2014} goes further,
  and states that the Bernstein-von Mises Theorem holds when we use
  $h_0$, the maximum marginal likelihood estimate of $h$.  There are
  regularity conditions; see \citet{PetroneRousseauScricciolo:2014},
  which also contains references for precise statements of the
  Bernstein-von Mises Theorem.  To conclude, if $n$ is sufficiently
  large, $95\%$ credible sets based on $\nu_{h_n,y_{(m)}}$ have
  asymptotic frequentist coverage probability equal to $.95$.
\end{remark}

We now discuss the role of regenerative simulation in our
development.  Broadly speaking, the \textsl{existence} of
regenerative sequences is guaranteed under very general
conditions---here we note not only the distinguished point technique
of \citet{MyklandTierneyYu:1995} mentioned earlier, but also the
fact that for any chain satisfying our minimal regularity condition
of Harris ergodicity, there exists a $j \geq 1$ such that there is a
minorizing pair $(s, Q)$ for the $j$-step Markov transition function
$K^j$ \citep[Section~5.2]{MeynTweedie:1993}.  However, it is often
very difficult to construct a \textsl{useful} minorization
condition, i.e.\ one that gives rise to regenerations that are
frequent enough so that law of large numbers and CLT approximations
are valid for reasonable sample sizes.  If we do succeed in
obtaining a useful regeneration sequence, then we can estimate
variances and construct confidence sets using the estimate given in
Part~$2$ of Theorem~\ref{thm:an}, and it is widely recognized that
estimation of variances using regeneration---when it is
feasible---outperforms estimation using other methodologies
\citep{FlegalJones:2010}.  Additionally, it has the advantage that
because we start the chain at a regeneration point
(i.e.\ $\theta_1 \sim Q$), the issue of burn-in does not even exist.

It is very interesting to note that we have used regenerative
simulation in a theoretical manner: our proof of asymptotic
normality of $n^{1/2}(h_n - h_0)$ (see~\eqref{eq:an-argmax-n})
requires only the existence of a regeneration sequence, and does not
require that we go through a laborious trial and error process to
construct one that is useful in the practical sense.  Very briefly,
to obtain asymptotic results regarding $h_n$, we need uniformity in
the convergence of $B_n$ to $B$.  Empirical process theory gives us
results on uniformity, but only in the iid setting, and regenerative
simulation bridges the gap between the Markov chain setting and the
iid setting.  Once we have established the asymptotic normality of
$n^{1/2}(h_n - h_0)$, we are free to estimate the asymptotic
variance and form confidence sets using other methods, for example
batching, which we now discuss.

Batching is implemented by breaking up the sequence $\theta_1,
\ldots, \theta_n$ into $M$ consecutive pieces of equal lengths
called batches.  For $m = 1, \ldots, M$, batch $m$ is used to
produce an estimate $h_n^{[m]}$ in the obvious way.  If $M$ is
fixed, then under the regularity conditions of
Theorem~\ref{thm:an},~\eqref{eq:an-argmax-n} states that for each
$m$, $(n/M)^{1/2} (h_n^{[m]} - h_0) \cd \mathcal{N} (0, \sigma^2)$,
where $\sigma^2 = E(N_1) v^2$.  If the batch length is large enough
relative to the ``mixing time'' of the chain, then the $h_n^{[m]}$'s
are approximately independent.  If the independence assumption was
exactly true rather than approximately true, then the sample
variance of $(n/M)^{1/2} h_n^{[1]}, \ldots, (n/M)^{1/2} h_n^{[M]}$
would be a valid estimator of $\sigma^2$.  Standard theoretical
results regarding batching deal with the situation in which $g$ is a
$P$-integrable function, and the Markov chain $\theta_1, \ldots,
\theta_n$ is used to estimate $\int g \, dP$ via $(1/n) \sum_{i=1}^n
g(\theta_i)$.  These results, which assume that $n^{1/2} \bigl(
(1/n) \sum_{i=1}^n g(\theta_i) - \int g \,dP \bigr) \cd \mathcal{N}
(0, \sigma^2(g))$, state that under regularity conditions which
include $M \rightarrow \infty$ at a certain rate, the batch-based
estimate of $\sigma^2(g)$ is strongly consistent; see
\citet{FlegalHaranJones:2008} and also \citet{JonesEtal:2006}, who
recommend using $M = n^{1/2}$.  Our situation is different in that
our estimate $h_n$ is not an average, but is the argmax of a
function based on $\theta_1, \ldots, \theta_n$.  Nevertheless, the
method applies, with the minor modification that when we form the
``sample variance,'' the centering value is based on $h_n$ rather
than on the average of the $h_n^{[m]}$'s.  As is clear from the
description above, batch-based estimates of variance are very easy
to program.  However, it is generally acknowledged that they are
outperformed by estimates based on regeneration or spectral methods.

\subsection{Convergence of Estimate of Posterior Expectation}
\label{sec:pe}
This section concerns the Bayesian framework discussed earlier, in
which $\{ \nu_h, \, h \in \mathcal{H} \}$ is a family of prior
densities on $\theta$; for each $h$, $\nu_{h,y}$ is the posterior
corresponding to $\nu_h$; $h_1 \in \mathcal{H}$ is fixed but
arbitrary, and $\theta_1, \theta_2, \ldots$ is an ergodic Markov
chain with invariant distribution $\nu_{h_1,y}$.  Suppose that $g$
is a real-valued function of $\theta$ and consider $I_g(h) = \int
g(\theta) \nu_{h,y}(\theta) \, d\theta$, the posterior expectation
of $g(\theta)$ given $Y = y$, when the prior is $\nu_h$.  We have
\begin{equation}
  \label{eq:main-conv2}
  \frac{1}{n} \sum_{i=1}^n g(\theta_i) \frac{\nu_h(\theta_i) }
  {\nu_{h_1} (\theta_i)} \cas \int g(\theta) \frac{\nu_h(\theta)}
  {\nu_{h_1}(\theta)} \nu_{h_1,y}(\theta) \, d\theta =
  \frac{m_{y}(h)} {m_{y}(h_1)} I_g(h),
\end{equation}
in which the first equality follows from~\eqref{eq:post} and
cancellation of the likelihood.  Therefore,
\begin{equation}
  \label{eq:ihat}
  \hat{I}_g (h) \coloneqq \frac{(1/n) \sum_{i=1}^n g(\theta_i)
  [\nu_h(\theta_i) / \nu_{h_1}(\theta_i)]} {(1/n) \sum_{i=1}^n
  [\nu_h(\theta_i) / \nu_{h_1}(\theta_i)]} \cas \frac{[m_{y}(h) /
  m_{y}(h_1)] I_g(h)} {m_{y}(h) / m_{y}(h_1)} = I_g(h),
\end{equation}
where the convergence of the numerator and the denominator in the
expression for $\hat{I}_g (h)$ follow from~\eqref{eq:main-conv2}
and~\eqref{eq:main-conv}, respectively.  In the original expression
given in~\eqref{eq:ihat-orig}, $\hat{I}_g (h)$ is a weighted average
of the $g(\theta_i)$'s (with weights all equal to $1/n$ if $\nu_h =
\nu_{h_1}$, and becoming more disparate as $\nu_h$ and $\nu_{h_1}$
become more dis-similar).  The definition of $\hat{I}_g (h)$ given
in~\eqref{eq:ihat} clearly matches the original expression, so we
see that $\hat{I}_g (h)$ may be represented either as a weighted
average or as a ratio of two ordinary averages.  To establish almost
sure uniform convergence and functional weak convergence results for
$\hat{I}_g (h)$, we will work with the latter representation,
because doing so will enable us to use tools from empirical process
theory.  With this in mind, recall that in the present framework
$f_h = \nu_h / \nu_{h_1}$.  We will work with the classes of
functions $\mathcal{F} = \{ f_h, \, h \in \mathcal{H} \}$ and
$\mathcal{G} = \{ g f_h, \, h \in \mathcal{H} \}$.  We will later
assume that the sequence $\theta_1, \theta_2, \ldots$ is a Markov
chain satisfying certain conditions, and
Theorem~\ref{thm:functional-wc-mc} pertains to that case; however,
in order to give an overview of our results, it is convenient to
first assume that the $\theta_i$'s form an iid sequence: $\theta_i
\iid P \coloneqq \nu_{h_1,y}$.  Recall that $P_n$ is the empirical
measure that gives mass $1/n$ to each of $\theta_1, \ldots,
\theta_n$, and that for a signed measure $\mu$ and a function $f$,
$\mu(f)$ denotes $\int f \, d\mu$.  In the present specialized
Bayesian context, $f_h \geq 0$; thus the $L_1(P)$ norm of $f_h$ is
simply $\int f_h \, dP$.  Our goal is to establish that under
certain conditions:
\begin{enumerate}[itemsep=0mm,topsep=2mm,leftmargin=4.5mm]
\item We have the Glivenko-Cantelli results
  \begin{equation*}
    \sup_{h\in\mathcal{H}} |(P_n - P)(f_h)| \cas 0 \qquad \text{and}
    \qquad \sup_{h\in\mathcal{H}} |(P_n - P)(g f_h)| \cas 0.
  \end{equation*}
\item We have the ``Donsker results''
  \begin{equation}
    \label{eq:donsker-fg}
    n^{1/2} (P_n - P)(f_{\cdot}) \cd \mathbb{F}(\cdot) \qquad
    \text{and} \qquad n^{1/2} (P_n - P)(g f_{\cdot}) \cd
    \mathbb{G}(\cdot),
  \end{equation}
  where $\mathbb{F}$ and $\mathbb{G}$ are mean $0$ Gaussian
  processes indexed by $\mathcal{H}$.
\end{enumerate}
By applying the delta method to the function $q(u, v) = u/v$, we
then obtain the Glivenko-Cantelli and Donsker results
\begin{enumerate}[itemsep=0mm,topsep=2mm,leftmargin=4.5mm]
\item[3.] \itemEqstar{\sup_{h\in\mathcal{H}} |\hat{I}_g (h) -
  I_g(h)| \cas 0,}
\item[4.] \itemEq{\label{eq:donsker-ihat}}{ n^{1/2} \bigl( \hat{I}_g
  (\cdot) - I_g(\cdot) \bigr) \cd \mathbb{I}_g (\cdot),} where
  $\mathbb{I}_g$ is a mean $0$ Gaussian process indexed by
  $\mathcal{H}$.
\end{enumerate}
We now give some definitions we will need in order to explain what
is meant by~\eqref{eq:donsker-fg} and~\eqref{eq:donsker-ihat}.
Define $X_n = n^{1/2} (P_n - P)$.  Let $\mathcal{V}$ be any set of
real-valued functions defined on $\Theta$ and let $l^{\infty}
(\mathcal{V})$ denote the space of bounded functions from
$\mathcal{V}$ to $\Real$ equipped with the supremum norm.  Assume
that
\begin{equation*}
  \sup_{V\in\mathcal{V}} |V(\theta) - P(V)| < \infty \qquad
  \text{for every } \theta \in \Theta.
\end{equation*}
Under this condition the empirical process $\{ X_n(V), \, V \in
\mathcal{V} \}$ can be viewed as a map from $\Theta^n$ into
$l^{\infty} (\mathcal{V})$.  Any measurable function $Z \colon
\Theta^n \rightarrow l^{\infty} (\mathcal{V})$ induces a
distribution on $l^{\infty} (\mathcal{V})$.  Although the functions
we will be working with will in general be measurable, in order to
properly state the relevant definitions and theorems from empirical
process theory, in our definitions we will deal with functions which
are not necessarily measurable.  For an arbitrary map $M$ from an
arbitrary probability space $(\Omega, \mathcal{E}, \mu)$ to the
extended real line $\bar{\Real}$, $E^*(M)$ denotes the outer
integral of $M$ with respect to $\mu$.  (The outer integral is
defined by $E^*(M) = \inf \{ \int Y \, d\mu \colon Y \, \text{is }
\mathcal{E}\text{-measurable}, \, Y \ge M \}$.)  Suppose $Z_1, Z_2,
\ldots$ and $Z$ are maps into $l^{\infty} (\mathcal{V})$, and that
$Z$ is measurable.  We say that $Z_n$ converges weakly to $Z$, and
we write $Z_n \cd Z$, if $E^*(\phi(Z_n)) \rightarrow E(\phi(Z))$ for
every bounded, continuous, real function $\phi$ on $l^{\infty}
(\mathcal{V})$.

We now return to the empirical process $X_n = n^{1/2} (P_n - P)$.  A
class $\mathcal{V}$ is called a Donsker class if $X_n \cd X$ in
$l^{\infty} (\mathcal{V})$, where the limit $X$ is a mean $0$
Gaussian process with covariance function
\begin{equation*}
  \Cov\bigl( X(V_1), X(V_2) \bigr) = P(V_1 V_2) - P(V_1) P(V_2),
  \qquad V_1, V_2 \in \mathcal{V},
\end{equation*}
and has paths which are uniformly continuous with respect to the
semi-metric $\rho_P$ on $l^{\infty} (\mathcal{V})$ defined by
$\rho_P^2(f_1, f_2) = \Var_P\bigl( f_1(\theta) - f_2(\theta)
\bigr)$.  Sometimes we will say $\mathcal{V}$ is $P$-Donsker, to
emphasize the dependence on $P$.

We say that a class $\mathcal{V}$ of measurable functions $V \colon
\Theta \rightarrow \Real$ is $P$-measurable if for every $n$ and
every vector $(e_1, \ldots, e_n) \in \Real^n$, the function
\begin{equation*}
  (\theta_1, \ldots, \theta_n) \mapsto \sup_{V\in\mathcal{V}}
  \bigg| \sum_{i=1}^n e_i V(\theta_i) \bigg|
\end{equation*}
is measurable on the completion of $(\Theta^n, \mathcal{B}^n, P^n)$.

Because $\mathcal{F}$ and $\mathcal{G}$ are simply parametric
families indexed by $h \in \mathcal{H}$, we will slightly abuse
terminology and take the two convergence statements
in~\eqref{eq:donsker-fg} to mean $X_n \cd X$ in $l^{\infty}
(\mathcal{F})$ and $X_n \cd X$ in $l^{\infty} (\mathcal{G})$,
respectively.  The limit $\mathbb{F}$ is a mean $0$ Gaussian process
indexed by $h \in \mathcal{H}$ and covariance function
\begin{equation*}
  \Cov \bigl( \mathbb{F}(h'), \mathbb{F}(h'') \bigr) = P(f_{h'}
  f_{h''}) - P(f_{h'}) P(f_{h''}) \qquad \text{for any } h', h'' \in
  \mathcal{H}.
\end{equation*}
Similarly, $\mathbb{G}$ is a mean $0$ Gaussian process indexed by $h
\in \mathcal{H}$ and covariance function
\begin{equation*}
  \Cov \bigl( \mathbb{G}(h'), \mathbb{G}(h'') \bigr) = P(g^2 f_{h'}
  f_{h''}) - P(g f_{h'}) P(g f_{h''}) \qquad \text{for any } h', h''
  \in \mathcal{H},
\end{equation*}
and we will discuss the covariance function of the limit
$\mathbb{I}_g$ in~\eqref{eq:donsker-ihat} later.  For $\delta > 0$,
let $\mathcal{F}_{\delta} = \{ \phi - \psi \colon \phi, \psi \in
\mathcal{F}, \, {\| \phi - \psi \|}_{P,2} < \delta \}$ and let
$\mathcal{F}_{\infty}^2 = \{ \xi^2 : \xi \in \mathcal{F}_{\infty}
\}$.

Before we state the next theorem, we need to lay down preparations
for its fourth part, which regards functional weak convergence of
the process $n^{1/2} \bigl( \hat{I}_g(\cdot) - I_g(\cdot) \bigr)$.
Let $C(\mathcal{H})$ be the space of all continuous functions $x
\colon \mathcal{H} \rightarrow \Real$, with the topology induced by
the sup norm metric $\rho$: for $x, y \in C(\mathcal{H})$, $\rho(x,
y) = {\| x - y \|}_{\infty} = \sup_h |x(h) - y(h)|$.  Clearly,
functional weak convergence of $n^{1/2} \bigl( \hat{I}_g(\cdot) -
I_g(\cdot) \bigr)$ cannot take place in a space of the type
$l^{\infty}(\mathcal{V})$ for some set of functions $\mathcal{V}$,
and in fact, as we will see, the weak convergence will take place in
the space $C(\mathcal{H})$.  (As usual, if $\mu_n, \, n = 1, 2,
\ldots$ and $\mu$ are probability measures on $C(\mathcal{H})$, we
say that $\mu_n \cd \mu$ if $\int \Phi \, d\mu_n \rightarrow \int
\Phi \, d\mu$ for all functions $\Phi \colon C(\mathcal{H})
\rightarrow \Real$ which are bounded and continuous.)

We now define the expression for the covariance function and give
motivation for its form.  For any $h', h'' \in \mathcal{H}$, the
multivariate CLT states that
\begin{equation}
  \label{eq:mv-clt}
  \begin{pmatrix}
    U_1 \\
    U_2 \\
    U_3 \\
    U_4
  \end{pmatrix}
  \coloneqq n^{1/2}
  \begin{pmatrix}
    P_n(g f_{h'}) - P(g f_{h'}) \\
    P_n(f_{h'}) - P(f_{h'}) \\
    P_n(g f_{h''}) - P(g f_{h''}) \\
    P_n(f_{h''}) - P(f_{h''})
  \end{pmatrix}
  \cd \mathcal{N} \bigl( 0, \Sigma(h', h'') \bigr),
\end{equation}
where $\Sigma(h', h'')$ is the $4 \times 4$ matrix given by
$\Sigma(h', h'')_{ij} = \Cov(U_i, U_j), \, i, j = 1, 2, 3, 4$.
Consider the function $\phi \colon \Real^4 \rightarrow \Real^2$
defined by $\phi(u_1, u_2, u_3, u_4) = \linebreak \commentt{xxxI
forced a line break here}(u_1 / u_2, u_3 / u_4)$.  Then, if we apply
the delta method to~\eqref{eq:mv-clt} using $\phi$, we get
\begin{equation}
  \label{eq:clt-I}
  n^{1/2}
  \begin{pmatrix}
    \hat{I}_g(h') - I_g(h') \\
    \hat{I}_g(h'') - I_g(h'')
  \end{pmatrix}
  \cd \mathcal{N} \bigl( 0, M(h', h'') \bigr),
\end{equation}
where $M(h', h'') = (\nabla \phi)^{\top} \Sigma(h', h'') \nabla
\phi$, and $\nabla \phi$ (viewed as a $4 \times 2$ matrix) is
evaluated at the vector of means $\bigl( P(g f_{h'}), P(f_{h'}), P(g
f_{h''}), P(f_{h''}) \bigr)$.  The matrix $M(h', h'')$ describes the
covariance structure for the process $\mathbb{I}_g(\cdot)$.
(Expressions for $\nabla \phi$ and $M(h', h'')$ are given in
\citet{Park:2015}.)
\begin{theorem}
  \label{thm:functional-wc-iid}
  Assume that $\theta_1, \ldots, \theta_n$ are independent and
  identically distributed according to $P$.
  \renewcommand{\labelenumi}{\theenumi}
  \begin{enumerate}[itemsep=0mm,topsep=1.5mm,leftmargin=3.5mm]
  \item
    \begin{enumerate}
    \item Suppose that $f_{\cdot}(\cdot) \colon \mathcal{H} \times
      \Theta \rightarrow \Real$ is continuous in $h$ for $P$-almost
      all $\theta$.  If $\sup_{h\in\mathcal{H}} f_h$ is measurable
      and $\int \sup_{h\in\mathcal{H}} f_h \, dP < \infty$, then
      $\mathcal{F}$ is $P$-Glivenko-Cantelli.
    \item Suppose that $(g f_{\cdot})(\cdot) \colon \mathcal{H}
      \times \Theta \rightarrow \Real$ is continuous in $h$ for
      $P$-almost all $\theta$.  If $\sup_{h\in\mathcal{H}} |g f_h|$
      is measurable and $\int \sup_{h\in\mathcal{H}} |g f_h| \, dP <
      \infty$, then $\mathcal{G}$ is $P$-Glivenko-Cantelli.
    \end{enumerate}
  \item Assume the conditions of Part~$1$ of the theorem, and also
    that for every $\theta \in \Theta$, $\nabla_h f_h$ exists and is
    continuous on $\mathcal{H}$.  Then
    \begin{equation}
      \label{eq:ucI}
      \sup_{h\in\mathcal{H}} |\hat{I}_g (h) - I_g(h)| \cas 0.
    \end{equation}
  \item
    \begin{enumerate}[itemsep=1mm,topsep=1.5mm]
    \item Suppose that the classes $\mathcal{F}$,
      $\mathcal{F}_{\delta}, \, \delta > 0$, and
      $\mathcal{F}_{\infty}^2$ are all $P$-measurable.  Assume also
      that for $P$-almost all $\theta \in \Theta$, $\nabla_h f_h$
      exists and is continuous on $\mathcal{H}$.  If (1)
      $\sup_{h\in\mathcal{H}} \| \nabla_h f_h \|$ is measurable and
      (2) the functions $f_h, \, h \in \mathcal{H}$ and
      $\sup_{h\in\mathcal{H}} \| \nabla_h f_h \|$ are all square
      integrable with respect to $P$, then the class $\mathcal{F}$
      is $P$-Donsker.
    \item Suppose that the classes $\mathcal{G}$,
      $\mathcal{G}_{\delta}, \, \delta > 0$, and
      $\mathcal{G}_{\infty}^2$ are all $P$-measurable.  Assume also
      that for $P$-almost all $\theta \in \Theta$, $\nabla_h (g
      f_h)$ exists and is continuous on $\mathcal{H}$.  If (1)
      $\sup_{h\in\mathcal{H}} \| \nabla_h (g f_h) \|$ is measurable
      and (2) the functions $g f_h, \, h \in \mathcal{H}$ and
      $\sup_{h\in\mathcal{H}} \| \nabla_h (g f_h) \|$ are all square
      integrable with respect to $P$, then the class $\mathcal{G}$
      is $P$-Donsker.
    \end{enumerate}
  \item Under the conditions of Part~$3$ of the theorem, we have
    \begin{equation*}
      n^{1/2} \bigl( \hat{I}_g(\cdot) - I_g(\cdot) \bigr) \cd
      \mathbb{I}_g (\cdot) \qquad \text{in } C(\mathcal{H}),
    \end{equation*}
    where $\mathbb{I}_g$ is a Gaussian process indexed by
    $\mathcal{H}$ with mean $0$ and covariance function
    \begin{align*}
      & \Cov(\mathbb{I}_g(h'), \mathbb{I}_g(h'')) \\ & = \frac{
      P(g^2 f_{h'} f_{h''}) - P(g f_{h'} f_{h''}) \Bigl( \frac{ P(g
      f_{h''}) } {P(f_{h''})} + \frac{ P(g f_{h'}) } {P(f_{h'})}
      \Bigr) + \frac{ P(g f_{h'}) P(g f_{h''}) }{P(f_{h'})
      P(f_{h''})} P( f_{h'} f_{h''}) } {P(f_{h'} ) P(f_{h''} )}.
    \end{align*}
  \end{enumerate}
\end{theorem}

\vspace{3mm} %
Part~$1$(a) is, of course, simply a restatement of
Theorem~\ref{thm:gc-iid}; we have repeated it here only to clarify
the structure of our results.  The $P$-measurability conditions
cannot be omitted.  However, in all the problems we have
encountered, the relevant functions are not only measurable, but are
actually continuous.

In Remark~\ref{rem:conf-bands}, which follows the statement of
Theorem~\ref{thm:functional-wc-mc}, we develop a construction of
confidence bands for $I_g(\cdot)$ and we explain why
Theorem~\ref{thm:functional-wc-mc} shows that these bands are valid
globally.  Theorem~\ref{thm:functional-wc-mc} pertains to Markov
chains, but the same construction and arguments can be applied to
the iid case---we use Theorem~\ref{thm:functional-wc-iid} instead of
Theorem~\ref{thm:functional-wc-mc}.

\vspace{3mm} %
The next result is a version of Theorem~\ref{thm:functional-wc-iid}
that applies to Markov chains.  Recall that $N_r = \tau_r -
\tau_{r-1}$ is the length of the $r^{\text{th}}$ tour and that
$S_r^{(h)}$ is defined by~\eqref{eq:Sh}.  Similarly, define
$T_r^{(h)} = \sum_{i=\tau_{r-1}}^{\tau_r-1} g(\theta_i)
f_h(\theta_i), \, r = 1, 2, \ldots$.  Let $\mathscr{F} = \{
S_1^{(h)}, \, h \in \mathcal{H} \}$ and $\mathscr{G} = \{ T_1^{(h)},
\, h \in \mathcal{H} \}$.  Part~$3$ of
Theorem~\ref{thm:functional-wc-mc} asserts that under certain
conditions the classes $\mathscr{F}$ and $\mathscr{G}$ are Donsker,
and before stating the theorem, it is necessary to be very clear
regarding what these classes are, and what ``Donsker'' means.  Let
$\sP$ be the distribution of the Markov chain $\theta_1, \theta_2,
\ldots$.  For any $h \in \mathcal{H}$, $S_1^{(h)}$ is a function
mapping the measure space $(\Theta^{\infty}, \mathcal{B}^{\infty},
\sP)$ into $\Real_+$.  To see this it may be helpful to imagine that
we are dealing with the very simple case of a regenerative chain
which has an ``proper atom'' at a singleton.  That is, there exists
a point $\alpha \in \Theta$ which has positive probability under the
invariant measure.  Thus, with probability one the chain returns to
$\alpha$ infinitely often, and the times of return to $\alpha$ are
regeneration times $\tau_0, \tau_1, \tau_2, \ldots$.  In this case
(with probability one) the sequence $\theta_1, \theta_2, \ldots$
itself determines $\tau_0$ and $\tau_1$.  Then, $S_1^{(h)} \colon
\Theta^{\infty} \rightarrow \Real_+$ is defined by
$S_1^{(h)}(\theta_1, \theta_2, \ldots) = \sum_{i=\tau_0}^{\tau_1-1}
f_h(\theta_i)$, and we have a similar definition for $T_1^{(h)}$.
Chains which have a proper atom at a singleton are quite rare, and
we consider them only for exposition.  We remark on the case of a
general regenerative Markov chain at the end of the proof of
Theorem~\ref{thm:functional-wc-mc}.  To clarify, $\mathscr{F}$ and
$\mathscr{G}$ are classes of functions on $\Theta^{\infty}$, in
contrast to $\mathcal{F}$ and $\mathcal{G}$, which are classes of
functions on $\Theta$.  These classes will be $\sP$-Donsker, and we
note that $\sP$ is a distribution on the infinite product space
$\Theta^{\infty}$, to be distinguished from $P$, which is a
distribution on $\Theta$.

As we will see, Parts~$3$ and~$4$ of
Theorem~\ref{thm:functional-wc-mc} are functional CLT's that
concerns certain stochastic processes indexed by $h \in
\mathcal{H}$.  In order to motivate them, we need to first
understand the version of these parts of the theorem that pertains
to the very simple situation in which we are considering a single
value of $h$.  Thus, let $h \in \mathcal{H}$ be fixed.  We now
consider CLT's for averages formed from the sequences $S_1^{(h)},
S_2^{(h)}, \ldots$ and $T_1^{(h)}, T_2^{(h)}, \ldots$.  We have
$E(S_1^{(h)}) = E_P(f_h(\theta)) E(N_1)$ and $E(T_1^{(h)}) =
E_P(g(\theta) f_h(\theta)) E(N_1)$ (see~\eqref{eq:EY}).
Under~A\ref{ass:A1} and the conditions $E_P(f_h^{2+\epsilon}
(\theta)) < \infty$ and $E_P[(g f_h)^{2+\epsilon} (\theta)] <
\infty$, the expectations $E\bigl[ (S_1^{(h)})^2 \bigr]$, $E\bigl[
(T_1^{(h)})^2 \bigr]$, and $E(N_1^2)$ are all finite (Theorem~$2$ of
\citeauthor{HobertEtal:2002} \citeyear{HobertEtal:2002}).
Therefore, the simple multivariate CLT gives
\begin{equation}
  \label{eq:multivariate-clt}
  R^{1/2}
  \begin{pmatrix}
    \bigl( \sum_{r=1}^R T_r^{(h)} \bigr) / R - E_P(g(\theta)
    f_h(\theta)) E(N_1) \\[1mm]
    \bigl( \sum_{r=1}^R S_r^{(h)} \bigr) / R - E_P( f_h(\theta))
    E(N_1) \\[1mm]
    \bigl( \sum_{r=1}^R N_r \bigr) / R - E(N_1)
  \end{pmatrix}
  \cd \mathcal{N}(0, V_h),
\end{equation}
where $V_h = \Cov\bigl( (T_1^{(h)}, S_1^{(h)}, N_1)^{\top} \bigr)$.
We apply the delta method to~\eqref{eq:multivariate-clt} three
times, using the functions $q_1(u, v, w) = v/w$, $q_2(u, v, w) =
u/w$, and $q_3(u, v, w) = u/v$ to obtain three CLT's:
\begin{equation}
  \label{eq:three-clts-1}
  \begin{split}
    R^{1/2} \Biggl( \frac{\sum_{r=1}^R S_r^{(h)}} {\sum_{r=1}^R N_r}
    - E_P(f_h(\theta)) \Biggr) & \cd \mathcal{N} \bigl( 0, (\nabla
    q_1)^{\top} V_h \nabla q_1 \bigr), \\
    R^{1/2} \Biggl( \frac{\sum_{r=1}^R T_r^{(h)}} {\sum_{r=1}^R N_r}
    - E_P(g(\theta) f_h(\theta)) \Biggr) & \cd \mathcal{N} \bigl( 0,
    (\nabla q_2)^{\top} V_h \nabla q_2 \bigr), \\
    R^{1/2} \Biggl( \frac{\sum_{r=1}^R T_r^{(h)}} {\sum_{r=1}^R
    S_r^{(h)}} - I_g(h) \Biggr) & \cd \mathcal{N} \bigl( 0, (\nabla
    q_3)^{\top} V_h \nabla q_3 \bigr).
  \end{split}
\end{equation}
With the relationships $n = \sum_{r=1}^R N_r$, $\sum_{r=1}^R
S_r^{(h)} = \sum_{i=1}^n f_h(\theta_i)$, $\sum_{r=1}^R T_r^{(h)} =
\sum_{i=1}^n g(\theta_i) f_h(\theta_i)$, and the fact that $n/R \cas
E(N_1)$,~\eqref{eq:three-clts-1} may be restated as
\begin{equation}
  \label{eq:three-clts-2}
  \begin{split}
    n^{1/2} \biggl( \frac{\sum_{i=1}^n f_h(\theta_i)}{n} -
    E_P(f_h(\theta)) \biggr) & \cd \mathcal{N}\bigl( 0, E(N_1)
    (\nabla q_1)^{\top} V_h \nabla q_1 \bigr), \\
    n^{1/2} \biggl( \frac{\sum_{i=1}^n g(\theta_i) f_h(\theta_i)}{n}
    - E_P(g(\theta) f_h(\theta)) \biggr) & \cd \mathcal{N}\bigl( 0,
    E(N_1) (\nabla q_2)^{\top} V_h \nabla q_2 \bigr), \\
    n^{1/2} \biggl( \frac{\sum_{i=1}^n g(\theta_i) f_h(\theta_i)}
    {\sum_{i=1}^n f_h(\theta_i)} - I_g(h) \biggr) & \cd
    \mathcal{N}\bigl( 0, E(N_1) (\nabla q_3)^{\top} V_h \nabla q_3
    \bigr)
  \end{split}
\end{equation}
(with the understanding that here, $n$ is random).  Of course, under
geometric ergodicity and the moment conditions $E_P(f_h^{2+\epsilon}
(\theta)) < \infty$ and $E_P[(g f_h)^{2+\epsilon} (\theta)] <
\infty$, asymptotic normality of the three quantities on the left
side of~\eqref{eq:three-clts-2} is already known (corollary to
Theorem~18.5.3 of \citeauthor{IbragimovLinnik:1971}
\citeyear{IbragimovLinnik:1971}).  The point of
obtaining~\eqref{eq:three-clts-2} as we did above is that the method
enables us to get functional versions of the three statements
in~\eqref{eq:three-clts-2} (i.e.\ weak convergence of the three
quantities on the left side of~\eqref{eq:three-clts-2} as processes
in $h$) if we can show that the classes $\mathscr{F}$ and
$\mathscr{G}$ are Donsker.  This is precisely what Part~$3$ of
Theorem~\ref{thm:functional-wc-mc} asserts.  The theorem will refer
to the following conditions.
\begin{list}{B\arabic{itemnum} }{
  \usecounter{itemnum}
  \setlength{\topsep}{2mm}
  \setlength{\itemsep}{0mm}
  \setlength{\leftmargin}{7.4mm}
  }
\item For every $h \in \mathcal{H}$, there exists $\epsilon > 0$
  such that $E_P(f_h^{2+\epsilon} (\theta)) < \infty$.
  \label{ass:B1}
\item For every $h \in \mathcal{H}$, there exists $\epsilon > 0$
  such that $E_P[(g f_h)^{2+\epsilon} (\theta)] < \infty$.
  \label{ass:B2}
\end{list}
\begin{theorem}
  \label{thm:functional-wc-mc}
  Assume that $\theta_1, \theta_2, \ldots$ is a Harris ergodic
  Markov chain with invariant distribution $P$ for which there
  exists a regeneration sequence $1 = \tau_0 < \tau_1 < \tau_2 <
  \cdots$ satisfying $E(\tau_1 - \tau_0) < \infty$.
  \renewcommand{\labelenumi}{\theenumi}
  \begin{enumerate}[itemsep=0mm,topsep=1.5mm,leftmargin=3.5mm]
  \item
    \begin{enumerate}
    \item Suppose that $f_{\cdot} (\cdot) \colon \mathcal{H} \times
      \Theta \rightarrow \Real$ is continuous in $h$ for $P$-almost
      all $\theta$.  Suppose also that $\sup_h S_1^{(h)}$ is
      measurable and integrable.  Then~\eqref{eq:gc-mc} holds.
    \item Suppose that $(g f_{\cdot}) (\cdot) \colon \mathcal{H}
      \times \Theta \rightarrow \Real$ is continuous in $h$ for
      $P$-almost all $\theta$.  Suppose also that $\sup_h
      |T_1^{(h)}|$ is measurable and integrable.  Then in analogy
      with~\eqref{eq:gc-mc}, we have
      \begin{equation*}
        \sup_h \bigg| \frac{1}{n} \sum_{i=1}^n g(\theta_i)
        f_h(\theta_i) - E_P(g(\theta) f_h(\theta)) \bigg| \cas 0.
      \end{equation*}
    \end{enumerate}
  \item Assume the conditions of Part~$1$ of the theorem, and also
    that for every $\theta \in \Theta$, $\nabla_h f_h$ exists and is
    continuous on $\mathcal{H}$.  Then
      \begin{equation}
        \label{eq:conv-supIhat}
        \sup_{h\in\mathcal{H}} |\hat{I}_g (h) - I_g(h)| \cas 0.
      \end{equation}
  \item
    \begin{enumerate}
    \item Suppose that the classes $\mathscr{F}$,
      $\mathscr{F}_{\delta}, \, \delta > 0$, and
      $\mathscr{F}_{\infty}^2$ are all $\sP$-measurable.  Suppose
      also that for almost all $\theta \in \Theta$, $\nabla_h f_h$
      exists and is continuous on $\mathcal{H}$.
      Under~A\ref{ass:A1},~B\ref{ass:B1}, and the condition that
      $\sup_{h\in\mathcal{H}} \| \nabla_h S_1^{(h)} \|$ is
      measurable and square integrable with respect to $\sP$, the
      class $\mathscr{F}$ is $\sP$-Donsker.
    \item Suppose that the classes $\mathscr{G}$,
      $\mathscr{G}_{\delta}, \, \delta > 0$, and
      $\mathscr{G}_{\infty}^2$ are all $\sP$-measurable.  Suppose
      also that for almost all $\theta \in \Theta$, $\nabla_h (g
      f_h)$ exists and is continuous on $\mathcal{H}$.
      Under~A\ref{ass:A1},~B\ref{ass:B2}, and the condition that
      $\sup_{h\in\mathcal{H}} \| \nabla_h T_1^{(h)} \|$ is
      measurable and square integrable with respect to $\sP$, the
      class $\mathscr{G}$ is $\sP$-Donsker.
    \end{enumerate}
  \item Under the conditions of Part~$3$ of the theorem, we have
    \begin{equation}
      \label{eq:wcIhatmc}
      R^{1/2} \bigl( \hat{I}_g(\cdot) - I_g(\cdot) \bigr) \cd
      \mathbb{I}_g^* (\cdot) \qquad \text{in } C(\mathcal{H}),
    \end{equation}
    where $\mathbb{I}_g^*$ is a Gaussian process indexed by
    $\mathcal{H}$ with mean $0$ and covariance function
    \begin{align*}
       \Cov\bigl( \mathbb{I}_g^*(h'), \mathbb{I}_g^*(h'')
       \bigr) & = \Bigl[ \sP\bigl( S_1^{(h')} \bigr) \sP\bigl(
                  S_1^{(h'')} \bigr) \Bigr]^{-1} \Biggl[ \sP\bigl(
                  T_1^{(h')} T_1^{(h'')} \bigr) \\
              & \hspace{11mm} - \sP\bigl( S_1^{(h')} T_1^{(h'')}
                  \bigr) \biggl( \frac{ \sP (T_1^{(h'')}) }
                  {\sP(S_1^{(h'')})} + \frac{ \sP(T_1^{(h')}) }
                  {\sP(S_1^{(h')})} \biggr) \\
              & \hspace{11mm} + \frac{ \sP(T_1^{(h')})
                  \sP(T_1^{(h'')}) }{\sP(S_1^{(h')})
                  \sP(S_1^{(h'')})} \sP\bigl( S_1^{(h')} S_1^{(h'')}
                  \bigr) \Biggr].
    \end{align*}
    Consequently,
    \begin{equation}
      \label{eq:wcIhatiid}
      n^{1/2} \bigl( \hat{I}_g(\cdot) - I_g(\cdot) \bigr) \cd
      \tilde{\mathbb{I}}_g (\cdot) \qquad \text{in } C(\mathcal{H}),
    \end{equation}
    where $\tilde{\mathbb{I}}_g$ is a Gaussian process indexed by
    $\mathcal{H}$ with mean $0$ and covariance function
    \begin{equation*}
      \Cov\bigl( \tilde{\mathbb{I}}_g(h'), \tilde{\mathbb{I}}_g(h'')
      \bigr) = E(N_1) \Cov\bigl( \mathbb{I}_g^*(h'),
      \mathbb{I}_g^*(h'') \bigr).
    \end{equation*}
    In~\eqref{eq:wcIhatmc} $\hat{I}_g(h)$ is interpreted as
    $\hat{I}_g(h) = \bigl( \sum_{r=1}^R T_r^{(h)} \bigr) /
    \sum_{r=1}^R S_r^{(h)}$, and the limit is as $R \rightarrow
    \infty$, whereas in~\eqref{eq:wcIhatiid} $\hat{I}_g(h)$ and the
    limit are interpreted differently: $\hat{I}_g(h) = \bigl(
    \sum_{i=1}^n g(\theta_i) f_h(\theta_i) \bigr) / \sum_{i=1}^n
    f_h(\theta_i)$, and $n = \sum_{r=1}^R N_r$ is random.
  \end{enumerate}
\end{theorem}
\begin{remark}
  \label{rem:conf-bands}
  Here we discuss how to form globally valid confidence bands for
  $I(\cdot)$ (we drop the subscript ``$g$'' to lighten the
  notation).  We would like to proceed as follows.  Having
  established that $n^{1/2} \bigl( \hat{I}(\cdot) - I(\cdot) \bigr)
  \cd \tilde{\mathbb{I}}(\cdot)$, we find the distribution of
  $\sup_h |\tilde{\mathbb{I}}(h)|$.  If $s_{\alpha}$ is the $(1 -
  \alpha)$-quantile of this distribution, then the band $\hat{I}(h)
  \pm n^{-1/2} s_{\alpha}$ has asymptotic coverage probability equal
  to $1 - \alpha$.  Unfortunately, except for very unusual cases,
  the distribution of $\sup_h |\tilde{\mathbb{I}}(h)|$ cannot be
  obtained analytically.  Spectral methods can be used for the
  problem of forming confidence intervals for $I(h)$ for a single
  value of $h$, but not for the problem of forming confidence bands.
  We know of no way to use regenerative simulation to construct
  confidence bands.  However, the method of batching works, as
  follows.

  For a positive integer $M$, the sequence $\theta_1, \ldots,
  \theta_n$ is broken up into $M$ consecutive pieces, each of length
  $n/M$ (we are ignoring divisibility issues).  For $m = 1, \ldots,
  M$, let $\hat{I}^{(m)}(h)$ be the estimate of $I(h)$ based on
  batch $m$, and let
  \begin{equation*}
    \mathcal{I}_m = \sup_h \Bigl( \frac{n}{M} \Bigr)^{1/2}
    |\hat{I}^{(m)}(h) - \hat{I}(h)|, \qquad \bar{\mathcal{I}}_m =
    \sup_h \Bigl( \frac{n}{M} \Bigr)^{1/2} |\hat{I}^{(m)}(h) -
    I(h)|.
  \end{equation*}
  (The difference between $\mathcal{I}_m$ and $\bar{\mathcal{I}}_m$
  is that the latter is not computable, because it involves the
  unknown function $I(\cdot)$.)  Let $\bar{\mathcal{I}}_{[1]} \leq
  \bar{\mathcal{I}}_{[2]} \leq \cdots \leq \bar{\mathcal{I}}_{[M]}$
  be the order statistics of the sequence $\bar{\mathcal{I}}_1,
  \ldots, \bar{\mathcal{I}}_M$ and, similarly, let
  $\mathcal{I}_{[1]} \leq \mathcal{I}_{[2]} \leq \cdots \leq
  \mathcal{I}_{[M]}$ be the order statistics of the sequence
  $\mathcal{I}_1, \ldots, \mathcal{I}_M$.  Now suppose that $M
  \rightarrow \infty$ in such a way that $n/M \rightarrow \infty$.
  Below is the outline of an argument which shows that the band
  $\hat{I}(h) \pm n^{-1/2} \mathcal{I}_{[(1-\alpha)M]}$ has coverage
  probability that is asymptotically equal to $1 - \alpha$.
  \begin{enumerate}[itemsep=0mm,topsep=1mm,leftmargin=4.5mm]
  \item For every $m$, we have $\bar{\mathcal{I}}_m \cd \sup_h
    |\tilde{\mathbb{I}}(h)|$ by Theorem~\ref{thm:functional-wc-mc},
    and if the distribution of $\sup_h |\tilde{\mathbb{I}}(h)|$ is
    continuous, then $\bar{\mathcal{I}}_{[(1-\alpha)M]}$ converges
    in distribution to $\delta_{s_{\alpha}}$, the point mass at
    $s_{\alpha}$.
  \item Therefore the (uncomputable) band $\hat{I}(h) \pm n^{-1/2}
    \bar{\mathcal{I}}_{[(1-\alpha)M]}$ has coverage probability that
    converges to $1 - \alpha$.
  \item The difference between $\mathcal{I}_m$ and
    $\bar{\mathcal{I}}_m$ is small uniformly in $m$; more precisely,
    we have $\max_{1 \leq m \leq M} |\mathcal{I}_m -
    \bar{\mathcal{I}}_m| \cP 0$.  Therefore the band $\hat{I}(h) \pm
    n^{-1/2} \mathcal{I}_{[(1-\alpha)M]}$ also has coverage
    probability that converges to $1 - \alpha$.
  \end{enumerate}
  Details are given in \citet{Park:2015}.
\end{remark}
\begin{remark}
  \label{rem:st}
  We have seen that for any $h_1 \in \mathcal{H}$, if $\theta_1,
  \theta_2, \ldots$ is a Markov chain with invariant distribution
  $\nu_{h_1,y}$ then, under certain regularity conditions, the
  estimates $B_n(h)$ and $\hat{I}_g(h)$ are consistent and
  asymptotically normal.  These estimates can be unstable, however,
  if $h$ is far from $h_1$, and there may not exist a single value
  of $h_1$ that gives rise to estimates that are stable for all $h
  \in \mathcal{H}$.  Serial tempering (\citet{MarinariParisi:1992,
  GeyerThompson:1995}; see also \citet{Geyer:2011} for a review, and
  \citet{Tan-Zhiqiang:2014} for recent developments) can be very
  effective in handling this problem.  A very brief description of
  the method in the present context is as follows.  We select $m$
  points $h_1, \ldots, h_m \in \mathcal{H}$; these should be taken
  to ``cover'' $\mathcal{H}$ in the sense that every $h$ in
  $\mathcal{H}$ is ``close'' to at least one of the $h_j$'s.  Let
  $\mathcal{L} = \{ 1, \ldots, m \}$; the elements of $\mathcal{L}$
  are called ``labels.''  For each $j \in \mathcal{L}$, let $\Phi_j$
  be a Markov transition function with invariant distribution
  $\nu_{h_j,y}$.  A Markov chain running on the state space
  $\mathcal{L} \times \Theta$ is generated as follows.  If the
  current state of the chain is $(j, \theta)$, a new label $j'$ is
  generated, and $\theta'$ is generated from the distribution
  $\Phi_{j'}(\theta, \cdot)$.  The mechanism for generating the
  labels is set up in such a way that the $\theta$-sequence has
  invariant distribution $\sum_{j=1}^m \alpha_j \nu_{h_j,y}$, where
  the $\alpha_j$'s are all nearly equal to $1/m$.  From the
  $\theta$-sequence, the quantities $B(h)$ and $I_g(h)$ can be
  estimated in a stable manner for any $h$ which is ``close'' to at
  least one of the $h_j$'s, or more precisely, for any $h$ such that
  $\nu_h$ is ``close'' to at least one of $\nu_{h_1}, \ldots,
  \nu_{h_m}$.  \textsl{The results of this paper do not require that
  the sequence $\theta_1, \theta_2, \ldots$ have invariant
  distribution equal to $\nu_{h_1,y}$ for some $h_1 \in
  \mathcal{H}$, and in fact the invariant distribution can be a
  mixture $\sum_{j=1}^m \alpha_j \nu_{h_j,y}$, for judiciously
  chosen $h_1, \ldots, h_m$, as described above, for example.}
\end{remark}

\section{Illustrations}
\label{sec:illustrations}
Here we present two illustrations.  The first deals with the
so-called Latent Dirichlet Allocation model, which is used for
organizing and searching electronic documents.  The version of the
model we discuss is indexed by a two-dimensional hyperparameter.
Our focus will be on obtaining globally-valid confidence sets for a
certain posterior expectation of interest.  For the data set we
study, the amount of time it takes to run the Markov chain is a
significant issue because each cycle has length $7788$.  We will use
the results of Section~\ref{sec:pe} to determine the minimal Markov
chain length that is needed to obtain acceptably narrow confidence
regions.  The second illustration deals with a model for Bayesian
variable selection in linear regression.  For this situation our
interest will be on hyperparameter selection, and we will use the
results of Section~\ref{sec:argmax}.  We will see that for the data
set we use, a very modest Markov chain length is all that is needed
to produce narrow confidence sets for the empirical Bayes choice of
the hyperparameters.

\subsection{Sensitivity Analysis in the Latent Dirichlet Allocation Model}
\label{sec:salda}
Probabilistic topic modelling is an area of machine learning that
deals with methods for understanding, summarizing, and searching
large electronic archives.  Traditional keyword-based searches are
very fast, but have important deficiencies.  Suppose we are
interested in searching for all statistical papers that deal with
censored data.  A search using the keywords ``censored data'' will
not return papers that use the expression ``incomplete data''.  In
topic-based searches, we do a search based on a concept or topic.  A
topic is not an expression; it is, by definition, a distribution
over a set of expressions.  Thus the topic mentioned above gives a
lot of mass to expressions like ``Kaplan-Meier'', ``censored data'',
and ``incomplete data'', and little mass to expressions like
``spectral decomposition''.

Latent Dirichlet Allocation (LDA,
\citeauthor{BleiNgJordan:2003}~\citeyear{BleiNgJordan:2003}) is by far
the most used topic model.  We will consider the version of the model
that deals only with individual words, as opposed to expressions
consisting of several words.  Suppose we have a corpus of documents,
for example a set of articles from \textsl{The New York Times}, and
these span several different topics, such as sports, medicine,
politics, etc.  The words in the documents come from a vocabulary
$\mathcal{V}$, which is a set consisting of $V$ words $u_1, \ldots,
u_V$.  For each document, the data we have for that document is a
sequence of length $V$ consisting of the number of times that word
$u_v$ occurs, for $v = 1, \ldots, V$.  In LDA, we imagine that for
each word in each document, there is a latent (i.e.\ unobserved)
variable indicating a topic from which that word is drawn.  LDA
enables us to make inference on these latent variables, and therefore,
on the topics that are covered by each document as a whole.
Therefore, LDA enables us to cluster together documents which are
similar, i.e.\ documents which share common topics.  By its very
nature, LDA is completely automatic in how it defines the topics:
these are distributions over the vocabulary, and are themselves latent
variables.  To be more precise, in LDA there is no such thing as a
topic called ``sports''.  Instead, there is a distribution on
$\mathcal{V}$ which gives most of its mass to words like ``homerun'',
``marathon'', and ``NBA''.  A human is then free to call this
distribution ``sports'' if he/she wishes.

We now give more detail.  The vocabulary $\mathcal{V}$ is taken to
be the union of all the words in all the documents of the corpus,
after removing uninformative words (like ``the'' and ``of'').  There
are $D$ documents in the corpus, and for $d = 1, \ldots, D$,
document $d$ has $n_d$ words, $w_{d1}, \ldots, w_{dn_d}$.  The order
of the words is viewed as uninformative, so is neglected.  Each word
is represented as an index $1 \times V$ vector with a $1$ at the
$s^{\text{th}}$ element, where $s$ denotes the term selected from
the vocabulary.  Thus, document $d$ is represented by the vector
$\bw_d = (w_{d1}, \ldots, w_{dn_d})$ and the corpus is represented
by the vector $\bw = (\bw_1, \ldots, \bw_D)$.  The number of topics,
$K$, is finite and known.  By definition, a topic is a point in
$\simplex_V$, the $(V - 1)$-dimensional simplex.  For $d = 1,
\ldots, D$, for each word $w_{di}$, $z_{di}$ is an index $1 \times
K$ vector which represents the latent variable that denotes the
topic from which $w_{di}$ is drawn.  The distribution of $z_{d1},
\ldots, z_{dn_d}$ will depend on a document-specific variable
$\theta_d$ which indicates a distribution on the topics for document
$d$.  We will use $\Dir_L(a_1, \ldots, a_L)$ to denote the
finite-dimensional Dirichlet distribution on the $L$-dimensional
simplex.  Also, we will use $\Mult_L(b_1, \ldots, b_L)$ to denote
the multinomial distribution with number of trials equal to $1$ and
probability vector $(b_1, \ldots, b_L)$.  We will form a $K \times
V$ matrix $\bbeta$, whose $t^{\text{th}}$ row is the $t^{\text{th}}$
topic (how $\bbeta$ is formed will be described shortly).  Thus,
$\bbeta$ will consist of vectors $\beta_1, \ldots, \beta_K$, all
lying in $\simplex_V$.  Formally, LDA is described by the following
hierarchical model, in which $\eta, \alpha \in (0, \infty)$ are
hyperparameters:
\begin{enumerate}[itemsep=0mm,topsep=2mm,leftmargin=4.5mm]
\item $\beta_t \iid \Dir_V(\eta, \ldots, \eta), \, t = 1, \ldots,
  K$.
  \label{eq:LDA-d}
\item $\theta_d \iid \Dir_K(\alpha, \ldots, \alpha), \, d = 1, \ldots,
  D$, and the $\theta_d$'s are independent of the $\beta_t$'s.
  \label{eq:LDA-c}
\item Given $\theta_1, \ldots, \theta_D$, $z_{di} \iid
  \Mult_K(\theta_d), \, i = 1, \ldots, n_d, \, d = 1, \ldots, D$, and
  the $D$ vectors $(z_{11}, \ldots, z_{1n_1}), \ldots, (z_{D1},
  \ldots, z_{Dn_D})$ are independent.
  \label{eq:LDA-b}
\item Given $\bbeta$ and the $z_{di}$'s, $w_{di}$ are independently
  drawn from the row of $\bbeta$ indicated by $z_{di}, \, i = 1,
  \ldots, n_d, \, d = 1, \ldots, D$.
  \label{eq:LDA-a}
\end{enumerate}

From the model statement, we see that there is a latent topic variable
for every word that appears in the corpus.  Thus it is possible that a
document spans several topics.  However, because there is a single
$\theta_d$ for document $d$, the model encourages different words in
the same document to have the same topic.  Also note that the
hierarchical nature of LDA encourages different documents to share the
same topics.  This is because $\bbeta$ is chosen once, at the top of
the hierarchy, and is shared among the $D$ documents.  Let $\btheta =
(\theta_1, \ldots, \theta_D)$, $\bz_d = (z_{d1}, \ldots, z_{dn_d})$
for $d = 1, \ldots, D$, $\bz = (\bz_1, \ldots, \bz_D)$, and let $\bpsi
= (\bbeta, \btheta, \bz)$.  The model is indexed by the hyperparameter
vector $h = (\eta, \alpha)$.  For any given $h$, lines~$1$--$3$ induce
a prior distribution on $\bpsi$, which we denote by $\nu_h$.  Line~$4$
gives the likelihood.  The words $\bw$ are observed, and we are
interested in $\nu_{h,\bw}$, the posterior distribution of $\bpsi$
given $\bw$ corresponding to $\nu_h$.

The hyperparameter $h$ has a strong effect on the distribution of the
parameters of the model.  For example, when $\eta$ is large, the
topics tend to be probability vectors which spread their mass evenly
among many words in the vocabulary, whereas when $\eta$ is small, the
topics tend to put most of their mass on only a few words.  Also, when
$\alpha$ is large, each document tends to involve many different
topics; on the other hand, in the limiting case where $\alpha
\rightarrow 0$, each document involves a single topic, and this topic
is randomly chosen from the set of all topics.

In the literature, the following choices for $h = (\eta, \alpha)$
have been presented: $h_{\text{GS}} = (0.1, 50/K)$, used in
\citet{GriffithsSteyvers:2004}; $h_{\text{A}} = (0.1, 0.1)$, used in
\citet{AsuncionEtal:2009}; and $h_{\text{RS}} = (1/K, 1/ K)$, used
in the \texttt{Gensim} topic modelling package
\citep{RehurekSojka:2010}, a well-known package used in the topic
modelling community.  These choices are ad-hoc, and not based on any
principle; nevertheless, they do get used.
\citet{BleiNgJordan:2003} propose $h_0 = \argmax_h m_{\bw}(h)$, as
we do, but their approach for estimating $h_0$ is quite a bit
different from ours, and involves a combination of the EM algorithm
and ``variational inference.''  Very briefly, $\bw$ is viewed as
``observed data,'' and $\bpsi$ is viewed as ``missing data.''
Because the ``complete data likelihood'' $p_h(\bpsi, \bw)$ is
available, the EM algorithm is a natural candidate for estimating
$\argmax_h m_{\bw}(h)$, since $m_{\bw}(h)$ is the ``incomplete data
likelihood.''  But the E-step in the algorithm is infeasible because
it requires calculating an expectation with respect to the
intractable distribution $\nu_{h,\bw}$.  \citet{BleiNgJordan:2003}
substitute an approximation to this expectation.  Unfortunately,
because there are no useful bounds on the approximation, and because
the approximation is used at every iteration of the algorithm, there
are no results regarding the theoretical properties of this method.
Determination of the hyperparameter is currently an open problem in
LDA modelling \citep{WallachEtal:2009}.

We illustrate our methodology on a corpus of documents from the
English Wiki\-pedia, originally created by \citet{George:2015}.
When a Wikipedia article is created, it is typically tagged to one
or more categories, one of which is the ``primary category.''  The
corpus consists of $8$ documents from the category
\textsl{Leopardus}, $8$ from the category \textsl{Lynx}, and $7$
from \textsl{Prionailurus}, and we took $K = 3$, as in
\citet{George:2015}.  There are $303$ words in the vocabulary, and
the total number of words in the corpus is $7788$.  The data set is
relatively small.  However, it is challenging to analyze because the
topics are very close to each other, so in the posterior
distribution there is a great deal of uncertainty regarding the
latent topic indicator variables, and this is why we chose this data
set.

A reader of a given article may wish to look at related articles, so
a question of interest is whether the topics for two given documents
are nearly the same.  One way to word this question precisely is to
ask what is the posterior probability that $\| \theta_i - \theta_j
\| \leq \epsilon$, where $i$ and $j$ are the indices of the
documents in question and $\epsilon$ is some user-specified small
number.  Here, $\| \cdot \|$ denotes ordinary Euclidean distance.
This posterior probability will of course depend on $h$, and we
would like to view the estimates of the posterior probability as $h$
varies, together with (simultaneous) error margins.

\begin{figure}[h]
  \centering
  \vspace{-24mm} %
  \includegraphics[width=.8\linewidth]{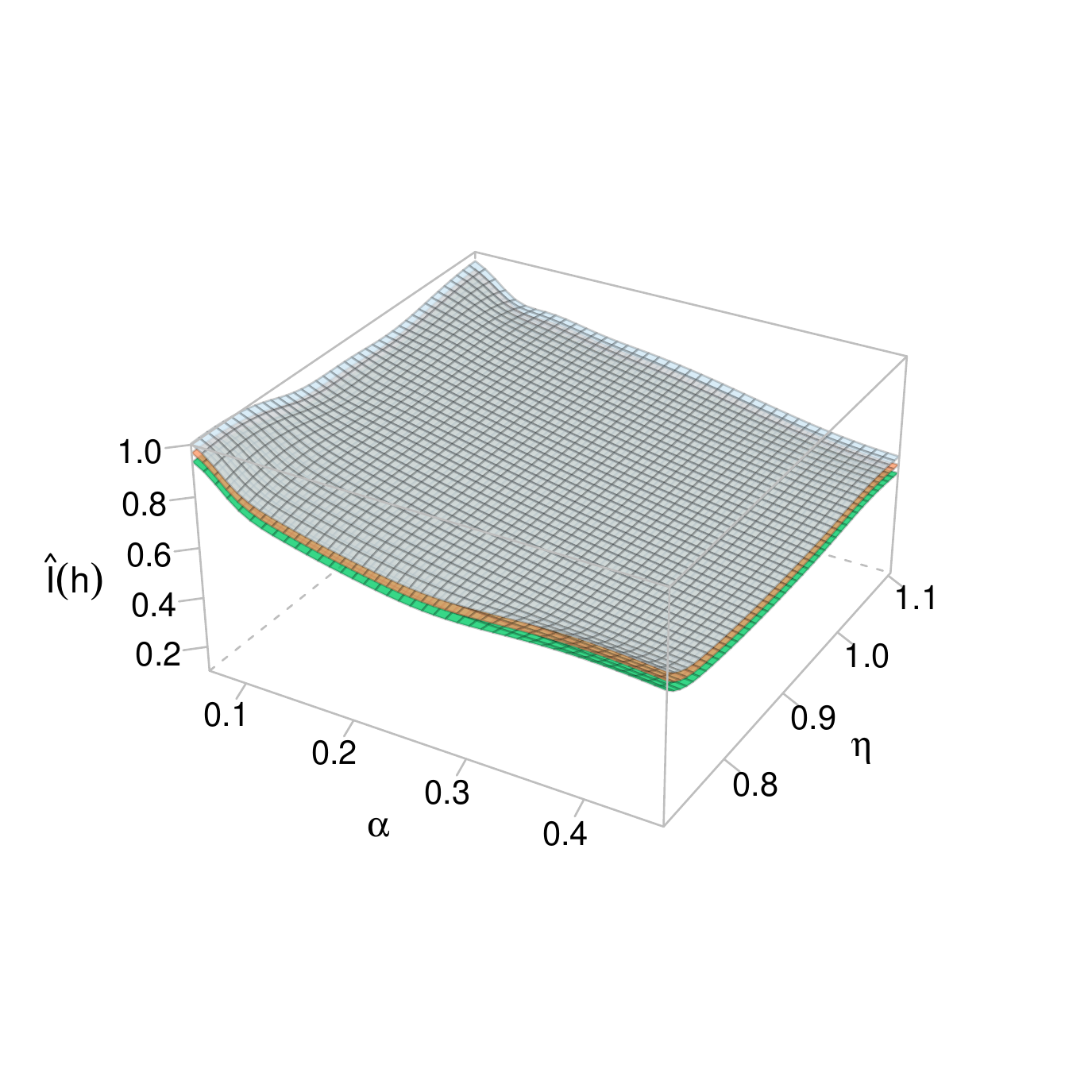}
  \vspace{-23mm} %
  \caption{Estimates with confidence region for $I(h) =
  \nu_{h,\bw}(\| \theta_7 - \theta_8 \| \leq .05)$, the posterior
  probability that the topics for documents~$7$ and~$8$ of the
  Wikipedia corpus are ``very close.''  The plot shows that this
  posterior probability varies considerably with $h$, and suggests
  that care be taken in choosing the hyperparameter.}
  \label{fig:post-prob-cr}
\end{figure}

To this end, we used the methodology developed in
Section~\ref{sec:pe} for simultaneous estimation of posterior
expectations (here the posterior expectations of the indicator of a
set).  The warning given in Remark~\ref{rem:st} regarding the high
variance of the simple single-chain estimate~\eqref{eq:main-conv}
applies, and we use instead a serial tempering chain (cf.\
Remark~\ref{rem:st}), the details of which are given in the next
paragraph.  We consider documents~$7$ and~$8$, which are the
articles ``Pampas cat'' and ``Pantanal cat'' under the Wikipedia
category Leopardus, and we are interested in the posterior
probability of the event $\| \theta_7 - \theta_8 \| \leq .05$.  Our
estimate of $\argmax_h m_{\bw}(h)$ is $h_n = (\eta_n, \alpha_n) =
(.915, .245)$, and the estimate of the posterior probability under
the empirical Bayes choice of $h$ is $\nu_{h_n,\bw}(\| \theta_7 -
\theta_8 \| \leq .05) = .7039$.  For the other choices of $h$ we
have $\nu_{h_{\text{GS}},\bw}(\| \theta_7 - \theta_8 \| \leq .05) =
.1619$, $\nu_{h_{\text{A}},\bw}(\| \theta_7 - \theta_8 \| \leq .05)
= .1498$, and $\nu_{h_{\text{RS}},\bw}(\| \theta_7 - \theta_8 \|
\leq .05) = .1298$, and we see that all three are far from the
estimate based on the empirical Bayes choice of $h$.  We also
calculated the ratio of the marginal likelihood of $h_n$ to the
marginal likelihood of each of $h_{\text{GS}}$, $h_{\text{A}}$, and
$h_{\text{RS}}$ and noted that each ratio is astronomically large.
Therefore, none of these values of $h$ are deemed even remotely
plausible, and as these choices of $h$ do not have any theoretical
basis, there is no credibility to posterior probability estimates
based on them.  Figure~\ref{fig:post-prob-cr} gives a plot of the
estimate of $\nu_{h,\bw}(\| \theta_7 - \theta_8 \| \leq .05)$,
together with a globally valid confidence set of level $.95$ over a
relatively small region centered at $h_n$.  The figure shows that
the posterior probabilities vary greatly with $h$, ranging from
$.553$ to $.972$, even over a small $h$-region, underscoring the
fact that the choice of hyperparameter should be made carefully.

Our serial tempering chain is based on the ``augmented collapsed
Gibbs sampler'' developed in \citet{George:2015}, and which runs on
the entire set of latent variables $(\bbeta, \btheta, \bz)$.  A
single cycle of this Markov chain runs over $7788$ nodes.  To form
the confidence region we used the construction described in
Remark~\ref{rem:conf-bands}.  We took the grid size for the chain
(``$m$'' in Remark~\ref{rem:conf-bands}) to be $105$, with the $105$
reference values evenly spaced over the $h$-region.  With this
choice the chain gives very stable estimates.  The length of the
chain was $500{,}000$, and the number of batches was $707$ (roughly
the square root of the chain length).  With this chain length the
confidence region is adequately narrow, and with a length of only
$50{,}000$ it was not.

\subsection{Hyperparameter Choice for Bayesian Variable Selection in
Linear Regression}
\label{sec:bvslr}
The most commonly used setup for variable selection in Bayesian
linear regression is described as follows.  We have a response
vector $Y = (Y_1, \ldots, Y_m)^{\top}$ and a set of potential
predictors $X_1, \ldots, X_q$, each a vector of length $m$.  Every
subset of predictors is identified with a binary vector $\gamma =
(\gamma_1, \ldots, \gamma_q)^{\top} \in \{ 0, 1 \}^q$, where
$\gamma_j = 1$ if $X_j$ is included in the model and $\gamma_j = 0$
otherwise.  For every $\gamma$, we have a model given by
\begin{equation*}
  Y = 1_m \beta_0 + X_{\gamma} \beta_{\gamma} + \epsilon,
\end{equation*}
where $1_m$ is the vector of $m$ $1$'s, $X_{\gamma}$ is the design
matrix whose columns consist of the predictor vectors corresponding
to $\gamma$, $\beta_{\gamma}$ is the vector of coefficients for that
subset, and $\epsilon \sim \mathcal{N}_m(0, \sigma^2 I)$.  For this
setup, the unknown parameter is $\theta = (\gamma, \sigma, \beta_0,
\beta_{\gamma})$, which includes the indicator of the subset of
variables that go into the regression model.  The prior on $\theta$
is a hierarchy in which we first select the variables that go into
the regression model, then a ``non-informative prior'' is given to
$(\sigma^2, \beta_0)$, and given $\gamma$ and $\sigma$, we choose
$\beta_{\gamma}$ from some proper distribution.  The specific
instance of this model that we will consider is indexed by two
hyperparameters, $w \in (0, 1)$ and $g > 0$, and is given in detail
as follows:
\begin{subequations}
  \label{eq:vsblm}
  \begin{alignat}{2}
    \label{eq:vsblm-a}
  & \text{given } \gamma, \sigma, \beta_0, \beta_{\gamma}, \quad & Y                   & \sim \mathcal{N}_{m}(1_m \beta_0 + X_{\gamma} \beta_{\gamma}, \sigma^2 I),               \\[1.5mm]
    \label{eq:vsblm-b}
  & \text{given } \gamma, \sigma, \quad                          & \beta_{\gamma}      & \sim \mathcal{N}_{q_{\gamma}} \bigl( 0, g \sigma^2 (X^{\top}_{\gamma} X_{\gamma})^{-1} \bigr), \\[1.5mm]
    \label{eq:vsblm-c}
  &                                                              & (\sigma^2, \beta_0) & \sim p(\beta_0, \sigma^2) \propto 1 / \sigma^2,                                       \\[1.5mm]
    \label{eq:vsblm-d}
  & \quad                                                        & \gamma              & \sim p(\gamma) = w^{q_{\gamma}} (1 - w)^{q - q_{\gamma}}.
  \end{alignat}
\end{subequations}
The prior on $\gamma$ given by~\eqref{eq:vsblm-d} is the so-called
independence Bernoulli prior, in which every variable goes into the
model with probability $w$, independently of all the other
variables.  In~\eqref{eq:vsblm-b}, $q_{\gamma} = \sum_{j=1}^q
\gamma_j$ is the number of predictors that go in the regression, and
the prior on $\beta_{\gamma}$ is Zellner's $g$-prior
\citep{Zellner:1986}.  Because $(\sigma^2, \beta_0)$ is given an
improper prior (line~\eqref{eq:vsblm-c}), the prior on $\theta$ is
improper; however, it turns out that the posterior distribution of
$\theta$ is proper.  Models of the type~\eqref{eq:vsblm} were
introduced by \citet{MitchellBeauchamp:1988} and have been studied
in dozens of papers; see \citet{LiangEtal:2008} for a review.

The hyperparameter $h = (w, g)$ plays a critical role: if $w$ is
small and $g$ is large, the prior $\nu_h$ concentrates its mass on
models with few variables and large coefficients, while if $w$ is
large and $g$ is small, $\nu_h$ concentrates its mass on models with
many variables and small coefficients.  (To appreciate the
importance of the role played by $h$, note that
\citet{GeorgeFoster:2000} have shown that for the slightly different
version of~\eqref{eq:vsblm} in which $\sigma^2$ is assumed known,
$h$ can be chosen so that the highest posterior probability model is
exactly the best model under the AIC/$C_p$, BIC, or RIC criteria.)
Thus, $h$ effectively determines the method that is used to carry
out variable selection, so it is important to choose it properly.

Unless $q$ is relatively small ($q$ less than $20$ or $25$), the
posterior distribution of $\theta = (\gamma, \sigma, \beta_0,
\beta_{\gamma})$ is intractable, because to compute it we need to
calculate $2^q$ integrals \citep{GeorgeFoster:2000}.
\citet{SmithKohn:1996} developed a Markov chain algorithm which runs
only on $\gamma$, the other variables being integrated out.  Their
chain is a simple Gibbs sampler which runs on the vector $(\gamma_1,
\ldots, \gamma_q)^{\top}$, updating one component at a time.  This
chain does not fit into our framework, which requires a Markov chain
that runs on $\theta = (\gamma, \sigma, \beta_0, \beta_{\gamma})$.
\citet{Buta:2010} developed a Markov chain, based on the
\citet{SmithKohn:1996} chain, which runs over $(\gamma, \sigma,
\beta_0, \beta_{\gamma})$.  (She proved that for her Markov chain,
the rate of convergence to the posterior distribution of $\theta$ is
exactly the same as the rate of convergence to the posterior
distribution of $\gamma$ for the \citet{SmithKohn:1996} chain, where
convergence is in terms of the absolute deviation norm.)  We will
use the chain developed by \citet{Buta:2010} for the analysis below.

To implement the methods of this paper, we need a ``ratio of
densities $\nu_{h_1} / \nu_{h_2}$'' (cf.\
equation~\eqref{eq:main-conv}).  Note that the prior distributions
are not absolutely continuous with respect to the product of
counting measure on $\{ 0, 1 \}^q$ and Lebesgue measure on $(0,
\infty) \times \Real_+ \times \Real^{q+1}$ (the dimension of
$\beta_{\gamma}$ is not fixed).  The ``ratio of densities $\nu_{h_1}
/ \nu_{h_2}$'' then needs to be replaced by the Radon-Nikodym
derivative.  To be precise, let $\bar{\nu}_h$ be the distribution on
$\theta$ induced by~\eqref{eq:vsblm-d},~\eqref{eq:vsblm-c},
and~\eqref{eq:vsblm-b}.  Then~\eqref{eq:main-conv} becomes
\begin{equation*}
  \frac{1}{n} \sum_{i=1}^n \biggl[ \frac{d\bar{\nu}_h}
  {d\bar{\nu}_{h_1}} \biggr] (\theta_i) \cas \int \biggl[
  \frac{d\bar{\nu}_h} {d\bar{\nu}_{h_1}} \biggr] (\theta) \,
  \bar{\nu}_{h_1,y} (d\theta) = \frac{m_y(h)} {m_y(h_1)}.
\end{equation*}
The Radon-Nikodym derivative was obtained in \cite{Doss:2007} and is
given by
\begin{equation*}
  \biggl[ \frac{d\bar{\nu}_{h_1}} {d\bar{\nu}_{h_2}} \biggr]
  (\theta) = \biggl( \frac{w_1}{w_2} \biggr)^{q_{\gamma}} \biggl(
  \frac{1 - w_1}{1 - w_2} \biggr)^{q-q_{\gamma}} \times \, \frac{
  \phi_{q_{\gamma}} \bigl( \beta_{\gamma}; 0, g_1 \sigma^2
  (X'_{\gamma} X_{\gamma})^{-1} \bigr) } { \phi_{q_{\gamma}} \bigl(
  \beta_{\gamma}; 0, g_2 \sigma^2 (X'_{\gamma} X_{\gamma})^{-1}
  \bigr) },
\end{equation*}
where $\phi_d (u; a, V)$ is the density of the $d$-dimensional
normal distribution with mean $a$ and covariance $V$, evaluated at
$u$.

\begin{figure}[h]
  \includegraphics[width=.497\linewidth]{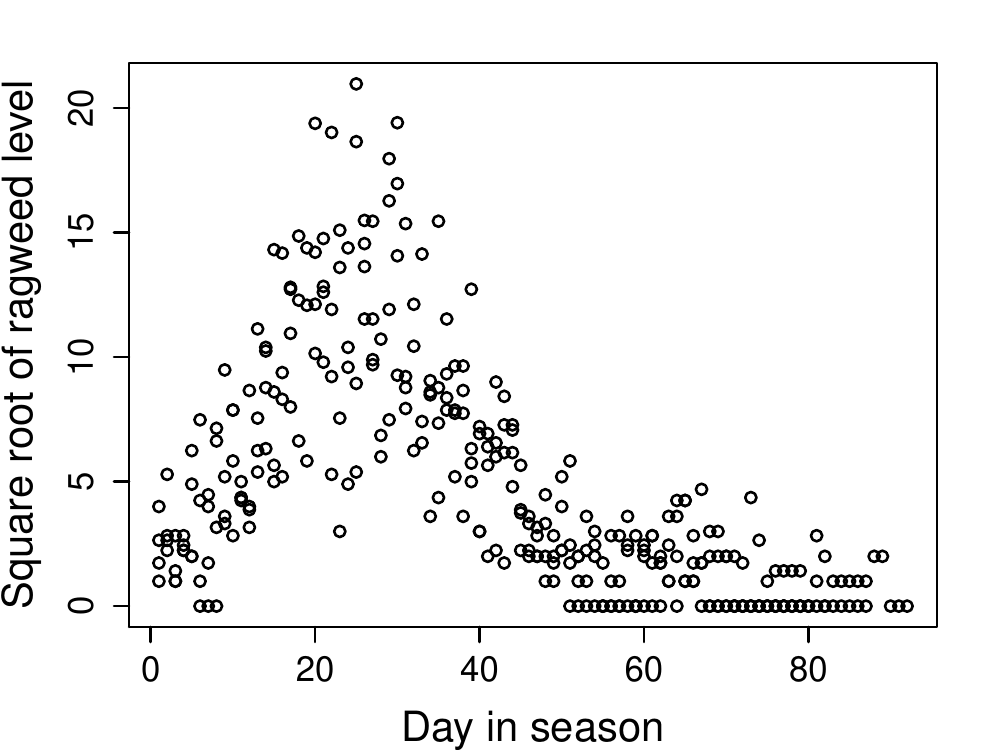}
  \includegraphics[width=.497\linewidth]{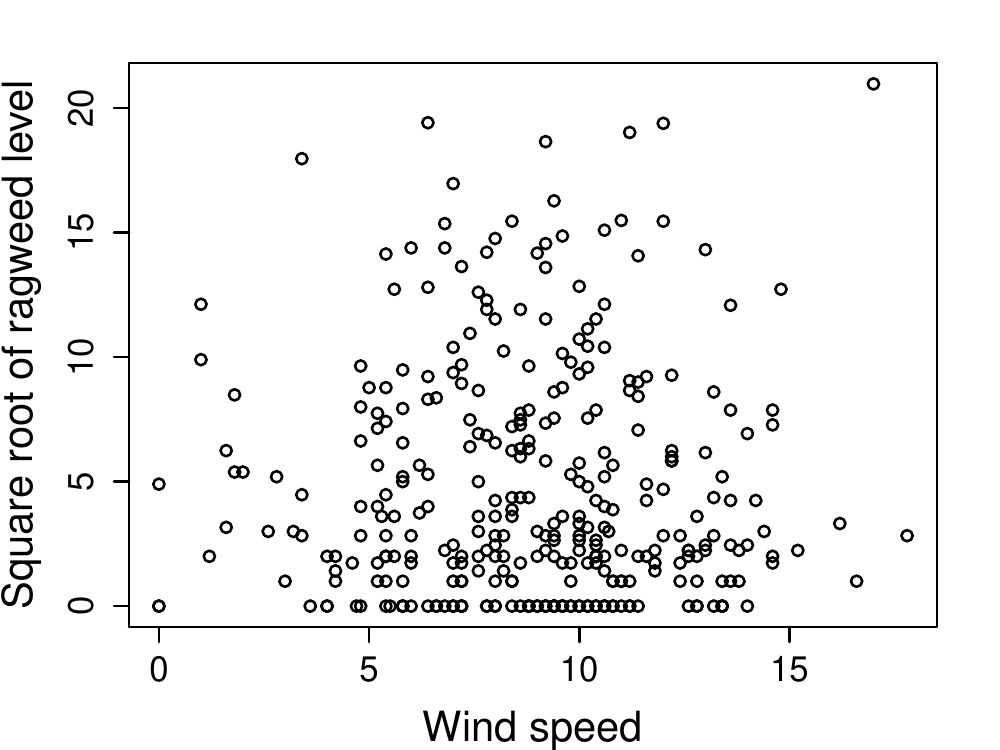}
  \caption{Scatterplots of response against each of two predictors
  for the ragweed data set.}
  \label{fig:ragweed-vs-predictors}
\end{figure}

\vspace{3mm} %
For our illustration we consider the ragweed data
of~\citet{StarkEtal:1997}, who were interested in determining how
meteorological variables can be used to forecast ragweed pollen
levels.  The response variable is the ragweed level (grains/m$^3$)
for $335$ days in Kalamazoo, Michigan, USA\@.  Although the data set
contains other predictors, we restrict our analysis to two:
\textsf{day} (day number in the current ragweed pollen season) and
\textsf{wind} (wind speed forecast in knots for following day).
Following~\citet{RuppertWandCarroll:2003}, we take the square root
of the ragweed level as the response.
Figure~\ref{fig:ragweed-vs-predictors} gives separate plots of the
response versus each of the two predictors.  From the figure we see
that the effect of \textsf{day} is certainly nonlinear, but whether
\textsf{wind} acts nonlinearly is not clear.

We fit each of the two predictors nonparametrically via cubic
regression splines involving $10$ equally spaced knots.  Hence the
model we use has the form
\begin{align*}
  Y_i & = \textstyle \beta_0 + \alpha_1 \textsf{day}_i + \alpha_2
          \textsf{day}_i^2 + \alpha_3 \textsf{day}_i^3 +
          \sum_{t=1}^{10} \alpha_{t+3} (\textsf{day}_i -
          \tilde{d}_t)_+^3 \\[1mm]
      & \hspace{8mm} + \textstyle \beta_1 \textsf{wind}_i + \beta_2
          \textsf{wind}_i^2 + \beta_3 \textsf{wind}_i^3 +
          \sum_{t=1}^{10} \beta_{t+3} (\textsf{wind}_i -
          \tilde{w}_t)_+^3 + \epsilon_i,
\end{align*}
for $i = 1, \ldots, 335$, where $\tilde{d}_1 < \cdots <
\tilde{d}_{10}$ represent the knots for the \textsf{day} explanatory
variable, $\tilde{w}_1 < \cdots < \tilde{w}_{10}$ the knots for the
\textsf{wind} explanatory variable, and $(x)_+ = \max\{ 0, x \}$.
Note that there are $26$ coefficients that could be set to $0$, of
which $20$ correspond to knots along the domain of the two
predictors.  Our plan is to carry out the following two steps:
\begin{enumerate}[itemsep=0mm,topsep=1mm,leftmargin=4.5mm]
\item We form a point estimate and confidence region for $\argmax_h
  m_y(h)$ by running a Markov chain.
\item We estimate the posterior distribution of $\theta$ when the
  prior is $\nu_{h_n}$, where $h_n$ is the estimate of $\argmax_h
  m_y(h)$ obtained in Step~$1$, by running another Markov chain.
\end{enumerate}

\begin{figure}[h]
  \setlength{\belowcaptionskip}{-2.1mm}
  \includegraphics[width=.497\linewidth]{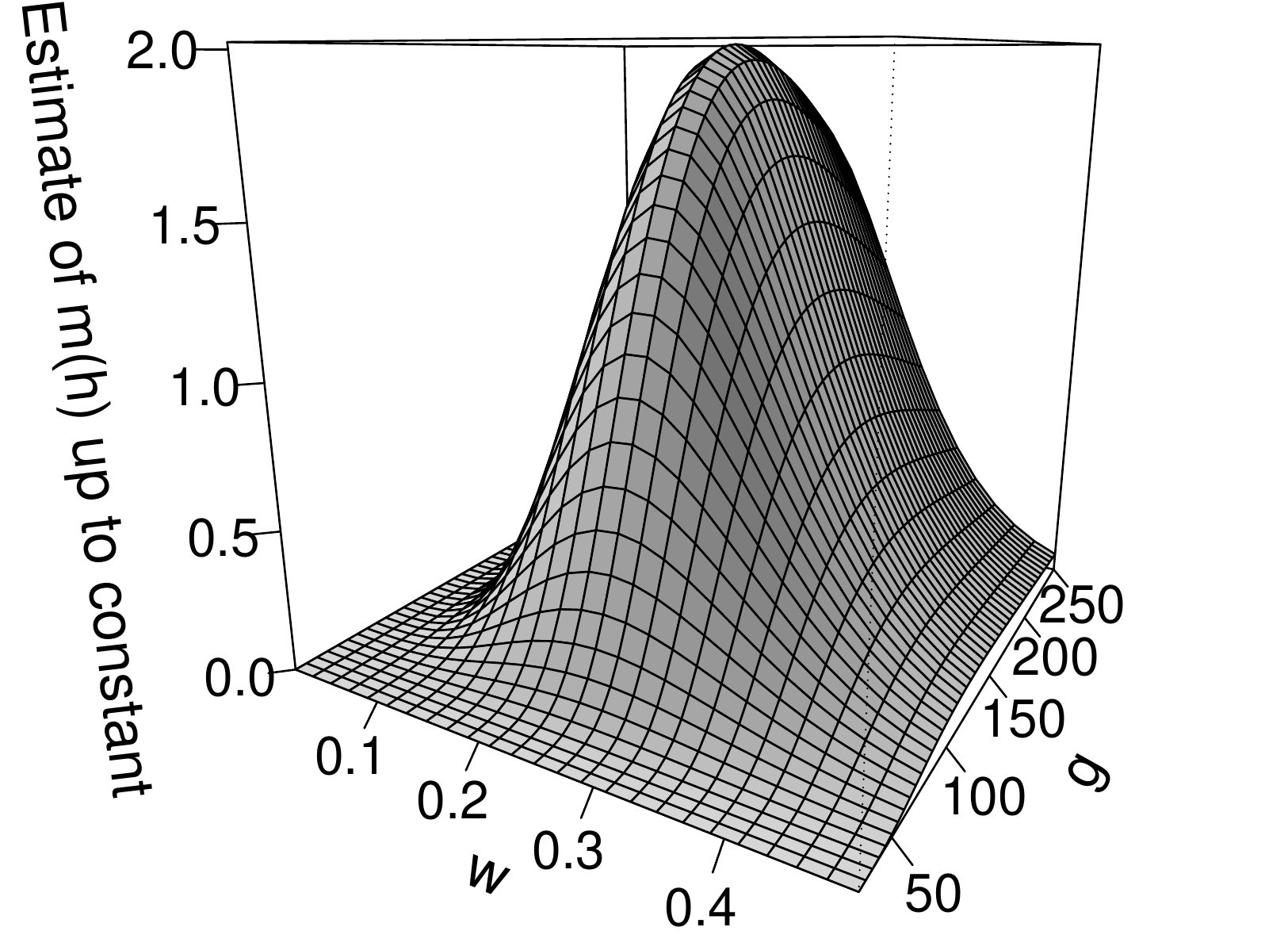}
  \includegraphics[width=.37\linewidth]{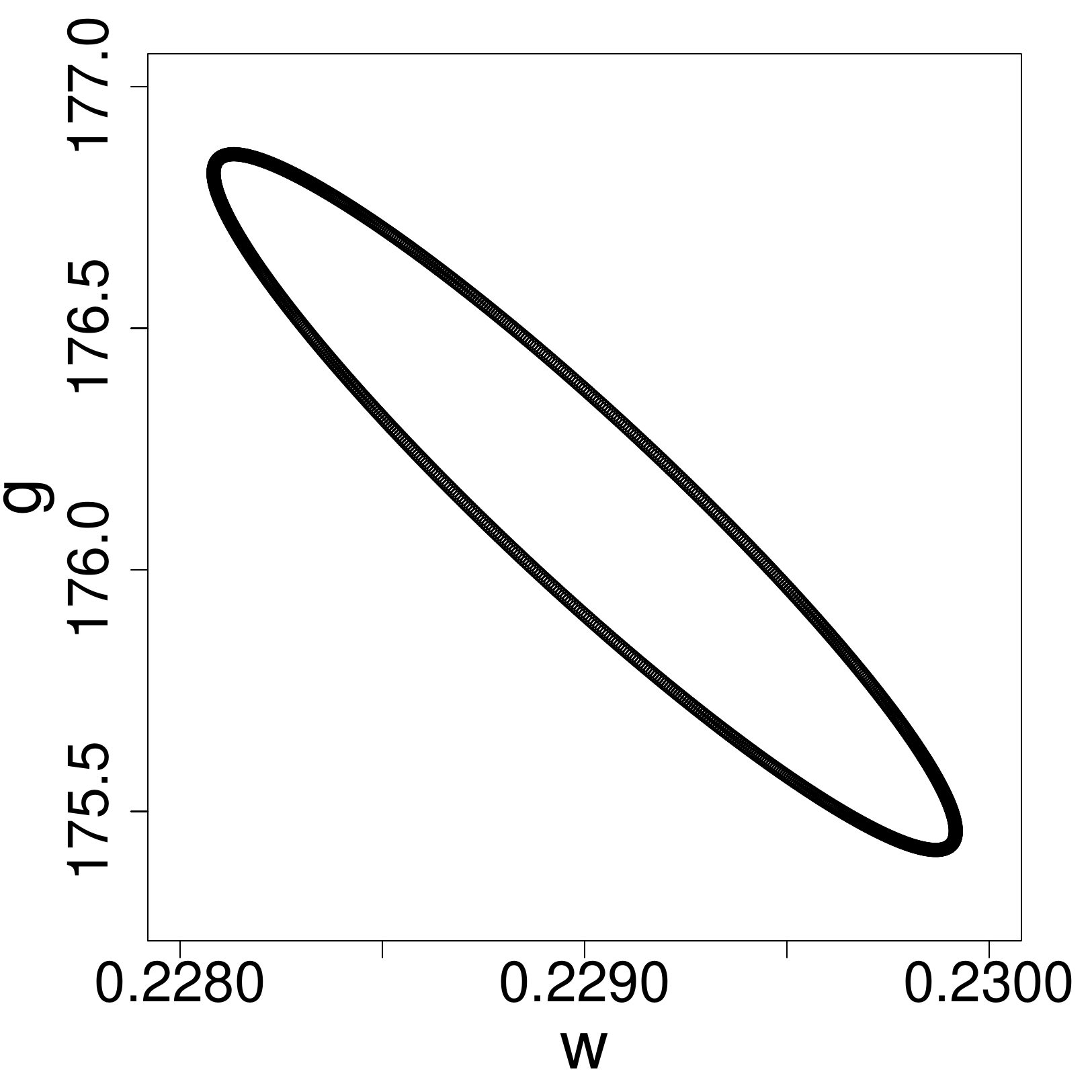}
  \caption{Left Panel: Estimate of the marginal likelihood $m_y(h)$
  (up to a multiplicative constant).  The argmax is $(w_n, g_n) =
  (.23, 176)$, and the small value of $w_n$ suggests a sparse model.
  Right Panel: Confidence region for $\argmax_h m_y(h)$.  The tight
  region indicates that the small Markov chain length used is
  adequate.}
  \label{fig:cellipse}
\end{figure}

For Step~$1$ we ran a Markov chain of length $40{,}000$, using $h_1
= (.3, 100)$, from which we formed the surface $B_n(h)$, shown on
the left panel of Figure~\ref{fig:cellipse}.  The argmax of the
surface is $(.23, 176)$, and the $95\%$ confidence region for
\linebreak $\argmax_h m_y(h)$ is the ellipse shown in the right
panel of Figure~\ref{fig:cellipse}.  For Step~$2$, we ran a new
Markov chain, of length $10^5$.  For this chain, the highest
probability model is the model which selects the variables
\textsf{wind}, $\textsf{day}^2$, $\textsf{day}^3$, $(\textsf{day} -
\tilde{d}_3)_+^3$, and $(\textsf{day} - \tilde{d}_5)_+^3$.
Interestingly, this model is the same as the model selected by the
lasso, when we choose the tuning parameter by cross-validation.

Let $\mathcal{E}$ denote the ellipse.  Our theory tells us that we
are $95\%$ confident that $\argmax_h B(h) \in \mathcal{E}$, so we
should run chains with posterior distributions $\nu_{h,y}, \, h \in
\mathcal{E}$, and determine the highest posterior probability models
for all $h \in \mathcal{E}$.  By checking a few points on the
boundary of the ellipse, we saw that the ellipse is narrow enough so
that the highest probability model is the same for all $h \in
\mathcal{E}$.  Had this not been the case, we would have run the
Step~$1$ chain for more cycles, getting a ellipse that is more
narrow.

The value of $w$ that is selected is small, which reflects sparsity:
a small model is adequate for fitting the data.
We now put our approach in the context of the existing literature.
\citet{LiangEtal:2008} review methods for selecting $g$ in the
version of model~\eqref{eq:vsblm} in which $w$ is fixed at $1/2$.
The literature has several data-independent choices (e.g.\ $g =
\max(m, q^2))$, but these generally do not perform well.  As a
data-dependent choice, they propose $\hat{g} = \argmax_g m_y(g)$,
and to obtain it suggest an EM algorithm in which the model
indicator $\gamma$ is viewed as missing data.  Unfortunately, the
M-step in the algorithm involves a sum of $2^q$ terms.  Unless $q$
is relatively small, complete enumeration is not possible, and
\citet{LiangEtal:2008} propose summing only over the most
significant terms.  However, determining which terms these are may
be very difficult in some problems.  Our approach provides a
feasible way of obtaining the maximizer of the likelihood, and this
for the model in which both $w$ and $g$ are unknown.

\pagebreak[4]
\section*{APPENDIX}
\setcounter{equation}{0}
\renewcommand{\theequation}{A.\arabic{equation}}
\renewcommand{\thesection}{A}

\section*{Proof of Theorem~\ref{thm:gc-PakesPollard}}
In order to prove Theorem~\ref{thm:gc-PakesPollard}, we need a few
definitions and results from empirical process theory.  An envelope
$U$ is any function satisfying $|V| \le U$ for all $V \in
\mathcal{V}$.  For example, $\sup_{h\in\mathcal{H}} f_h$ is an
envelope for the class $\mathcal{F} = \{ f_h, \, h \in \mathcal{H}
\}$.
\begin{definition}[Definition 2.7 of \citet{PakesPollard:1989}]
  \label{S-def2.7}
  We say that the class $\mathcal{F}$ is Euclidean for the envelope
  $F$ if there exist positive constants $A$ and $b$ with the
  following property: if $0 < \epsilon \le 1$ and if $Q$ is a
  measure for which $\int F \, dQ < \infty$, then there are
  functions $f_1, \ldots, f_n$ in $\mathcal{F}$ such that
  \begin{enumerate}[itemsep=0mm,topsep=0mm]
  \item $n \le A \epsilon^{-b}$
  \item The class $\mathcal{F}$ is covered by the union of the
    closed balls (in the $L_1(Q)$ metric) with radius $\epsilon \int
    F \, dQ$ and centers $f_1, \ldots, f_n$; in other words, for
    each $f$ in $\mathcal{F}$, there is an $f_i, \, i = 1, \ldots,
    n$, with $\int |f - f_i| \, dQ \le \epsilon \int F \, dQ$.
  \end{enumerate}
  The constants $A$ and $b$ may not depend on $Q$.
\end{definition}
\begin{lemma}[Lemma 2.8 of \citet{PakesPollard:1989}]
  \label{S-lemma2.8}
  If $\mathcal{F}$ is Euclidean for the envelope $F$ and if $\int F
  \, dP < \infty$, then ${\| P_n - P \|}_{\mathcal{F}}$ converges to
  $0$ almost surely.
\end{lemma}
\begin{lemma}[Lemma 2.13 of \citet{PakesPollard:1989}]
  \label{S-lemma2.13}
  Let $\mathcal{F} = \{ f(\cdot, h) \colon h \in \mathcal{H} \}$ be
  a class of functions on $\Theta$ indexed by a bounded subset
  $\mathcal{H}$ of $\Real^k$.  If there exist an $\alpha > 0$ and a
  nonnegative function $\phi(\cdot)$ such that $|f(\theta, h) -
  f(\theta, h')| \le \phi(\theta) \| h - h' \|^{\alpha}$ for $\theta
  \in \Theta$ and $h, h' \in \mathcal{H}$, then $\mathcal{F}$ is a
  Euclidean class with the envelope $F(\cdot) = |f(\cdot, h^*)| + M
  \phi(\cdot)$, where $h^*$ is an arbitrary point of $\mathcal{H}$
  and $M = ( 2 k^{1/2} \sup_{\mathcal{H}} \| h - h^* \|)^{\alpha}$.
\end{lemma}

\vspace{-3mm} %
\paragraph{Proof of Theorem~\ref{thm:gc-PakesPollard}}
Recall that we have assumed that for almost all $\theta$, $\nabla_h
f_h(\theta)$ is continuous in $h \in \mathcal{H}$, and that
$\mathcal{H}$ is compact.  Therefore, there exists a set $D$, with
$P(D) = 1$, such that for all $\theta \in D$ we have $\sup_h \|
\nabla_h f_h(\theta) \| < \infty$.  For $\theta \in D$ and any $h,
h' \in \mathcal{H}$, we have
\begin{equation*}
  |f(\theta, h) - f(\theta, h')| \le \| \nabla_h f_h(\theta)
  \rfloor_{h=\tilde{h}(\theta)} \| \, \| h - h' \| \le \sup_h \|
  \nabla_h f_h(\theta) \| \, \| h - h' \|,
\end{equation*}
where $\tilde{h}(\theta)$ lies between $h$ and $h'$.  Let $h^*$ be
an arbitrary point of $\mathcal{H}$.  Define $F \colon \Theta
\rightarrow \bar{\Real}$ as follows:
\begin{equation*}
  F(\theta) =
    \begin{cases}
      |f(\theta, h^*)| + M \sup_h \| \nabla_h f_h(\theta) \| & \text{if } \theta \in D, \\[1mm]
      \ \infty                                               & \text{if } \theta \notin D,
    \end{cases}
\end{equation*}
where $M = 2 k^{1/2} \sup_h \| h - h^* \|$.  By
Lemma~\ref{S-lemma2.13} with $\phi(\theta) = \sup_h \| \nabla_h
f_h(\theta) \|$ and $\alpha = 1$, $\mathcal{F}$ is Euclidean with
envelope $F$.  We have
\begin{equation*}
  \int F \, dP = \int_D F \, dP + \int_{D^c} F \, dP = \int_D
  |f_{h^*}| \, dP + M \int_D \sup_h \| \nabla_h f_h \| \, dP <
  \infty,
\end{equation*}
since $\int f_h \, dP < \infty$ for any $h \in \mathcal{H}$, and
$\int \sup_h \| \nabla_h f_h \| \, dP < \infty$ by assumption.
Therefore, by Lemma~\ref{S-lemma2.8}, ${\| P_n - P
\|}_{\mathcal{F}}$ converges to $0$ almost surely, i.e.\ the class
$\mathcal{F}$ is $P$-Glivenko-Cantelli.  \qed

\section*{Proof of Lemma~\ref{lem:conv-emp-argmax}}
Denote $h_0 = \argmax_h f(h)$ and $h_n = \argmax_h f_n(h)$.  Let
$\epsilon > 0$ and let $B_{h_0}^{\epsilon}$ be the open ball
centered at $h_0$ and with radius $\epsilon$.  Since $h_0$ is the
unique maximizer of $f$, for any $h \notin B_{h_0}^{\epsilon}$,
$f(h) < f(h_0)$, and since $H \setminus B_{h_0}^{\epsilon}$ is
compact, $f$ achieves its maximum on $H \setminus
B_{h_0}^{\epsilon}$, say at $h_*$, i.e.\ $\sup_{h \in H \setminus
B_{h_0}^{\epsilon}} f(h) = f(h_*) < f(h_0)$.  Let $\delta = f(h_0) -
f(h_*)$.  By uniform convergence, there exists $n_0$ such that for
all $n \geq n_0, \sup_h |f_n(h) - f(h)| < \delta/2$, and in
particular, for all $n \geq n_0, f_n(h_0) > f(h_0) - \delta/2$.  We
have
\begin{equation}
  \label{S-eq:fnhn}
  f_n(h_n) \geq f_n(h_0) > f(h_0) - \delta/2.
\end{equation}
At the same time, for all $n \geq n_0, f_n(h) < f(h) + \delta/2$ for
all $h \in H \setminus B_{h_0}^{\epsilon}$.  Now if $h_n$ was in $H
\setminus B_{h_0}^{\epsilon}$, we would have
\begin{equation*}
  f_n(h_n) < f(h_n) + \delta/2 \leq f(h_*) + \delta/2 = f(h_0) -
  \delta + \delta/2 = f(h_0) - \delta/2,
\end{equation*}
which contradicts~\eqref{S-eq:fnhn}.  Therefore, we conclude that
$h_n \in B_{h_0}^{\epsilon}$.  \qed

\section*{Proof of Theorem~\ref{thm:an}} 
\textit{Proof of Part~1.} \ Recall that $n = \tau_R$ is the total
number of cycles required to achieve $R$ regenerations, and note
that $R \rightarrow \infty$ implies $n \rightarrow \infty$.  We
expand $\nabla_h B_n(h_n)$ around $h_0$:
\begin{equation*}
  \nabla_h B_n(h_n) = \nabla_h B_n(h_0) + \nabla_h^2 B_n(h^*) (h_n -
  h_0),
\end{equation*}
where $h^*$ is between $h_n$ and $h_0$.  Since $\nabla_h B_n(h_n) =
0$ and $\nabla_h B(h_0) = 0$, we have
\begin{align*}
  R^{1/2} (h_n -
  h_0) & = - \bigl( \nabla_h^2 B_n(h^*) \bigr)^{-1} R^{1/2} \nabla_h
           B_n(h_0) \\
       & = - \bigl( \nabla_h^2 B_n(h^*) \bigr)^{-1} R^{1/2} \bigl(
           \nabla_h B_n(h_0) - \nabla_h B(h_0) \bigr).
\end{align*}
Our plan is to show that $\nabla_h^2 B_n(h^*) \cas J(h_0)$ and that
\begin{equation}
  \label{S-eq:d2bhs}
  R^{1/2} \bigl( \nabla_h B_n(h_0) - \nabla_h B(h_0) \bigr) \cd
  \mathcal{N}(0, \tau^2(h_0)),
\end{equation}
as this will prove the theorem.  To show $\nabla_h^2 B_n(h^*) \cas
J(h_0)$, we first note that
\begin{equation}
  \label{S-eq:2ndder-bound}
  \| \nabla_h^2 B_n(h^*) - \nabla_h^2 B(h_0) \| \leq \| \nabla_h^2
  B_n(h^*) - \nabla_h^2 B(h^*) \| + \| \nabla_h^2 B(h^*) -
  \nabla_h^2 B(h_0) \|.
\end{equation}
Since all the conditions of Theorem~\ref{thm:gc-mc} are satisfied,
$\sup_h |B_n(h) - B(h)| \cas 0$, which by
Lemma~\ref{lem:conv-emp-argmax} entails $h_n \cas h_0$, so by
continuity of $\nabla_h^2 B(h)$ at $h_0$, we conclude that the
second term on the right side of~\eqref{S-eq:2ndder-bound} converges
to $0$ almost surely.

We now consider the first term on the right side
of~\eqref{S-eq:2ndder-bound} and we use arguments similar to those
used in the proof of Theorem~\ref{thm:gc-mc} to show that this term
converges to $0$ almost surely.  For any $h \in \mathcal{H}$,
\begin{equation}
  \label{S-eq:d2b}
  \begin{split}
    \nabla_h^2 B_n(h) - \nabla_h^2
    B(h) & = \frac{\bigl( \sum_{r=1}^R \nabla_h^2 S_r^{(h)} \bigr) /
             R }{\bigl( \sum_{r=1}^R N_r \bigr) / R} - \nabla_h^2
             E_P(f_h(\theta)) \\
         & = \frac{\bigl( \sum_{r=1}^R \nabla_h^2 S_r^{(h)} \bigr) /
             R }{\bigl( \sum_{r=1}^R N_r \bigr) / R} - E_P\bigl(
             \nabla_h^2 f_h(\theta) \bigr) \\
         & = \frac{\bigl( \sum_{r=1}^R \nabla_h^2 S_r^{(h)} \bigr) /
             R }{\bigl( \sum_{r=1}^R N_r \bigr) / R} - \frac{E(N_1)
             E_P\bigl( \nabla_h^2 f_h(\theta) \bigr)}{E(N_1)}, \\
  \end{split}
\end{equation}
where the second equality in~\eqref{S-eq:d2b} follows by
assumption~\eqref{eq:interchange-d-i}.  By~A\ref{ass:A5},
Theorem~\ref{thm:gc-iid} implies that
\begin{equation*}
  \sup_h \bigg| \frac{1}{R} \sum_{r=1}^R \nabla_h^2 S_r^{(h)} -
  E(N_1) E_P\bigl( \nabla_h^2 f_h(\theta) \bigr) \bigg| \cas 0,
\end{equation*}
and since $(1/R) \sum_{r=1}^R N_r \cas E(N_1)$, we obtain
\begin{equation*}
  \sup_h \Bigg| \frac{\bigl( \textstyle \sum_{r=1}^R \nabla_h^2
  S_r^{(h)} \bigr) / R} {\bigl( \sum_{r=1}^R N_r \bigr) / R} -
  \frac{E(N_1) E_P\bigl( \nabla_h^2 f_h(\theta) \bigr)} {E(N_1)}
  \Bigg| \cas 0,
\end{equation*}
i.e.\ $\sup_h | \nabla_h^2 B_n(h) - \nabla_h^2 B(h) | \cas 0$.  This
shows that the first term on the right side
of~\eqref{S-eq:2ndder-bound} converges to $0$ almost surely, which
now implies that $\| \nabla_h^2 B_n(h^*) - \nabla_h^2 B(h_0) \| \cas
0$.  Therefore,
\begin{equation}
  \label{S-eq:d2bnhs}
  \nabla_h^2 B_n(h^*) \cas \nabla_h^2 B(h_0) = J(h_0).
\end{equation}

We now consider the left side of~\eqref{S-eq:d2bhs}.  We have
\begin{multline*}
  R^{1/2} \bigl( \nabla_h B_n(h_0) - \nabla_h B(h_0) \bigr) =
  R^{1/2} \Biggl( \frac{\sum_{r=1}^R \nabla_h S_r^{(h_0)}}
  {\sum_{r=1}^R N_r} - E_P(\nabla_h f_{h_0}(\theta)) \Biggr) \\ =
  \frac{R^{1/2}} {\bigl( \sum_{r=1}^R N_r \bigr) / R} \biggl( \frac{
  \sum_{r=1}^R \nabla_h S_r^{(h_0)} - \sum_{r=1}^R N_r E_P(\nabla_h
  f_{h_0}(\theta)) } {R} \biggr).
\end{multline*}
Now in view of~A\ref{ass:A1} and~A\ref{ass:A2}, Theorem~$2$ of
\citet{HobertEtal:2002} implies that \linebreak \commentt{xxxI
forced a line break here}$E\bigl( \| \nabla_h S_1^{(h_0)} \|^2
\bigr) < \infty$ and $E(N_1^2) < \infty$.  Also, $\bigl(
\sum_{r=1}^R N_r \bigr) / R \cas E(N_1)$.  Therefore, by the CLT,
$R^{1/2} \bigl( \nabla_h B_n(h_0) - \nabla_h B(h_0) \bigr) \cd
\mathcal{N} \bigl( 0, \tau^2(h_0) \bigr)$, and together
with~\eqref{S-eq:d2bnhs}, this implies~\eqref{eq:an-argmax-R}.

\vspace{3mm} %
\noindent %
\textit{Proof of Part~2.}  \ That $J_n(h_n) \cas J(h_0)$ follows by
an argument virtually identical to the argument used to show that
$\nabla_h^2 B_n(h^*) \cas \nabla_h^2 B(h_0) = J(h_0)$.  Since
$J(h_0)$ is nonsingular, we obtain
\begin{equation*}
  \bigl[ J_n(h_n) \bigr]^{-1} \cas \bigl[ J(h_0) \bigr]^{-1}.
\end{equation*}

We now proceed to show that $\tau_n^2(h_n) \cas \tau^2(h_0)$, and we
do this by working with quantities $\rho^2(h)$ and $\rho_n^2(h)$
which are the same as $\tau^2(h)$ and $\tau_n^2(h)$, respectively,
except that they do not include the terms $[E(N_1)]^{-2}$ and
$\bar{N}^{-2}$, respectively: Define
\begin{equation*}
  \rho^2(h) = E\Bigl( \bigl[ \nabla_h S_1^{(h)} - N_1 E_P(\nabla_h
  f_h(\theta)) \bigr] \bigl[ \nabla_h S_1^{(h)} - N_1 E_P(\nabla_h
  f_h(\theta)) \bigr]^{\top} \Bigr),
\end{equation*}
and
\begin{equation*}
  \rho_n^2(h) = \frac{1}{R} \sum_{r=1}^{R} \bigl( \nabla_h S_r^{(h)}
  - N_r \nabla_h \bar{S}^{(h)} / \bar{N} \bigr) \bigl( \nabla_h
  S_r^{(h)} - N_r \nabla_h \bar{S}^{(h)} / \bar{N} \bigr)^{\top}.
\end{equation*}
We will show that $\rho_n^2(h_n) \cas \rho^2(h_0)$, which will show
that $\tau_n^2(h_n) \cas \tau^2(h_0)$.  To show that $\rho_n^2(h_n)
\cas \rho^2(h_0)$, we express $\rho_n^2(h_n) - \rho^2(h_0)$ as the
sum of four differences, and we show that each of these converges to
$0$ almost surely.  As in Remark~\ref{rem:int-cond-2}, we will
assume that $\dim(\mathcal{H}) = 1$.  We do this only for notational
simplicity, as all our results and arguments are valid without this
restriction.

The first difference is $D_1 \coloneqq (1/R) \sum_{r=1}^{R} \bigl(
\nabla_h S_r^{(h_n)} \bigr)^2 - E\bigl[ \bigl( \nabla_h S_1^{(h_0)}
\bigr)^2 \bigr]$.  Letting
\begin{align*}
  D_{11} & = \frac{1}{R} \sum_{r=1}^{R} \bigl( \nabla_h S_r^{(h_n)}
             \bigr)^2 - E\bigl[ \bigl( \nabla_h S_1^{(h_n)} \bigr)^2
             \bigr], \\
  D_{12} & = E\bigl[ \bigl( \nabla_h S_1^{(h_n)} \bigr)^2 \bigr] -
             E\bigl[ \bigl( \nabla_h S_1^{(h_0)} \bigr)^2,
\end{align*}
we have $|D_1| \leq |D_{11}| + |D_{12}|$.  By~A\ref{ass:A7},
Theorem~\ref{thm:gc-iid} implies that $D_{11} \cas 0$.  Consider now
$D_{12}$.  Clearly $\bigl( \nabla_h S_r^{(h_n)} \bigr)^2 \cas \bigl(
\nabla_h S_r^{(h_0)} \bigr)^2$.  By~A\ref{ass:A7}, we may apply the
dominated convergence theorem to conclude that $E\bigl[ \bigl(
\nabla_h S_1^{(h_n)} \bigr)^2 \bigr] \cas E\bigl[ \bigl( \nabla_h
S_1^{(h_0)} \bigr)^2 \bigr]$, i.e.\ $D_{12} \cas 0$.  Therefore $D_1
\cas 0$.

The second difference is
\begin{equation*}
  D_2 \coloneqq \frac{\nabla_h \bar{S}^{(h_n)}}{\bar{N}} \frac{1}{R}
  \sum_{r=1}^{R} \nabla_h S_r^{(h_n)} N_r - E_P(\nabla_h
  f_{h_0}(\theta)) E\bigl( \nabla_h S_1^{(h_0)} N_1 \bigr).
\end{equation*}
We have
\begin{align*}
  |D_2| & \leq \biggl| \frac{\nabla_h \bar{S}^{(h_n)}}{\bar{N}}
            \frac{1}{R} \sum_{r=1}^{R} \nabla_h S_r^{(h_n)} N_r -
            E_P(\nabla_h f_{h_n}(\theta)) E\bigl( \nabla_h
            S_1^{(h_n)} N_1 \bigr) \biggr| \\
        & \qquad + \biggl| E_P(\nabla_h f_{h_n}(\theta)) E \bigl(
            \nabla_h S_1^{(h_n)} N_1 \bigr) - E_P(\nabla_h
            f_{h_0}(\theta)) E\bigl( \nabla_h S_1^{(h_0)} N_1 \bigr)
            \biggr| \\
        & \coloneqq |D_{21}| + |D_{22}|,
\end{align*}
in self-defining notation.  Consider $D_{21}$.  From~A\ref{ass:A7},
$E\bigl( \sup_h |\nabla_h S_1^{(h)}| \bigr) < \infty$, and together
with the SLLN, this gives $\big|\nabla_h \bar{S}^{(h_n)} / \bar{N} -
E\bigl( \nabla_h S_1^{(h_n)} \bigr) / E(N_1) \big| \cas 0$, i.e.
\begin{equation}
  \label{S-eq:AA}
  \bigg| \frac{\nabla_h \bar{S}^{(h_n)}}{\bar{N}} - E_P(\nabla_h
  f_{h_n}(\theta)) \bigg| \cas 0.
  \end{equation}
Now
\begin{equation}
  \label{S-eq:BB}
  E\Bigl( \sup_h |\nabla_h S_1^{(h)}N_1| \Bigr) \leq \biggl(
  E\biggl[ \Bigl( \sup_h |\nabla_h S_1^{(h)}| \Bigr)^2 \biggr]
  E(N_1^2) \biggr)^{1/2}
\end{equation}
by the Cauchy-Schwartz inequality.  The first expectation on the
right side of~\eqref{S-eq:BB} is finite by~A\ref{ass:A7}, and
$E(N_1^2) < \infty$ by Theorem~$2$ of \citet{HobertEtal:2002}.
Therefore,
\begin{equation}
  \label{S-eq:CC}
  \biggl| \frac{1}{R} \sum_{r=1}^{R} \nabla_h S_r^{(h_n)} N_r - E
  \bigl( \nabla_h S_1^{(h_n)} N_1 \bigr) \biggr| \cas 0.
\end{equation}
From~\eqref{S-eq:AA} and~\eqref{S-eq:CC} we see that $D_{21} \cas
0$.  From~A\ref{ass:A6} and finiteness of \linebreak \commentt{xxxI
forced a line break here}$E\bigl( \sup_h |\nabla_h S_1^{(h)} N_1|
\bigr)$, we may apply dominated convergence to see that $D_{22} \cas
0$, and so conclude that $D_2 \cas 0$.  Let $D_3$ denote the third
difference.  Since $D_3 = D_2$, we have $D_3 \cas 0$ also.

The fourth difference is
\begin{equation*}
  D_4 = \biggl( \frac{\nabla_h \bar{S}^{(h_n)}}{\bar{N}} \biggr)^2
  \frac{1}{R} \sum_{r=1}^{R} N_r^2 - \bigl[ E_P(\nabla_h
  f_{h_0}(\theta)) \bigr]^2 \hspace{.3mm} E(N_1^2).
\end{equation*}
We showed earlier that $\nabla_h \bar{S}^{(h_n)} / \bar{N} \cas
E_P(\nabla_h f_{h_0}(\theta))$.  The SLLN gives \linebreak
\commentt{xxxI forced a line break here}$(1/R) \sum_{r=1}^{R} N_r^2
\cas E(N_1^2)$ (finiteness of $E(N_1^2)$ is a consequence of
Theorem~$2$ of \citet{HobertEtal:2002}).  Therefore $D_4 \cas 0$.
\qed

\vspace{3mm} %
Before we prove Theorem~\ref{thm:functional-wc-iid}, we need to give
some background material on empirical processes.  The
Pollard-Koltchinskii Theorem \citep{Pollard:1982,
Koltchinskii:1981}, stated as Theorem~\ref{S-thm:donsker} below,
gives sufficient conditions for a class of functions to be Donsker.
In order to state it, we need to introduce additional terminology.
The covering number $N(\epsilon, \mathcal{V}, \| \cdot \|)$ is the
minimum number of open balls of radius $\epsilon$ using the norm $\|
\cdot \|$ whose union covers the class $\mathcal{V}$.  In all of our
development we will use the $L_1$ norm or the $L_2$ norm.  The
uniform entropy integral is
\begin{equation}
  \label{S-eq:1}
  J(\mathcal{V}) = \int_0^1 \sqrt{\log \sup_{Q\in\mathcal{D}}
  N\bigl( \epsilon {\| U \|}_{Q,2}, \mathcal{V}, L_2(Q) \bigr)} \,
  d\epsilon,
\end{equation}
where $\mathcal{D}$ is the set of all finitely discrete probability
measures on $(\Theta, \mathcal{B})$ and ${\| U \|}_{Q,2}^2 = \int
U^2 \, dQ$.
\begin{theorem}[Theorem~8.19 in \citet{Kosorok:2008}]
  \label{S-thm:donsker}
  Let $\mathcal{F}$ be a class of measurable functions with envelope
  $F$ and for which $J(\mathcal{F}) < \infty$.  Suppose that the
  classes $\mathcal{F}_{\delta}, \, \delta > 0$ and
  $\mathcal{F}_{\infty}^2$ are all $P$-measurable.  If $F^2$ is
  measurable and integrable, then $\mathcal{F}$ is $P$-Donsker.
\end{theorem}
The condition $J(\mathcal{F}) < \infty$ in
Theorem~\ref{S-thm:donsker} can be verified by applying a simple
upper bound to the covering number
(inequality~\eqref{S-eq:bound-cn}) and Lemma~\ref{S-lemma:uec}
below.
\begin{lemma}
  \label{S-lemma:uec}
  Let $g \colon \Real^+ \rightarrow \Real^+$ be a nonincreasing
  function.  Suppose that $g(\epsilon) \le C \epsilon^{-c}$ for some
  constants $C > 0$ and $c > 0$.  Then $\int_0^1 \sqrt{\log
  (g(\epsilon))} \, d\epsilon < \infty$.
\end{lemma}

\subsection*{Proof of Lemma~\ref{S-lemma:uec}}
We have $\log (g(\epsilon)) \le \log(C) + c \log(1 / \epsilon)$.
Therefore
\begin{equation*}
  \epsilon \log (g(\epsilon)) \le \epsilon \log(C) + c \, \epsilon
  \log(1 / \epsilon) \rightarrow 0 \qquad \text{as } \epsilon
  \searrow 0.
\end{equation*}
This convergence implies that there exists $\delta > 0$ such that
$\epsilon \log (g(\epsilon)) \le 1$ whenever $\epsilon \in (0,
\delta)$.  Without loss of generality, we suppose that $\delta < 1$.
We have
\begin{align*}
  \int_0^1 \sqrt{\log (g(\epsilon))} \,
  d\epsilon & = \int_0^{\delta} \sqrt{\log (g(\epsilon))} \,
                d\epsilon + \int_{\delta}^1 \sqrt{\log
                (g(\epsilon))} \, d\epsilon \\
            & \le \int_0^{\delta} \epsilon^{-1/2} \, d\epsilon +
                \int_{\delta}^1 \sqrt{\log (g(\delta))} \,
                d\epsilon \\
            & = 2 \sqrt{\delta} + (1 - \delta) \sqrt{\log
                (g(\delta))} < \infty. &  \pushright{\hspace{22.3mm} \qed}
\end{align*}

Let $\mathcal{V}$ be a set of functions defined on $\Theta$ with
envelope $U$, let $p > 0$, and let $Q$ be a probability measure on
$\Theta$.  Suppose that $Q(U^p) < \infty$; we can then define the
norm on $\mathcal{V}$ given by
\begin{equation*}
  {\| \phi \|}_{Q,p,U} = \bigg( \frac{Q(|\phi|^p)} {Q(U^p)}
  \bigg)^{1/p}.
\end{equation*}
Suppose additionally that $\mathcal{V}$ is Euclidean for $U$, and
let $A$ and $b$ be the positive constants appearing in the
definition of Euclidean (Definition~\ref{S-def2.7}).  If $\epsilon
\in (0, 1]$ and $p > 1$, then
\begin{equation}
  \label{S-eq:bound-cn}
  N(\epsilon, \mathcal{V}, {\| \cdot \|}_{Q,p,U}) \le A (2 /
  \epsilon)^{pb}
\end{equation}
\citep[p.~789]{NolanPollard:1987}.  We now return to the class
$\mathcal{F} = \{ f_h, \, h \in \mathcal{H} \}$.  In the proof of
Theorem~\ref{thm:gc-PakesPollard} we showed that if for $P$-almost
all $\theta \in \Theta$, $\nabla_h f_h$ exists and is continuous on
$\mathcal{H}$, then for any point $h' \in \mathcal{H}$ the class
$\mathcal{F}$ is Euclidean with envelope
\begin{equation}
  \label{S-eq:envelope}
  F(\theta) = f_{h'}(\theta) + M \sup_{h\in\mathcal{H}} \|
  \nabla_h f_h(\theta) \|,
\end{equation}
where $M = 2 k^{1/2} \sup_{h\in\mathcal{H}} \| h - h' \|$ (recall
that $k$ is the dimension of $\mathcal{H}$).  Thus, by
\eqref{S-eq:bound-cn} with $p = 2$, for $\epsilon \in (0, 1]$, for
any probability measure $Q$ satisfying $Q(F^2) < \infty$ we have
\begin{equation*}
  N(\epsilon, \mathcal{F}, {\| \cdot \|}_{Q,2,F}) \le A (2 /
  \epsilon)^{2b}.
\end{equation*}
For any probability measure $Q$ and $\epsilon \in (0, 1]$ we have
\begin{equation*}
  N \bigl( \epsilon {\| F \|}_{Q,2}, \mathcal{F}, L_2(Q) \bigr) =
  N(\epsilon, \mathcal{F}, {\| \cdot \|}_{Q, 2, F}) \leq A (2 /
  \epsilon)^{2b}.
\end{equation*}
Therefore,
\begin{equation*}
  g(\epsilon) \coloneqq \sup_{Q\in\mathcal{D}} N \bigl( \epsilon {\|
  F \|}_{Q,2}, \mathcal{F}, L_2(Q) \bigr) \le A (2 / \epsilon)^{2b},
\end{equation*}
so, Lemma~\ref{S-lemma:uec} with $C = A 2^{2b}$ and $c = 2 b$ gives
$\int_0^1 \sqrt{\log (g(\epsilon))} \, d\epsilon < \infty$, i.e.\
the condition $J(\mathcal{F}) < \infty$ in
Theorem~\ref{S-thm:donsker} is satisfied.  We summarize this in the
following theorem.
\begin{theorem}
  \label{S-thm:uec}
  If for $P$-almost all $\theta \in \Theta$, $\nabla_h f_h$ exists
  and is continuous on $\mathcal{H}$, then the class $\mathcal{F}$
  is Euclidean with envelope $F$ given by~\eqref{S-eq:envelope}, and
  $J(\mathcal{F}) < \infty$.
\end{theorem}

\section*{Proof of Theorem~\ref{thm:functional-wc-iid}}
\begin{enumerate}[itemsep=1mm,topsep=1.5mm]
\item Part~(a) is a verbatim restatement of Theorem~\ref{thm:gc-iid}
  and Part~(b) follows from Theorem~\ref{thm:gc-iid}.
\item In essence the result is trivial: for $P$-almost every
  sequence $\theta_1, \theta_2, \ldots$, \linebreak \commentt{xxxI
  forced a line break here}$(1/n) \sum_{i=1}^n g(\theta_i)
  f_h(\theta_i)$ converges to $P(g f_h)$ uniformly in $h$ and
  \linebreak[4] \commentt{xxxI forced a line break here}$(1/n)
  \sum_{i=1}^n f_h(\theta_i)$ converges to $P(f_h)$ uniformly in
  $h$, so in view of the continuity of the function $q(u, v) = u/v$
  we have
  \begin{equation*}
    \frac{(1/n) \sum_{i=1}^n g(\theta_i) f_h(\theta_i)} {(1/n)
    \sum_{i=1}^n f_h(\theta_i)} \ \text{ converges to } \ \frac{P(g
    f_h)} {P(f_h)} \qquad \text{uniformly in } h,
  \end{equation*}
  which is assertion~\eqref{eq:ucI}.  There is a detail we need to
  check, namely that $P(f_h)$ is bounded away from $0$.  Now by
  assumption, for every $\theta$, $\nabla_h f_h$ exists and is
  continuous in $h$; so in particular, for every $\theta$, $f_h$ is
  continuous in $h$.  Therefore, $P(f_h)$ is continuous in $h$ by
  the dominated convergence theorem, and since $\mathcal{H}$ is
  compact, $\inf_h P(f_h) > 0$.
\item We will show that the class $\mathcal{F}$ is $P$-Donsker by
  checking the conditions of Theorem~\ref{S-thm:donsker}.  By
  Theorem~\ref{S-thm:uec}, the class $\mathcal{F}$ is Euclidean with
  envelope $F$ given by~\eqref{S-eq:envelope}, and $J(\mathcal{F}) <
  \infty$.  Equation~\eqref{S-eq:envelope} expresses $F$ as a sum of
  two functions, $f_{h'}$ and $M \sup_{h\in\mathcal{H}} \| \nabla_h
  f_h \|$.  Since each of these is measurable and square-integrable
  with respect to $P$, we may conclude that $F^2$ is measurable and
  integrable with respect to $P$.  Therefore the conditions of
  Theorem~\ref{S-thm:donsker} are all satisfied, and we conclude
  that the class $\mathcal{F}$ is $P$-Donsker.  The proof that
  $\mathcal{G}$ is $P$-Donsker is essentially identical.
\item For $P$-almost every $\theta$, $f_h(\theta)$ is continuous in
  $h$, and as we saw in the proof of Part~2 of the present theorem,
  $P(f_h)$ is continuous in $h$; so with probability one, $n^{1/2}
  (P_n(f_h) - P(f_h)) \in C(\mathcal{H})$.  Because $\mathcal{H}$ is
  compact, $C(\mathcal{H}) \subset l^{\infty} (\mathcal{F})$ (a
  formal proof of this fact is given in \citet{Park:2015}).
  Therefore, weak convergence of $n^{1/2} (P_n(f_h) - P(f_h))$ in
  $l^{\infty} (\mathcal{F})$ implies weak convergence of $n^{1/2}
  (P_n(f_h) - P(f_h))$ in $C(\mathcal{H})$, where $C(\mathcal{H})$
  is endowed with the sup norm
  \citep[cf.][Theorem~1.3.10]{vanderVaartWellner:1996}; i.e.\
  $n^{1/2} (P_n - P) (f_{\cdot}) \cd \mathbb{F}(\cdot)$ in
  $C(\mathcal{H})$, where $\mathbb{F}(\cdot)$ is a mean $0$ Gaussian
  process.  Similarly, $n^{1/2} (P_n - P) (g f_{\cdot}) \cd
  \mathbb{G}(\cdot)$ in $C(\mathcal{H})$, where $\mathbb{G}(\cdot)$
  is a mean $0$ Gaussian process.  Define the map $\Phi \colon
  C(\mathcal{H}) \times C(\mathcal{H}) \rightarrow C(\mathcal{H})$
  by $(\Phi(x, y))(h) = x(h) / y(h)$ where, for definiteness, we
  define $0/0 = 0$.  It is not hard to check that $\Phi$ is Hadamard
  differentiable at the point $(P(g f_{\cdot}), P(f_{\cdot}))$ (for
  a definition of Hadamard differentiability see, e.g.,
  \citet[Section~3.9.1]{vanderVaartWellner:1996})---we use the fact
  $\inf_h P(f_h) > 0$, established in the proof of Part~2.  The
  result now follows from the functional delta method
  \citep[Theorem~3.9.4]{vanderVaartWellner:1996}.
\end{enumerate}

\section*{Proof of Theorem~\ref{thm:functional-wc-mc}}
\begin{enumerate}[itemsep=1mm,topsep=1.5mm]
\item That~\eqref{eq:gc-mc} holds was demonstrated in the proof of
  Theorem~\ref{thm:gc-mc}, and the proof of the corresponding
  statement for the functions $g f_h$ is completely analogous.
\item The proof of~\eqref{eq:conv-supIhat} is identical to the proof
  of Part~2 of Theorem~\ref{thm:functional-wc-iid}.
\item The proof is analogous to the proof of Part~3 of
  Theorem~\ref{thm:functional-wc-iid}.  For Part~(a), we consider
  $S_1^{(h)}$ and $\mathscr{F}$ instead of $f_h$ and $\mathcal{F}$,
  respectively.  Continuity in $h$ of $\nabla_h S_1^{(h)}$ for
  almost all sequences $\theta_1, \theta_2, \ldots$ follows from
  continuity in $h$ of $\nabla_h f_h$ for almost all $\theta \in
  \Theta$, since with probability one, $S_1^{(h)}$ is a finite sum.
  In addition, by~A\ref{ass:A1} and~B\ref{ass:B1}, $E[(S_1^{(h)})^2]
  < \infty$ for each $h \in \mathcal{H}$.  Since $\sup_h \| \nabla_h
  S_1^{(h)} \|$ is measurable and square integrable with respect to
  $\sP$, by Part~3 of Theorem~\ref{thm:functional-wc-iid} we see
  that the class $\mathscr{F}$ is $\sP$-Donsker.  The proof of
  Part~(b) is virtually identical.  The only changes are that we
  consider $g f_h$ instead of $f_h$, and obtain finiteness of
  $E[(T_1^{(h)})^2]$ for all $h \in \mathcal{H}$ as a consequence
  of~A\ref{ass:A1} and~B\ref{ass:B2}.
\item The proof is entirely parallel to that of Part~4 of
  Theorem~\ref{thm:functional-wc-iid}.
\end{enumerate}

\vspace{3mm} %
\noindent %
Prior to the statement of Theorem~\ref{thm:functional-wc-mc}, we
noted that when the chain has a proper atom at a singleton, then the
sequence $\theta_1, \theta_2, \ldots$ itself determines the
regeneration times $\tau_0, \tau_1, \ldots$, so that $S_1^{(h)}$ can
be viewed as a function mapping $\Theta^{\infty}$ to $\Real_+$.  The
minorization condition discussed in Section~\ref{sec:unif-conv}
(cf.~\eqref{eq:min-cond}) determines the so-called ``split chain''
$(\theta_1, \delta_1), (\theta_2, \delta_2), \ldots$, for which the
set $\Theta \times \{ 1 \}$ is a proper atom
\citep[Section~4.4]{Nummelin:1984}.  The functions $S_1^{(h)}, \, h
\in \mathcal{H}$ may then be viewed as maps $S_1^{(h)} \colon
(\Theta \times \{ 0, 1 \})^{\infty} \rightarrow \Real_+$, and the
situation is the same as the simple situation described earlier.

\section*{Verification of Condition~(\ref{eq:bound-on-sup})
for Exponential Families in Canonical Form}
We now show that if $f_h = \nu_h / \nu_{h_*}$ for some fixed $h_*
\in \mathcal{H}$, and if $\{ \nu_h, \, h \in \mathcal{H} \}$ is an
exponential family and $h$ is the canonical parameter, then
condition~\eqref{eq:bound-on-sup} holds.  It is clearly sufficient
to show that
\begin{equation}
  \label{S-eq:bound-on-sup-bf}
  \begin{split}
    \text{there exist } & h_1, \ldots, h_d \in \mathcal{H} \text{
    and constants } c_1, \ldots, c_d \text{ such that} \\
    & \sup_{h\in\mathcal{H}} \nu_h(\theta) \leq \sum_{i=1}^d c_i
    \nu_{h_i}(\theta) \qquad \text{ for all } \theta \in \Theta.
  \end{split}
\end{equation}
(In fact, we can take $f_h = \nu_h / q$ where $q \notin \{ \nu_h, \,
h \in \mathcal{H} \}$.  So for example, instead of using a Markov
chain with invariant distribution $\nu_{h_*,y}$, we can use a serial
tempering chain, whose invariant distribution is a mixture of the
posteriors $\nu_{h_{*1},y}, \ldots, \nu_{h_{*m},y}$ for $h_{*1},
\ldots, h_{*m} \in \mathcal{H}$; see Remark~\ref{rem:st}.)  Recall
that $k$ denotes the dimension of $h$.  We will slightly abuse
notation and write $\omega$ instead of $h$, and $\Omega$ instead of
$\mathcal{H}$.  This is to avoid notational clashes, e.g.\ writing
$h = (h_1, \ldots, h_k)$ and at the same time having $h_1, \ldots,
h_d \in \mathcal{H}$.  We assume that the $\nu_{\omega}$'s form a
$k$-parameter exponential family with dominating measure $\mu$.
Thus for $\omega \in \Omega$, $\nu_{\omega}$ is a density with
respect to $\mu$, having the form $\nu_{\omega}(\theta) = \exp\bigl(
\sum_{i=1}^k \omega_i T_i(\theta) - A(\omega) \bigr)$, where the
$T_i$'s and $A$ are real-valued functions.  The set of all $\omega$
such that $\int \exp\bigl( \sum_{i=1}^k \omega_i T_i(\theta) \bigr)
\, d\mu(\theta) < \infty$ is called the natural parameter space, and
we assume that $\Omega$ is a compact subset of the interior of the
natural parameter space.  It is well known that $A(\omega) =
\log\bigl( \int \exp\bigl( \sum_{i=1}^k \omega_i T_i(\theta) \bigr)
\, d\mu(\theta) \bigr) $ is infinitely differentiable in the
interior of the natural parameter space, and in particular is
continuous there.  We will prove~\eqref{S-eq:bound-on-sup-bf} for
the case $k = 2$, the case $k > 2$ being no more difficult.

When $k = 2$, we have $\omega = (\omega_1, \omega_2)$.  We let $U =
\exp\bigl[ \omega_1 T_1(\theta) + \omega_2 T_2(\theta) \bigr]$ for
notational brevity.  Without loss of generality we take the compact
set $\Omega$ to be $[\omega_{1l}, \omega_{1u}] \times [\omega_{2l},
\omega_{2u}]$.  For any fixed $\omega \in [\omega_{1l}, \omega_{1u}]
\times [\omega_{2l}, \omega_{2u}]$, we have
\begin{equation*}
  U \leq
  \begin{cases}
    \exp\bigl[ \omega_{1u} T_1(\theta) + \omega_{2u} T_2(\theta)
    \bigr] & \text{if } T_1(\theta) \geq 0 \text{ and } T_2(\theta)
    \geq 0, \\
    \exp\bigl[ \omega_{1u} T_1(\theta) + \omega_{2l} T_2(\theta)
    \bigr] & \text{if } T_1(\theta) \geq 0 \text{ and } T_2(\theta)
    < 0, \\
    \exp\bigl[ \omega_{1l} T_1(\theta) + \omega_{2u} T_2(\theta)
    \bigr] & \text{if } T_1(\theta) < 0 \text{ and } T_2(\theta)
    \geq 0, \\
    \exp\bigl[ \omega_{1l} T_1(\theta) + \omega_{2l} T_2(\theta)
    \bigr] & \text{if } T_1(\theta) < 0 \text{ and } T_2(\theta) <
    0.
  \end{cases}
\end{equation*}
Therefore,
\begin{equation}
  \label{S-eq:bound-exp2}
  \begin{split}
  U & \leq \exp\bigl[ \omega_{1u} T_1(\theta) + \omega_{2u}
        T_2(\theta) \bigr] + \exp\bigl[ \omega_{1u} T_1(\theta) +
        \omega_{2l} T_2(\theta) \bigr] \\
    & \hspace{6mm} + \exp\bigl[ \omega_{1l} T_1(\theta) +
        \omega_{2u} T_2(\theta) \bigr] + \exp\bigl[ \omega_{1l}
        T_1(\theta) + \omega_{2l} T_2(\theta) \bigr]
  \end{split}
\end{equation}
for all $\theta \in \Theta$.  Let $c = \sup_{\omega\in\Omega}
\exp[-A(\omega)]$, which is finite, since $A$ is continuous and
$\Omega$ is compact.  Let
\begin{align*}
  & c_1 = c \exp[A(\omega^{(1)}) ], \hspace{.5mm} \omega^{(1)} =
    (\omega_{1u}, \omega_{2u}), \hspace{1.8mm} c_2 = c
    \exp[A(\omega^{(2)})], \hspace{.5mm} \omega^{(2)} = (\omega_{1u},
    \omega_{2l}), \\
  & c_3 = c \exp[A(\omega^{(3)}) ], \hspace{.5mm} \omega^{(3)} =
    (\omega_{1l}, \omega_{2u}), \hspace{1.8mm} c_4 = c
    \exp[A(\omega^{(4)})], \hspace{.5mm} \omega^{(4)} = (\omega_{1l},
    \omega_{2l}).
\end{align*}
By~\eqref{S-eq:bound-exp2} we get
\begin{equation*}
  \sup_{\omega\in\Omega} \nu_{\omega}(\theta) \leq c_1
  \nu_{\omega^{(1)}} (\theta) + c_2 \nu_{\omega^{(2)}} (\theta) +
  c_3 \nu_{\omega^{(3)}} (\theta) + c_4 \nu_{\omega^{(4)}} (\theta)
  \qquad \text{for all } \theta \in \Theta.
\end{equation*}

\section*{Acknowledgments}
We thank the referees for their helpful comments.


\begin{thebibliography}{46}
\expandafter\ifx\csname natexlab\endcsname\relax\def\natexlab#1{#1}\fi
\expandafter\ifx\csname url\endcsname\relax
  \def\url#1{\texttt{#1}}\fi
\expandafter\ifx\csname urlprefix\endcsname\relax\def\urlprefix{URL }\fi

\bibitem[{Asuncion et~al.(2009)Asuncion, Welling, Smyth and
  Teh}]{AsuncionEtal:2009}
\text{Asuncion, A.}, \text{Welling, M.}, \text{Smyth, P.} and \text{Teh, Y.~W.}
  (2009).
\newblock On smoothing and inference for topic models.
\newblock In \textit{Proceedings of the Twenty-Fifth Conference on Uncertainty
  in Artificial Intelligence}. UAI '09, AUAI Press, Arlington, Virginia, United
  States.

\bibitem[{Berger(1994)}]{Berger:1994}
\text{Berger, J.~O.} (1994).
\newblock An overview of robust {B}ayesian analysis (with discussion).
\newblock \textit{Test} \textbf{3} 5--124.

\bibitem[{Blei et~al.(2003)Blei, Ng and Jordan}]{BleiNgJordan:2003}
\text{Blei, D.~M.}, \text{Ng, A.~Y.} and \text{Jordan, M.~I.} (2003).
\newblock Latent {D}irichlet allocation.
\newblock \textit{Journal of Machine Learning Research} \textbf{3} 993--1022.

\bibitem[{Buta(2010)}]{Buta:2010}
\text{Buta, E.} (2010).
\newblock \textit{Computational Approaches for Empirical Bayes Methods and
  Bayesian Sensitivity Analysis}.
\newblock Ph.D. thesis, University of Florida.

\bibitem[{Doss et~al.(2014)Doss, Flegal, Jones and Neath}]{DossEtal:2014}
\text{Doss, C.}, \text{Flegal, J.~M.}, \text{Jones, G.~L.} and \text{Neath,
  R.~C.} (2014).
\newblock Markov chain {M}onte {C}arlo estimation of quantiles.
\newblock \textit{Electronic Journal of Statistics} \textbf{8} 2448--2478.

\bibitem[{Doss(2007)}]{Doss:2007}
\text{Doss, H.} (2007).
\newblock Bayesian model selection: Some thoughts on future directions.
\newblock \textit{Statistica Sinica} \textbf{17} 413--421.

\bibitem[{Doss and Tan(2014)}]{DossTan:2014}
\text{Doss, H.} and \text{Tan, A.} (2014).
\newblock Estimates and standard errors for ratios of normalizing constants
  from \mbox{multiple} {M}arkov chains via regeneration.
\newblock \textit{Journal of the Royal Statistical Society,
  \normalfont{Series~B}} \textbf{76} 683--712.

\bibitem[{Flegal et~al.(2008)Flegal, Haran and Jones}]{FlegalHaranJones:2008}
\text{Flegal, J.~M.}, \text{Haran, M.} and \text{Jones, G.~L.} (2008).
\newblock Markov chain {M}onte {C}arlo: Can we trust the third significant
  figure?
\newblock \textit{Statistical Science} \textbf{23} 250--260.

\bibitem[{Flegal and Jones(2010)}]{FlegalJones:2010}
\text{Flegal, J.~M.} and \text{Jones, G.~L.} (2010).
\newblock Batch means and spectral variance estimators in {M}arkov chain
  {M}onte {C}arlo.
\newblock \textit{The Annals of Statistics} \textbf{38} 1034--1070.

\bibitem[{George(2015)}]{George:2015}
\text{George, C.~P.} (2015).
\newblock \textit{Latent Dirichlet Allocation: Hyperparameter Selection and
  Applications to Electronic Discovery}.
\newblock Ph.D. thesis, University of Florida.

\bibitem[{George and Foster(2000)}]{GeorgeFoster:2000}
\text{George, E.~I.} and \text{Foster, D.~P.} (2000).
\newblock Calibration and empirical {B}ayes variable selection.
\newblock \textit{Biometrika} \textbf{87} 731--747.

\bibitem[{Geyer(2011)}]{Geyer:2011}
\text{Geyer, C.~J.} (2011).
\newblock Importance sampling, simulated tempering, and umbrella sampling.
\newblock In \textit{Handbook of Markov Chain Monte Carlo} (S.~P. Brooks, A.~E.
  Gelman, G.~L. Jones and X.~L. Meng, eds.). Chapman \& Hall/CRC, Boca Raton,
  295--311.

\bibitem[{Geyer and Thompson(1995)}]{GeyerThompson:1995}
\text{Geyer, C.~J.} and \text{Thompson, E.~A.} (1995).
\newblock Annealing {M}arkov chain {M}onte {C}arlo with applications to
  ancestral inference.
\newblock \textit{Journal of the American Statistical Association} \textbf{90}
  909--920.

\bibitem[{Griffiths and Steyvers(2004)}]{GriffithsSteyvers:2004}
\text{Griffiths, T.~L.} and \text{Steyvers, M.} (2004).
\newblock Finding scientific topics.
\newblock \textit{Proceedings of the National Academy of Sciences} \textbf{101}
  5228--5235.

\bibitem[{Hobert et~al.(2002)Hobert, Jones, Presnell and
  Rosenthal}]{HobertEtal:2002}
\text{Hobert, J.~P.}, \text{Jones, G.~L.}, \text{Presnell, B.} and
  \text{Rosenthal, J.~S.} (2002).
\newblock On the applicability of regenerative simulation in {M}arkov chain
  {M}onte {C}arlo.
\newblock \textit{Biometrika} \textbf{89} 731--743.

\bibitem[{Ibragimov and Linnik(1971)}]{IbragimovLinnik:1971}
\text{Ibragimov, I.~A.} and \text{Linnik, Y.~V.} (1971).
\newblock \textit{Independent and Stationary Sequences of Random Variables}.
\newblock Wolters-Noordhoff, Groningen.

\bibitem[{Jones et~al.(2006)Jones, Haran, Caffo and Neath}]{JonesEtal:2006}
\text{Jones, G.~L.}, \text{Haran, M.}, \text{Caffo, B.~S.} and \text{Neath, R.}
  (2006).
\newblock Fixed-width output analysis for {M}arkov chain {M}onte {C}arlo.
\newblock \textit{Journal of the American Statistical Association} \textbf{101}
  1537--1547.

\bibitem[{Kadane and Wolfson(1998)}]{KadaneWolfson:1998}
\text{Kadane, J.} and \text{Wolfson, L.~J.} (1998).
\newblock Experiences in elicitation.
\newblock \textit{Journal of the Royal Statistical Society: Series D (The
  Statistician)} \textbf{47} 3--19.

\bibitem[{Koltchinskii(1981)}]{Koltchinskii:1981}
\text{Koltchinskii, V.~I.} (1981).
\newblock On the central limit theorem for empirical measures.
\newblock \textit{Theory of Probability and Mathematical Statistics}
  \textbf{24} 71--82.

\bibitem[{Kosorok(2008)}]{Kosorok:2008}
\text{Kosorok, M.~R.} (2008).
\newblock \textit{Introduction to Empirical Processes and Semiparametric
  Inference}.
\newblock Springer, New York.

\bibitem[{Levental(1988)}]{Levental:1988}
\text{Levental, S.} (1988).
\newblock Uniform limit theorems for {H}arris recurrent {M}arkov chains.
\newblock \textit{Probability Theory and Related Fields} \textbf{80} 101--118.

\bibitem[{Liang et~al.(2008)Liang, Paulo, Molina, Clyde and
  Berger}]{LiangEtal:2008}
\text{Liang, F.}, \text{Paulo, R.}, \text{Molina, G.}, \text{Clyde, M.~A.} and
  \text{Berger, J.~O.} (2008).
\newblock Mixtures of $g$-priors for {B}ayesian variable selection.
\newblock \textit{Journal of the American Statistical Association} \textbf{103}
  410--423.

\bibitem[{Marinari and Parisi(1992)}]{MarinariParisi:1992}
\text{Marinari, E.} and \text{Parisi, G.} (1992).
\newblock Simulated tempering: A new {M}onte {C}arlo scheme.
\newblock \textit{Europhysics Letters} \textbf{19} 451--458.

\bibitem[{Meyn and Tweedie(1993)}]{MeynTweedie:1993}
\text{Meyn, S.~P.} and \text{Tweedie, R.~L.} (1993).
\newblock \textit{Markov Chains and Stochastic Stability}.
\newblock Springer-Verlag, New York, London.

\bibitem[{Mitchell and Beauchamp(1988)}]{MitchellBeauchamp:1988}
\text{Mitchell, T.} and \text{Beauchamp, J.} (1988).
\newblock Bayesian variable selection in linear regression.
\newblock \textit{Journal of the American Statistical Association} \textbf{83}
  1023--1036.

\bibitem[{Mykland et~al.(1995)Mykland, Tierney and Yu}]{MyklandTierneyYu:1995}
\text{Mykland, P.}, \text{Tierney, L.} and \text{Yu, B.} (1995).
\newblock Regeneration in {M}arkov chain samplers.
\newblock \textit{Journal of the American Statistical Association} \textbf{90}
  233--241.

\bibitem[{Newton and Raftery(1994)}]{NewtonRaftery:1994}
\text{Newton, M.} and \text{Raftery, A.} (1994).
\newblock Approximate {B}ayesian inference with the weighted likelihood
  bootstrap (with discussion).
\newblock \textit{Journal of the Royal Statistical Society,
  \normalfont{Series~B}} \textbf{56} 3--48.

\bibitem[{Nolan and Pollard(1987)}]{NolanPollard:1987}
\text{Nolan, D.} and \text{Pollard, D.} (1987).
\newblock U-{P}rocesses: {R}ates of convergence.
\newblock \textit{The Annals of Statistics} \textbf{15} 780--799.

\bibitem[{Nummelin(1984)}]{Nummelin:1984}
\text{Nummelin, E.} (1984).
\newblock \textit{General Irreducible Markov Chains and Non-negative
  Operators}.
\newblock Cambridge University Press, London.

\bibitem[{Pakes and Pollard(1989)}]{PakesPollard:1989}
\text{Pakes, A.} and \text{Pollard, D.} (1989).
\newblock Simulation and the asymptotics of optimization estimators.
\newblock \textit{Econometrica} \textbf{57} 1027--1057.

\bibitem[{Park(2015)}]{Park:2015}
\text{Park, Y.} (2015).
\newblock \textit{A Markov Chain Monte Carlo Approach to Empirical Bayes
  Inference and Bayesian Sensitivity Analysis via Empirical Processes}.
\newblock Ph.D. thesis, University of Florida.

\bibitem[{Petrone et~al.(2014)Petrone, Rousseau and
  Scricciolo}]{PetroneRousseauScricciolo:2014}
\text{Petrone, S.}, \text{Rousseau, J.} and \text{Scricciolo, C.} (2014).
\newblock Bayes and empirical {B}ayes: do they merge?
\newblock \textit{Biometrika} \textbf{101} 285--302.

\bibitem[{Pollard(1982)}]{Pollard:1982}
\text{Pollard, D.} (1982).
\newblock A central limit theorem for empirical processes.
\newblock \textit{Journal of the Australian Mathematical Society,
  \normalfont{Series A}} \textbf{33} 235--248.

\bibitem[{{\v R}eh{\r u}{\v r}ek and Sojka(2010)}]{RehurekSojka:2010}
\text{{\v R}eh{\r u}{\v r}ek, R.} and \text{Sojka, P.} (2010).
\newblock Software framework for topic modelling with large corpora.
\newblock In \textit{{Proceedings of the LREC 2010 Workshop on New Challenges
  for NLP Frameworks}}. ELRA, Valletta, Malta.

\bibitem[{Roy and Hobert(2007)}]{RoyHobert:2007}
\text{Roy, V.} and \text{Hobert, J.~P.} (2007).
\newblock Convergence rates and asymptotic standard errors for {MCMC}
  algorithms for {B}ayesian probit regression.
\newblock \textit{Journal of the Royal Statistical Society,
  \normalfont{Series~B}} \textbf{69} 607--623.

\bibitem[{Ruppert et~al.(2003)Ruppert, Wand and
  Carroll}]{RuppertWandCarroll:2003}
\text{Ruppert, D.}, \text{Wand, M.} and \text{Carroll, R.} (2003).
\newblock \textit{Semiparametric Regression}.
\newblock Cambridge University Press, Cambridge.

\bibitem[{Smith and Kohn(1996)}]{SmithKohn:1996}
\text{Smith, M.} and \text{Kohn, R.} (1996).
\newblock Nonparametric regression using {B}ayesian variable selection.
\newblock \textit{Journal of Econometrics} \textbf{75} 317--343.

\bibitem[{Stark et~al.(1997)Stark, Ryan, McDonald and Burge}]{StarkEtal:1997}
\text{Stark, P.~C.}, \text{Ryan, L.~M.}, \text{McDonald, J.~L.} and
  \text{Burge, H.~A.} (1997).
\newblock Using meteorologic data to model and predict daily ragweed pollen
  levels.
\newblock \textit{Aerobiologia} \textbf{13} 177--184.

\bibitem[{Sung and Geyer(2007)}]{SungGeyer:2007}
\text{Sung, Y.~J.} and \text{Geyer, C.~J.} (2007).
\newblock Monte {C}arlo likelihood inference for missing data models.
\newblock \textit{The Annals of Statistics} \textbf{35} 990--1011.

\bibitem[{Tan and Hobert(2009)}]{TanHobert:2009}
\text{Tan, A.} and \text{Hobert, J.~P.} (2009).
\newblock Block {G}ibbs sampling for {B}ayesian random effects models with
  improper priors: convergence and regeneration.
\newblock \textit{Journal of Computational and Graphical Statistics}
  \textbf{18} 861--878.

\bibitem[{Tan(2014)}]{Tan-Zhiqiang:2014}
\text{Tan, Z.} (2014).
\newblock Self-adjusted mixture sampling and locally weighted histogram
  analysis.
\newblock Tech. rep., Technical Report, Department of Statistics, Rutgers
  University.

\bibitem[{van~der Vaart and Wellner(1996)}]{vanderVaartWellner:1996}
\text{van~der Vaart, A.~W.} and \text{Wellner, J.~A.} (1996).
\newblock \textit{Weak Convergence and Empirical Processes, With Applications
  to Statistics}.
\newblock Springer-Verlag, New York.

\bibitem[{Wallach et~al.(2009)Wallach, Murray, Salakhutdinov and
  Mimno}]{WallachEtal:2009}
\text{Wallach, H.~M.}, \text{Murray, I.}, \text{Salakhutdinov, R.} and
  \text{Mimno, D.} (2009).
\newblock Evaluation methods for topic models.
\newblock In \textit{Proceedings of the 26th Annual International Conference on
  Machine Learning}. ACM.

\bibitem[{Wellner(2005)}]{Wellner:2005}
\text{Wellner, J.} (2005).
\newblock Empirical processes: Theory and applications.
\newline\urlprefix\url{https://www.stat.washington.edu/people/jaw/RESEARCH/TAL%
KS/Delft/emp-proc-delft-big.pdf}

\bibitem[{Wolpert and Schmidler(2012)}]{WolpertSchmidler:2012}
\text{Wolpert, R.~L.} and \text{Schmidler, S.~C.} (2012).
\newblock $\alpha$-stable limit laws for harmonic mean estimators of marginal
  likelihoods.
\newblock \textit{Statistica Sinica} \textbf{22} 1233--1251.

\bibitem[{Zellner(1986)}]{Zellner:1986}
\text{Zellner, A.} (1986).
\newblock On assessing prior distributions and {B}ayesian regression analysis
  with $g$-prior distributions.
\newblock In \textit{Bayesian Inference and Decision Techniques: Essays in
  Honor of Bruno de Finetti} (P.~K. Goel and A.~Zellner, eds.). Elsevier, New
  York.

\end{thebibliography}

\end{document}